\documentclass[secnumarabic,12pt,tightenlines,eqsecnum,floats,aps,amsmath,amssymb,nofootinbib,superscriptaddress,showpacs,prd]{revtex4-1}

\usepackage{amsmath,amsthm,amssymb,amsfonts}
\usepackage{graphicx}
\usepackage{palatino}
\usepackage{supertabular}
\usepackage[mathcal]{euler}
\usepackage{mathcomp}
\usepackage{float}
\usepackage{units}
 \usepackage{hyperref}

\numberwithin{equation}{section}

\begin{document}

\title{Hamiltonian treatment of linear field theories in the presence of boundaries: a geometric approach}

\author{J. Fernando \surname{Barbero G.}}
\email[]{fbarbero@iem.cfmac.csic.es}
\affiliation{Instituto de
Estructura de la Materia, CSIC, Serrano 123, 28006 Madrid, Spain}
\affiliation{Grupo de Teor\'{\i}as de Campos y F\'{\i}sica Estad\'{\i}stica, Instituto Universitario Gregorio Mill\'an
Barbany, Universidad Carlos III de Madrid, Unidad Asociada al IEM-CSIC.}

\author{Jorge \surname{Prieto}}
\email[]{jorgeprietoarranz@gmail.com}
\affiliation{Instituto Gregorio
Mill\'an, Grupo de Modelizaci\'on y Simulaci\'on Num\'erica,
Universidad Carlos III de Madrid, Avda. de la Universidad 30, 28911
Legan\'es, Spain}

\author{Eduardo J. \surname{S. Villase\~nor}}
\email[]{ejsanche@math.uc3m.es}
\affiliation{Instituto Gregorio
Mill\'an, Grupo de Modelizaci\'on y Simulaci\'on Num\'erica,
Universidad Carlos III de Madrid, Avda. de la Universidad 30, 28911
Legan\'es, Spain}
\affiliation{Grupo de Teor\'{\i}as de Campos y F\'{\i}sica Estad\'{\i}stica, Instituto Universitario Gregorio Mill\'an Barbany, Universidad Carlos III de Madrid, Unidad Asociada al IEM-CSIC.}

\begin{abstract}
The purpose of this paper is to study in detail the constraint structure of the Hamiltonian and symplectic-Lagrangian descriptions for the scalar and electromagnetic fields in the presence of spatial boundaries. We carefully discuss the implementation of the geometric constraint algorithm of Gotay, Nester and Hinds with special emphasis on the relevant functional analytic aspects of the problem. This is an important step towards the rigorous understanding of general field theories in the presence of boundaries, very especially when these fail to be regular. The geometric approach developed in the paper is also useful with regard to the interpretation of the physical degrees of freedom and the nature of the constraints when both gauge symmetries and boundaries are present.
\end{abstract}

\maketitle

\section{Introduction}{\label{intro}}

The study of field theories in the presence of boundaries has received some attention in the past, in particular regarding the interpretation of boundary conditions as constraints in their Hamiltonian formulations. This has an obvious importance for quantization as  constraints must be incorporated in one way or another. Numerous investigations about this problem have appeared in the literature; see for example \cite{Zab,SheikhJabbari:1999xd,Ardalan:1999av,Chu:1999gi,Kim:1999qs,Lee:1999kj,Sniatycki:1993wb,Sniatycki:1999vu,Ashtekar:1999wa,Engle:2010kt} and references therein. Some of these works, in particular those devoted to study the integrability (existence and uniqueness problems) of the Yang-Mills equations in bounded domains,   take into consideration the relevant functional analytic issues \cite{Sniatycki:1993wb,Sniatycki:1999vu}. However the details of the Hamiltonian formulation, in particular its systematic derivation and the role of the constraints, are missing. These kind of questions, on the other hand, have been addressed by other authors for simpler field theories \cite{Zab,SheikhJabbari:1999xd}. The problem with these papers is that they consider these issues from a rather formal point of view that avoids the discussion of important mathematical points. We feel that this precludes the correct identification of the physical degrees of freedom and the geometric classification of the constraints, in particular the role of boundary conditions. It is the purpose of this paper to fill in these gaps.

We will consider two types of field theories in the presence of boundaries: free scalar fields and electromagnetism (a gauge theory). Although scalars provide the simplest field theories, their careful study is useful in order to highlight a number of non-trivial issues associated with infinite dimensional systems. In particular, the fact that the symplectic form --one of the main elements of the Hamiltonian formulation-- is weakly non-degenerate, the necessity to precisely describe the domains of the Hamiltonian vector fields that define the dynamics (very specially, their closure in the appropriate topologies), and the interpretation and classification of the constraint manifolds. These functional analytic details play a crucial role in the geometric constraint algorithm for systems with an infinite number of degrees of freedom \cite{GotayNesterHinds}. The electromagnetic field requires extra care in its treatment because it has a (degenerate) presymplectic form. It is hence useful in order to study issues related to the presence of gauge symmetries. The fact that both models are linear, partially simplifies the analysis that we carry out here, however, we feel that the correct understanding of these cases is a necessary first step towards the consideration of more complicated models such as non-linear field theories in the presence of boundaries (Yang-Mills, general relativity, brane models,...).

The identification and interpretation of physical degrees of freedom for field theories is a relatively subtle problem because functional analytic issues must be unavoidably taken into account. Naively one thinks of the configuration of the system as given by \textit{fields} thought of as functions from space (or spacetime) to the set of real or complex numbers (or some finite dimensional vector space when considering tensor fields). In many situations, little attention is paid to the concrete smoothness properties of these fields. A common attitude is to believe that for sensible physical systems it is always possible to specify the mathematical conditions that render all the formal computations correct and all the relevant objects well defined. Although this turns out to be the case in many occasions, important physical features can be missed if these conditions are not explicitly considered. One such issue is, precisely, the identification of the \textit{independent} degrees of freedom. There are situations, relevant for field theories, in which the precise description of the degrees of freedom may be important and unavoidable. For example, the holographic models, where they are associated with hypersurfaces (or boundaries) of spacetime, come immediately to mind. Another interesting example is provided by quantum black hole models, in particular those that make use of inner boundaries for their description \cite{Ashtekar:1997yu,Agullo:2008yv,Engle:2009vc}. A precise determination and understanding of the quantum states requires the obtention of the independent classical configurations of the system, i.e. of its \textit{ classical degrees of freedom}. Finally, the existence of boundary degrees of freedom is common in condensed matter systems (see, for example, \cite{Fröhlich1991369,Balachandran:1994vi}).

The interplay between the presence of boundaries and gauge symmetries is an interesting subject --that actually motivates part of the present work-- because it seems natural to interpret boundary conditions as constraints on the field configurations. Constraints in the Hamiltonian formulation come in different guises that receive different treatments, in particular with regard to quantization. Here the classification of constraints as second or first class plays an important role \cite{Dirac1}. From a quantum point of view a popular approach is the so called Dirac quantization that requires the quantization of concrete mathematical expressions for the first class constraints and the identification of the kernels of the resulting operators. The vectors in these kernels play the role of physical states. Of course there is always the possibility of attempting to describe (in a manageable form) the so called reduced phase space and avoid some problems, in particular those related to gauge invariance and the arbitrariness of the particular description of the constraint manifold in terms of constraint functions. However, although this is possible on general grounds, it is very hard to do in practice. In any case, understanding the precise geometrical nature of the constraint submanifold in phase space is a necessary first step towards this goal.

The standard approach to get the Hamiltonian description for singular Lagrangians relies on the celebrated Dirac algorithm \cite{Dirac1}. Although this method provides a convenient way to deal with mechanical systems with a finite number of degrees of freedom it is much harder to use for field theories, in particular, if functional analytic issues have to be taken into account. In the late sixties and seventies, these difficulties lead to an effort to achieve a geometric (i.e. intrinsic, coordinate independent, and global) understanding of Hamiltonian systems that culminated in the development of the Gotay-Nester-Hinds (GNH) geometric constraint algorithm \cite{GotayNesterHinds,Gotaythesis,Gotay:1978dv} that we will use in the present paper. This approach has many advantages because it provides a very clear geometric point of view that makes it possible to incorporate, in a natural way, analytic aspects that are important both for regular and singular systems. In particular the relevant functional spaces can be identified in a systematic way in each step of the algorithm. It also generalizes the Dirac algorithm in some respects, for instance, its starting point is $(\mathcal{M},\omega,\alpha)$ where $(\mathcal{M},\omega)$  is a presimplectic manifold and $\alpha$ is 1-form that defines the dynamics (usually, $\alpha$ is the exterior derivative of the Hamiltonian). Hence, it can be used not only to deal with the standard Hamiltonian formulation, where $\mathcal{M}$ is taken to be the primary constraint submanifold,  but also leads to a purely symplectic-Lagrangian approach that may be useful for some physical applications. In this paper we will consider both points of view.

The structure of the paper is the following. After this introduction, Section \ref{general} discusses the general approach that we will use. In particular we review the GNH algorithm, the geometric classification of the constraint submanifolds, and discuss the variational, Hamiltonian and symplectic-Lagrangian points of view. After this, Sections \ref{Scalar} and \ref{EM} study in detail the scalar and the electromagnetic fields in the presence of boundaries subject to several important types of boundary conditions. We will carefully discuss here the constraint manifolds --where the boundary conditions play a central role-- and their classification as first or second class. We end the paper in Section \ref{conclusions} with the discussion of the main results and our conclusions. We provide several appendices where relevant background material is explained in some detail for the convenience of the reader. Appendix \ref{Sobolevspaces} introduces the functional spaces that are used throughout the paper and fixes notation. Appendix \ref{functionalspaces} compiles the proofs of several results that are necessary for the implementation of the GNH algorithm for the electromagnetic field. Finally, Appendix \ref{abstractwave} discusses the abstract wave equation \cite{MarsdenHughes}, which provides the framework for a wide class of free field theories.

\section{General framework}\label{general}

The goal of this section is to describe, in some detail, the general framework that we will use in the examples provided by the scalar and electromagnetic fields in the presence of boundaries. We will follow three different approaches to the study of the dynamics of these models: the variational, the Hamiltonian and the symplectic-Lagrangian. As mentioned in the introduction, field theories with an infinite number of degrees of freedom must be described with the help of infinite dimensional manifolds. Here, following \cite{Gotaythesis}, we will use Banach manifolds (see Appendix \ref{Sobolevspaces}). Many mathematical subtleties in this case stem precisely from the need to use these infinite dimensional spaces. We carefully discuss these issues in the following.

\subsection{The variational Lagrangian approach}

The variational Lagrangian approach is an essentially analytic point of view that provides a way to obtain the evolution equations for a classical system in many important instances. It consists of the following steps:

    \begin{itemize}
    \item The identification of the configuration space $\mathcal{Q}$. For standard mechanical systems this is a finite dimensional differentiable manifold. In the field theories that we will discuss in this paper these configuration spaces will be  Banach manifolds (actually, real Hilbert spaces).

   \item The introduction of the Lagrangian $L$ as a real differentiable function\footnote{We will not consider here time-dependent Lagrangians (that should be defined on $T\mathcal{Q}\times \mathbb{R}$).} $L:\mathfrak{D}\rightarrow \mathbb{R}$ on a domain $\mathfrak{D}\subset T\mathcal{Q}$ (generically, in infinite dimensional systems, the topology of $\mathfrak{D}$ will not be the one induced  from $T\mathcal{Q}$).  In finite dimensional systems one usually has $\mathfrak{D}= T\mathcal{Q}$. A common feature shared by the concrete  models that we study in the following is the fact that their Lagrangians cannot be defined on the whole of $T\mathcal{Q}$. This happens because the field equations for these models involve spatial derivatives and require some additional smoothness conditions on the fields in $\mathcal{Q}$. The appropriate subsets of $T\mathcal{Q}$, suitable to describe the dynamics, are constructed  in terms of \textit{manifold domains} $\mathcal{D}$ of the configuration manifold $\mathcal{Q}$ \cite{ChernoffMarsden}. In this framework the domains $\mathfrak{D}$ are of the form $\mathfrak{D}=T_\mathcal{D}\mathcal{Q}=\bigcup_{Q\in \mathcal{D}} T_Q\mathcal{Q}$ and are (generalized) subbundles of $T\mathcal{Q}$. When $\mathcal{Q}$ is a Banach space the tangent bundle is the product $T\mathcal{Q}=\mathcal{Q}\times \mathcal{Q}$ and the manifold domains have the form $T_\mathcal{D}\mathcal{Q}:=\mathcal{D}\times \mathcal{Q}$, where $\mathcal{D}$ are Banach spaces in their own right that can be considered as dense subspaces of the configuration space $\mathcal{Q}$. We  will describe in detail the manifold domains for each of the models that we consider in the paper.

    \item Fixing two points $Q_1,Q_2\in \mathcal{Q}$ that define the initial and final configurations of the system at two time instants $t_1<t_2$. We consider a set of possible trajectories (parametrized curves) joining them $\Phi:[t_1,t_2]\subset\mathbb{R}\rightarrow \mathcal{Q}$, i.e. satisfying the conditions $\Phi(t_1)=Q_1\,,\Phi(t_2)=Q_2$. In practice we will need to impose some additional, physically motivated, smoothness conditions on $\Phi$ (by demanding, for example, that the trajectories and the corresponding velocities are continuous). These are required to make mathematical sense of the particular form of the action that defines the dynamics (if it involves times derivatives, for instance) and also to be able to characterize its stationary points through the Euler-Lagrange equations. For a finite dimensional mechanical system a standard and convenient choice \cite{AbrahamMarsden} is $\mathcal{C}(Q_1,Q_2,[t_1,t_2]):=\{\Phi\in C^2([t_1,t_2],\mathcal{Q})|\Phi(t_1)=Q_1\,,\Phi(t_2)=Q_2\}$.
        This space can be endowed with an appropriate Banach manifold structure \cite{Eells} and provides a very convenient setup to study the problem of finding the stationary points of the action. In the case of field theories, however, one is forced to use spaces of curves with a more complicated structure. This is a consequence of the fact that the Lagrangians involve field derivatives and it is often necessary to  simultaneously consider several types of Hilbert spaces with different topologies.

    \item Introduce the action $S:\mathcal{C}(Q_1,Q_2,[t_1,t_2])\rightarrow \mathbb{R}$ as the function.
    $$
    S(\Phi):=\int_{t_1}^{t_2}L(\Phi(t),\dot{\Phi}(t))\mathrm{d}t\,.
    $$
    We will follow the standard custom of referring to the action as a \textit{functional} to highlight the fact that it depends on curves.

    \item Finally, look for stationary points of this function. These are defined as those curves in the domain of $S$ where the action is differentiable and the differential is zero. The stationary curves, $\Phi\in\mathcal{C}(Q_1,Q_2,[t_1,t_2])$, are the physical trajectories connecting $Q_1$ at $t_1$ and $Q_2$ at $t_2$ corresponding to the dynamics defined by the action. These are solutions to second order differential equations.
    \end{itemize}

The second order differential equations that describe the dynamics of the system in this context can be studied with the standard analytic tools in order to derive existence, uniqueness and regularity results, find their solutions or, at least, get a sufficient understanding of their behavior.

\subsection{The Hamiltonian approach and the GNH algorithm}

The road from the Lagrangian to the Hamiltonian frameworks in classical mechanics --as explained in the standard textbooks-- is a well-trodden one. By defining a suitable momentum variable and performing a Legendre transform, one arrives at the Hamiltonian formulation where the equations of motion take a pleasingly simple \emph{canonical} form and the dynamics is encoded in a single function, the Hamiltonian.

Hamiltonian dynamics is defined in the cotangent bundle $T^*\mathcal{Q}$ of the configuration space. For Banach manifolds $T^*\mathcal{Q}$ carries a canonical, weakly non-degenerate, symplectic form $\Omega\in\Lambda^2(T^*\mathcal{Q})$ i.e. such that the vector bundle map
$$\flat:T(T^*\mathcal{Q})\rightarrow T^*(T^*\mathcal{Q}):X\mapsto \flat(X)=i_X\Omega$$
is injective. When $\flat$ is a linear bundle isomorphism the symplectic form is said to be strongly non-degenerate. This happens if $\mathcal{Q}$ is reflexive and, in particular, for finite dimensional mechanical systems \cite{ChernoffMarsden}. Notice, however, that even if $\mathcal{Q}$ is reflexive (as in the examples that we consider in the paper) it may be unavoidable to work in a manifold domain of $\mathcal{Q}$. This means that the phase space will actually be of the form $T^*_{\mathcal{D}}\mathcal{Q}=\bigcup_{Q\in \mathcal{D}} T^*_Q\mathcal{Q}\subset T^*\mathcal{Q}$ and, hence, the symplectic form that one has to use (the pullback of the canonical form in $T^*\mathcal{Q}$ to $T^*_{\mathcal{D}}\mathcal{Q}$) will generically be only weakly-nondegenerate.

From a geometric point of view \cite{AbrahamMarsden}, the transition from the Lagrangian to the Hamiltonian formulations is carried out by means of the so called fiber derivative $FL$ of the Lagrangian $L$:
$$FL:T_\mathcal{D}\mathcal{Q}\rightarrow T_\mathcal{D}^*\mathcal{Q}\,,\quad w\mapsto FL(w)\,,$$
defined by
$$
\langle v\,|\, FL(w)\rangle:=\left.\frac{\mathrm{d}}{\mathrm{d}t}\right|_{t=0} L(w+t v)
$$
where $\langle \cdot \,|\,  \cdot \rangle$ is the natural pairing between elements of $T\mathcal{Q}$ and $T^*\mathcal{Q}$ over the same base point. The fiber derivative is used to define the canonical momenta and the Hamiltonian. In fact, when it exists, the Hamiltonian $H$ can be written in a purely geometric, coordinate independent, way  as
\begin{eqnarray*}
H\circ FL(w)= \langle w \,|\,  FL(w)\rangle -L(w)\,,\quad w\in T_\mathcal{D}\mathcal{Q}\subset T\mathcal{Q}\,.
\end{eqnarray*}
For the so called \textit{regular (hyperregular) systems} the fiber derivative is a local (global) diffeomorphism. In the remaining cases the Lagrangian is said to be \textit{singular}.

In order to obtain a Hamiltonian description of the dynamics defined by a given Lagrangian one would naively consider the following two steps:
\begin{itemize}
\item The determination of the Hamiltonian vector field $X\in\mathfrak{X}(T_{\mathcal{D}}^*\mathcal{Q})$ associated with the Hamiltonian of the system. This is obtained as the solution to the equation
    \begin{equation}
    i_X \Omega=dH\,,
    \label{Hamiltoneq}
    \end{equation}
    where $i_X$ denotes the interior product of $X$ and $\Omega$.
\item The obtention of the integral curves of $X$ that describe the time evolution of the system (by projection onto the configuration space $\mathcal{Q}$).
\end{itemize}

In the case of finite-dimensional, (hyper)regular, mechanical systems equation (\ref{Hamiltoneq}) is rather trivial because the domain of the Hamiltonian is $T^*\mathcal{Q}$ and the canonical symplectic form can be easily seen to be strongly non-degenerate. However, for singular finite systems this is usually not the case because the  domain of the Hamiltonian is a proper subset of $T^*\mathcal{Q}$ and, hence, the pull-back of the canonical symplectic form to it may be degenerate. As a consequence, the resolution of (\ref{Hamiltoneq}) requires some attention.

A common situation that one encounters in regular field theories is that the symplectic form in $T^*_\mathcal{D}\mathcal{Q}$ is only \textit{weakly non-degenerate}. In this case one must study if the 1-form $\mathrm{d}H$ lies in the range of the $\flat$ map in order to be able to solve equation (\ref{Hamiltoneq}). If it does not, a possible approach to the problem is to restrict (\ref{Hamiltoneq}) and define the Hamiltonian dynamics of the system in an appropriate subset of the phase space. This is part of the content of the algorithm\footnote{Although, as we have mentioned in Section \ref{intro},  the original algorithm is developed to be used in a more general context, it can be easily adapted to the usual Hamiltonian framework, as we discuss here, by following the ideas introduced in \cite{Gotaythesis}.} developed by Gotay, Nester and Hinds in \cite{GotayNesterHinds, Gotay:1978dv} to deal with a wide class of field theories.

Before we discuss it, some comments are in order. If the range of the fiber derivative $FL(T_\mathcal{D}\mathcal{Q})$ is a proper submanifold of $T_\mathcal{D}^*\mathcal{Q}$, according to its definition, the Hamiltonian
$$
H:FL(T_{\mathcal{D}}\mathcal{Q})\subset T_\mathcal{D}^*\mathcal{Q} \rightarrow \mathbb{R}
$$
is only defined there. $FL(T_\mathcal{D}\mathcal{Q})$ is known in the literature as the \textit{primary constraint} manifold and is the starting point of the algorithm developed by Dirac in \cite{Dirac1}. An important element in Dirac's approach to the quantization of constrained systems was his insistence in working in the \textit{full} phase space $T^*\mathcal{Q}$. His main idea was to find conditions (constraints) defining physical configurations, turn a certain generalization of the Poisson brackets (the so called Dirac brackets) into commutators, quantize the constraints and select appropriate physical states by considering their kernels. The insistence on working in the full phase space required the extension of the Hamiltonian from the primary constraint surface to the full $T^*\mathcal{Q}$. This can usually be done in many ways. It is actually advantageous to consider a large family of possible extensions and then restrict them, if necessary, to guarantee the consistency of the dynamics on an appropriate subset of the primary constraint manifold. In practice, this is done by choosing a particular extension of the Hamiltonian and adding to it a linear combination of independent functions, all of them vanishing on this primary constraint manifold, with arbitrary coefficients (Lagrange multipliers). An inconvenient feature of this approach is that it is local and lacks a clear geometric justification.

The approach that we will follow --based in the Gotay-Nester-Hinds (GNH) algorithm-- will be slightly different, definitely more geometrical, and \textit{global}. Our goal will not be a Dirac-like description in the full phase space but, rather, the identification of a suitable geometric domain in $T_\mathcal{D}^*\mathcal{Q}$ (in practice, a new phase space) where the dynamics is well defined, together with the identification of the vector fields whose integral curves will give the time evolution of the system. This is enough for the description of the reduced phase space and is a possible starting point for quantization.

\bigskip

\noindent \textbf{GNH algorithm for finite dimensional systems:} We consider first the case in which  $\mathcal{Q}$ is a \textit{finite dimensional} smooth manifold (and hence we will assume that $\mathcal{D}=\mathcal{Q}$). Let the presymplectic system $(\mathcal{M},\omega,\mathrm{d}H)$ consisting on the primary constraint submanifold $$\mathcal{M}:=FL(T\mathcal{Q})\subset T^*\mathcal{Q},$$  the pull-back $\omega$ of the canonical symplectic form $\Omega$ in $T^*\mathcal{Q}$ to $\mathcal{M}$, and the uniquely defined Hamiltonian $H: \mathcal{M}\rightarrow \mathbb{R}$.

The goal of the GNH algorithm is to find the \emph{maximal} submanifold $\mathcal{N}\subset \mathcal{M}$ where the equation
\begin{eqnarray}
(i_{X}\omega-\mathrm{d}H)|_{\mathcal{N}}=0\,,\label{HVF}
\end{eqnarray}
can be solved and gives rise to first order evolution equations on $\mathcal{N}$, in the sense that the solutions to (\ref{HVF}) are vector fields $X:\mathcal{N}\rightarrow T\mathcal{N}$.

We start by defining $\mathcal{M}_1:=\mathcal{M}=FL(T\mathcal{Q})$ and considering the set
$$
\mathcal{M}_{2}:=\{m\in \mathcal{M}_1\,:\, \mathrm{d}H(m)\in \mathrm{range}(\flat(m)) = \flat( T_m\mathcal{M}_1)=: T_m\mathcal{M}_1^\flat\}\subset \mathcal{M}_1\,,
$$
where the map $\flat$ is defined here as $\flat(m): T_m\mathcal{M}_1\rightarrow T^*_m\mathcal{M}_1:\,\, X\mapsto i_X \omega$. It is clear that $\mathcal{M}_{2}$ is the set of points  for which the equation
 \begin{eqnarray}
i_{X}\omega-\mathrm{d}H=0\label{Xec}
\end{eqnarray}
can be solved. For convenience, in the finite dimensional case, we will assume $\mathcal{M}_1$, $\mathcal{M}_2$, and all $\mathcal{M}_k$ that will be defined below, to be an \textit{embedded} submanifolds of $T^*\mathcal{Q}$. The solution will be a map of the form $X:\mathcal{M}_{2}\rightarrow T\mathcal{M}_{1}$ and will be generically non-unique and not \textit{tangent} to  $\mathcal{M}_{2}$ (i.e. $X(m)\not\in T_m \mathcal{M}_{2}$). The failure to be tangent to $\mathcal{M}_{2}$ means that any consistent dynamics must be restricted to a new submanifold $\mathcal{M}_{3}\subset \mathcal{M}_{2}$ where $X$ is tangent to $\mathcal{M}_{2}$:
$$
\mathcal{M}_{3}:=\{m\in \mathcal{M}_2\,:\, \mathrm{d}H(m)\in  T_m\mathcal{M}_2^\flat\}\subset \mathcal{M}_2\subset \mathcal{M}_1\,.
$$
 If in addition to be tangent to $\mathcal{M}_{2}$, the vectors $X$ that solve at this stage the equation
 $$
 (i_{X}\omega-\mathrm{d}H)|_{ \mathcal{M}_2}=0
 $$
 are also tangent to $\mathcal{M}_{3}$ the dynamics can be consistently defined there, if not the procedure must be iterated until it gives a suitable manifold. The previous procedure amounts to building the sets
$$
\mathcal{M}_{k+1}:=\{m\in\mathcal{M}_k: \mathrm{d}H(m)\in T_m\mathcal{M}_k^\flat\}\subset \mathcal{M}_k\subset \mathcal{M}_{k-1}\subset\cdots \subset \mathcal{M}_1\,,
$$
obtaining  the general solution to
$$
 (i_{X}\omega-\mathrm{d}H)|_{ \mathcal{M}_k}=0
$$
and finding  (if it exists)  the smallest $n\geq 1$ such that $\mathcal{M}_{n+1}=\mathcal{M}_n$. If this is possible, the manifold $\mathcal{N}:=\mathcal{M}_n$ and the (generically non-unique) vector fields $X:\mathcal{N}\rightarrow T\mathcal{N}$ that solve (\ref{HVF}) constitute the Hamiltonian description of the system.

Notice that in the finite dimensional case we have assumed that no topological complications arise in the chain
$$
\mathcal{N}=\mathcal{M}_n\subset \mathcal{M}_{n-1}\subset\cdots \subset \mathcal{M}_1 = \mathcal{M}\subset T^*\mathcal{Q}
$$
as the topologies are the natural ones induced by the manifold structure of $T^*\mathcal{Q}$. It is important to point out that some of the possible arbitrariness in the vector fields $X$ that solves (\ref{Xec}) at the first stage of the algorithm may disappear during the process of finding the sets $\mathcal{M}_k$. If some of it remains at the end we will have a gauge freedom in our model. It is also interesting to highlight the different origins and geometrical meaning of the \textit{primary constraint} submanifold $\mathcal{M}_1:=FL(T\mathcal{Q})$ and the others $\mathcal{M}_{k}$, $k\geq 2$, that just beg to be called \textit{secondary constraint} submanifolds.

\bigskip

\noindent \textbf{GNH algorithm for infinite dimensional systems:} In the infinite-dimensional case  the situation is subtler due to a number of issues. To list a few of them:
\begin{itemize}
\item First, we need to choose the class of infinite dimensional real manifolds that we will use to model the system. In the following, we will assume that $\mathcal{Q}$ is a Banach manifold (and, hence $T\mathcal{Q}$ and $T^*\mathcal{Q}$ are also Banach manifolds). Notice that  norms in infinite dimensional vector spaces are not necessarily equivalent, hence, it is important to specify upon which particular Banach space a Banach manifold is modeled.
\item The preceding procedure is not guaranteed to terminate and its naive generalization to the infinite dimensional case may lead to difficulties. In fact this happens even for such simple models as the free scalar field \cite{Gotaythesis}. In particular, the final arena for the dynamics may end up not being a Banach manifold (although in many cases it is a Fr\'echet manifold). A way to avoid some of these problems, introduced by Gotay in \cite{Gotaythesis}, consists in relaxing the strict tangency requirement of the finite-dimensional case to increase the chances that the procedure ends in a finite number of steps. Specifically, the requirement $X:\mathcal{N}\rightarrow T\mathcal{N}$ will be relaxed to $X:\mathcal{N}\rightarrow \underline{\overline{T}\mathcal{N}}$ as explained below (see also the discussion of the integral curves of the abstract wave equation in Appendix \ref{abstractwave}).
\item The natural topologies of the intermediate spaces $\mathcal{M}_k$ that will appear in the construction must be carefully taken into account.
\item Notice that in the infinite dimensional case, even for regular systems, if the symplectic form is only weakly non-degenerate the resolution of equation (\ref{Xec}) will generically lead to restrictions on the domain of the vector fields $X$ necessary to guarantee that $\mathrm{d}H$ is in the image of the $\flat$ map.
\end{itemize}

To generalize the preceding construction to the infinite-dimensional case let us consider $(\mathcal{M},\omega,\mathrm{d}H)$ where $\mathcal{M}=FL(T_\mathcal{D}\mathcal{Q})\subset T_\mathcal{D}^*\mathcal{Q}$, $\omega$ is the pullback of the canonical sympectic form in $T^*\mathcal{Q}$ to $\mathcal{M}$, and $H:\mathcal{M}\rightarrow \mathbb{R}$ the Hamiltonian. We will assume $\mathcal{M}_1:=\mathcal{M}$
to be a Banach manifold modeled on a Banach space $F_1$ and also  that
$$
\mathcal{M}_{2}:=\{m\in \mathcal{M}_1\,:\, \mathrm{d}H(m)\in T_m\mathcal{M}_1^\flat\}\subset \mathcal{M}_1\,,
$$
can be endowed with a Banach manifold structure, with model Banach space $F_2$, such that the inclusion $\jmath_2:\mathcal{M}_2\rightarrow \mathcal{M}_1$ is smooth (with the topologies on $\mathcal{M}_1$ and $\mathcal{M}_2$ given by the respective Banach manifold structures). Notice that, although $\mathcal{M}_2\subset \mathcal{M}_1$ as sets, in general the topology of $\mathcal{M}_2$ is not the induced topology from $\mathcal{M}_1$. Hence, in general, $\mathcal{M}_2$ is not an embedded submanifold of $\mathcal{M}_1$. As in the finite dimensional case, the definition of $\mathcal{M}_2$ must be understood as the solvability condition for $X$ in the equation
\begin{equation}
i_X\omega-\mathrm{d}H=0\,.
\label{HVF1}
\end{equation}
The solutions   $X:\mathcal{M}_2\rightarrow T\mathcal{M}_1$ to (\ref{HVF1}) satisfy  $X(m)\in T_{j_2(m)}\mathcal{M}_1$. Now, there may be points $m\in \mathcal{M}_2$ for which  $X(m)\not\in T_{j_2(m)} \overline{\mathcal{M}}_2\subset T_{\jmath_2 (m)}\mathcal{M}_1$. Here $\overline{\mathcal{M}}_2:= \mathrm{cl}_{\mathcal{M}_1}(j_2\mathcal{M}_2)\subset \mathcal{M}_1$ is the topological closure of $j_2\mathcal{M}_2$ in $\mathcal{M}_1$, that we will assume to be an embedded submanifold of $\mathcal{M}_1$. If this is the case, we will say that the vector field $X$ does not define first order evolution equations on $\mathcal{M}_2$, and we will further restrict the set of points and the possible vector fields to
$$
\mathcal{M}_3:=\{m\in \mathcal{M}_2\,:\, \mathrm{d}H(m) \in \underline{\overline{T}\mathcal{M}_2}^\flat \}\,,
$$
where
$$
\underline{\overline{T}\mathcal{M}_2}:= T \overline{\mathcal{M}}_2|_{j_2(\mathcal{M}_2)} \subset T\mathcal{M}_1\,, \quad \underline{\overline{T}\mathcal{M}_2}^\flat:= \flat \left( \underline{\overline{T}\mathcal{M}_2}\right)\subset \flat(T\mathcal{M}_1)\,.
$$
We will assume that $\mathcal{M}_3$ can be endowed with a Banach manifold structure, with model Banach space $F_3$, such that the inclusion $\mathcal{M}_3\stackrel{\jmath_3}{\longrightarrow} \mathcal{M}_{2}$
is smooth. In general, given the Banach manifolds $\mathcal{M}_k$, with Banach model spaces $F_k$, we will assume that
$$
{\mathcal{M}}_{k+1}:=\{m\in\mathcal{M}_k\,:\, \mathrm{d}H(m)\in \underline{\overline{T}\mathcal{M}_k}^\flat\}
$$
can be endowed with Banach manifold structures with model Banach spaces $F_{k+1}$ in such a way that
\begin{eqnarray*}
\mathcal{M}_{k+1}\stackrel{\jmath_{k+1}}{\longrightarrow} \mathcal{M}_{k}\stackrel{\jmath_{k}}{\longrightarrow} \cdots\stackrel{\jmath_3}{\longrightarrow}\mathcal{M}_2\stackrel{\jmath_2}{\longrightarrow}\mathcal{M}_1
\end{eqnarray*}
is a chain of smooth injective immersions $\jmath_i:\mathcal{M}_i\rightarrow \mathcal{M}_{i-1}$. Here we have used the notation
$$
\underline{\overline{T}\mathcal{M}_k}:= T \overline{\mathcal{M}}_k|_{\jmath_2\circ\cdots \circ \jmath_k(\mathcal{M}_k)} \subset T\mathcal{M}_1\,, \quad \underline{\overline{T}\mathcal{M}_k}^\flat:=\flat\left(\underline{\overline{T}\mathcal{M}_k}\right) \subset \flat(T\mathcal{M}_1)
$$
where $\overline{\mathcal{M}}_k :=\mathrm{cl}_{\mathcal{M}_1}(\jmath_2\circ\cdots\circ \jmath_k(\mathcal{M}_k))\subset \mathcal{M}_1$.

If it exists, the smallest $n\geq 1$ such that $\mathcal{M}_{n+1}=\mathcal{M}_n\neq \emptyset$ provides the maximal generalized submanifold $\mathcal{N}:=\mathcal{M}_n\subset \mathcal{M}$, with smooth inclusion $\jmath=\jmath_2\circ\cdots\circ \jmath_n$,  and the (generically non-unique) vector fields $X:\mathcal{N}\rightarrow \underline{T\overline{\mathcal{N}}}\subset T\mathcal{M}$  that  constitute the Hamiltonian description of the system. If it does not, the system may be inconsistent or one could be forced to define the dynamics  outside the class of Banach manifolds. We will not need to consider these situations in the present paper.

\bigskip

\noindent \textbf{Geometric classification of the constraint submanifolds:} The generalized submanifold $\mathcal{N}\stackrel{\jmath}{\rightarrow}\mathcal{M}$ of the presymplectic manifold $(\mathcal{M},\omega)$ given by the GNH algorithm consists of those states which are physically realizable.

The intrinsic classification of the constraint submanifolds of a presymplectic manifold was developed by Tulczyjew \cite{Tulczyjew} and Sniatycki \cite{Sniatycki:1974}. This classification scheme generalizes the local classification of the submanifolds of a strongly symplectic manifold given by Dirac in terms of constrain functions. In particular, in this classification, constraint submanifolds can be first class, second class, isotropic, Lagrangian or mixed:
\begin{itemize}
\item First class submanifolds: $\mathcal{N}\stackrel{\jmath}{\rightarrow}\mathcal{M}$ is said to be a (generalized) \emph{first class submanifold} of a presimplectic manifold $(\mathcal{M},\omega)$ if $T\mathcal{N}^\perp\subset \underline{T\mathcal{N}}$.
\item Second class submanifolds: $\mathcal{N}\stackrel{\jmath}{\rightarrow}\mathcal{M}$  is said to be a (generalized) \emph{second class submanifold} of a presimplectic manifold $(\mathcal{M},\omega)$ if $T\mathcal{N}^\perp\cap  \underline{T\mathcal{N}}=\{0\}$.
    \item Isotropic submanifolds: $\mathcal{N}\stackrel{\jmath}{\rightarrow}\mathcal{M}$  is said to be a (generalized) \emph{isotropic submanifold} of a presimplectic manifold $(\mathcal{M},\omega)$ if $\underline{T\mathcal{N}}\subset T\mathcal{N}^\perp$.
          \item Lagrangian submanifolds: $\mathcal{N}\stackrel{\jmath}{\rightarrow}\mathcal{M}$  is said to be a (generalized) \emph{Lagrangian submanifold} of a presimplectic manifold $(\mathcal{M},\omega)$ if $\underline{T\mathcal{N}}= T\mathcal{N}^\perp$.
             \item Mixed submanifolds: $\mathcal{N}\stackrel{\jmath}{\rightarrow}\mathcal{M}$  is said to be a (generalized) \emph{mixed submanifold} of a presimplectic manifold $(\mathcal{M},\omega)$ in the rest of the cases.
\end{itemize}
Here we have used the notation:
$$
T\mathcal{N}^\perp:=\{Z\in T\mathcal{M}|_{\mathcal{N}}\,:\, \omega|_{\mathcal{N}}(Z,X)=0,\,\, \forall X\in\underline{T\mathcal{N}}\}\quad \textrm{ where } \quad \underline{T\mathcal{N}}:= \jmath_*(T\mathcal{N})\,.
$$

\subsection{The symplectic-Lagrangian point of view}

Although there are analytic and geometric elements both in the Lagrangian and Hamiltonian descriptions of mechanics that we have sketched above, it is clear that the Lagrangian variational approach has a very analytic flavor whereas the Hamiltonian approach is basically geometrical. The Hamiltonian framework  relies heavily on the geometry of the cotangent bundle $T^*\mathcal{Q}$ and the naturally defined geometric structures present there (in particular the symplectic form). It is important to point out that such canonical structures do not exist in the tangent bundle. However this does not prevent us from having a geometric description of the dynamics in the tangent bundle similar to the Hamiltonian one. The key idea is to transfer the canonical symplectic form from $T^*\mathcal{Q}$ to $T_\mathcal{D}\mathcal{Q}$ by pulling it back with the help of the fiber derivative to define $\Omega_L:=FL^* \Omega$. Note that this pullback is always well defined (though generically presymplectic)  regardless of the particular properties of $FL$. The role of the Hamiltonian function on the cotangent bundle is now played by a closely related function on $T_\mathcal{D}\mathcal{Q}$ known as the energy\footnote{Notice that $E=H\circ FL$.}
\begin{eqnarray*}
E:T_\mathcal{D}\mathcal{Q}\rightarrow \mathbb{R}:\, w\mapsto\langle w|FL(w)\rangle-L(w)\,.
\end{eqnarray*}
With these ingredients the process of defining the dynamics in geometric terms on the tangent bundle is similar  to the one that we followed in the Hamiltonian case. It suffices to exchange the roles of $(FL(T_\mathcal{D}\mathcal{Q}),\omega,H)$ with $(T_\mathcal{D}\mathcal{Q},\Omega_L,E)$ and use the GNH algorithm as explained above.

Notice the similarity between  the symplectic Lagrangian approach and the Hamiltonian approach, as far as the GNH algorithm is concerned. This can be seen especially at the level of the equations that must be solved to obtain the respective Hamiltonian vector fields. Of course the difficulties associated with their solution are of the same type and, hence, the techniques needed to deal with them can be applied in both frameworks. Other issues such as the second order problem \cite{Gotaythesis} must be eventually addressed but will not play a significant role in the present paper.

\section{The scalar field in the presence of boundaries}\label{Scalar}

This section is devoted to the detailed study of the free scalar field in a bounded space region.
By free we mean that the dynamics is given by the (linear) wave equation. The presence of boundaries requires the careful consideration of the conditions that the fields must satisfy on them in order to guarantee that their evolution is completely determined by initial data. As mentioned in the introduction, one of the goals of this paper is to clarify the possible interpretation of boundary conditions as \textit{constraints}. To this end it is necessary to  use the appropriate mathematical tools. In the present case these issues can be satisfactorily addressed within the framework provided by the differential geometry of infinite dimensional manifolds, in particular those modeled on Banach and Hilbert spaces. The geometric interpretation of the constraint manifolds, and their classification according to the traditional first/second class division, will require the discussion of a number of relatively subtle functional analytic issues that are not relevant for finite dimensional mechanical systems and, for this reason, are usually neglected in formal approaches to this subject.

In the following we will separately consider the three approaches explained in the preceding section: the variational, the Hamiltonian and the Lagrangian-symplectic. Before doing so we establish the basic set up for the problem. To this end let us consider $\mathbb{R}^n$ endowed with the Euclidean metric\footnote{The results of the paper can be easily generalized to curved spatial manifolds. We will refrain from doing so here as this generalization does not change our conclusions regarding the treatment of boundary conditions.} $e_{ij}$ (and covariant derivative $\nabla_i$ such that $\nabla_i e_{jk}=0$) and the corresponding volume $n$-form that we denote as $\mathrm{vol}_\Sigma$. Let us take an open, connected, bounded region $\Sigma\subset\mathbb{R}^n$ with a smooth boundary\footnote{We restrict ourselves to the $C^\infty$ case but most of our results can be extended, with minor modifications, to manifolds with less regular boundaries, for example Lipschitz \cite{Girault}.}  $\partial\Sigma$.

The scalar fields that we are interested in are real functions with domain $\Sigma$. The configurations of the system that we want to study will be scalar fields i.e. real functions subject to smoothness conditions that originate in the fact that they are required to be solutions (classical or weak) to partial differential equations (PDE's) involving both time and space derivatives. We will consider Dirichlet and Robin boundary conditions (the Neumann boundary conditions are contained in the latter).

Our starting point will be the Dirichlet Lagrangian\footnote{We are using the Minkowski metric with signature $(+,-,-,-)$.} $L_{\scriptscriptstyle{D}}:H^1_f(\Sigma)\times L^2(\Sigma)\rightarrow \mathbb{R}$,
defined by
\begin{eqnarray}
L_{\scriptscriptstyle{D}}(Q,V)&=&\frac{1}{2}\langle V,V\rangle_{L^2}-\frac{1}{2}\langle \vec{\nabla}Q,\vec{\nabla}Q\rangle_{\vec{L}^2}\,,\label{lagrangianoescalarD}
\end{eqnarray}
and the Robin Lagrangian $L_{\scriptscriptstyle{R}}:H^1(\Sigma)\times L^2(\Sigma)\rightarrow \mathbb{R}$, given by
\begin{eqnarray}
L_{\scriptscriptstyle{R}}(Q,V)&=&\frac{1}{2}\langle V,V\rangle_{L^2}-\frac{1}{2}\langle \vec{\nabla}Q,\vec{\nabla}Q\rangle_{\vec{L}^2}+
\int_{\partial\Sigma}\big(AQ|_{\partial\Sigma}+\frac{B}{2}Q^2|_{\partial\Sigma}\big)\mathrm{vol}_{\partial\Sigma}\,,
\label{lagrangianoescalarR}
\end{eqnarray}
where
$$
\langle u_1,u_2\rangle_{L^2}:=\int_{\Sigma} u_1 u_2\,\mathrm{vol}_\Sigma\,,\quad \langle \vec{u}_1,\vec{u}_2\rangle_{\vec{L}^2}:=\int_{\Sigma} \vec{u}_1 \cdot \vec{u}_2\,\mathrm{vol}_\Sigma\,.
$$
The domains that we have chosen are the ``largest'' natural ones with the appropriate mathematical structure, in particular they are Hilbert manifolds such that the Lagrangians are smooth functions on them. Here and in the following $H^s(\Sigma)$ denotes the $s$-Sobolev Hilbert space on $\Sigma$ and $$H^1_f(\Sigma):=\{u\in H^1(\Sigma): u|_{\partial\Sigma}=f \}$$
 where $u|_{\partial\Sigma}$ denotes the image of the trace operator $\gamma$ acting on $u$ (see Appendix \ref{Sobolevspaces}). The trace $\gamma:H^1(\Sigma)\rightarrow L^2(\partial\Sigma)$ is a bounded operator that, restricted to continuous functions gives their boundary values \cite{Adams}. Notice that $f$ must be an element of the image of $\gamma$ and $H_f^1(\Sigma)$ is also an affine Hilbert space (it is a closed affine subspace of a Hilbert space). We will take the functions $A,B\in C^\infty(\partial\Sigma)$ and require $B\leq0$ for technical reasons that will be clear later. If $A=B=0$ the $L_{\scriptscriptstyle{R}}$ reduces to the Neumann Lagrangian $L_{\scriptscriptstyle{N}}:H^1(\Sigma)\times L^2(\Sigma)\rightarrow \mathbb{R}$, hence we will not discuss this case separately.

The preceding Lagrangians describe non-homogeneous boundary conditions. No generality is lost in the analysis that we present here by restricting to the homogeneous case, hence in the following we will take $f=0$ for the Dirichlet case and $A=0$ for the Robin one.

In order to make contact with the general discussion of Section \ref{general} we introduce the manifold domains that we use in the following. First, as the $L^2(\Sigma)$ scalar product plays a central role in the definition of the Lagrangian is natural to start by choosing $\mathcal{Q}=L^2(\Sigma)$ as configuration space. However, the term $\langle \vec{\nabla}Q,\vec{\nabla}Q\rangle_{\vec{L}^2}$ in the Lagrangian forces us to restrict ourselves to a manifold domain $\mathcal{D}$ of $L^2(\Sigma)$ where the derivatives are well defined and belong to $L^2(\Sigma)$. This leads us to consider $\mathcal{D}_D=H^1_0(\Sigma)$ and $\mathcal{D}_R=H^1(\Sigma)$ for the Dirichlet and Robin cases, respectively. Hence, the velocity phase spaces have the form $T_{\mathcal{D}_D}\mathcal{Q}=H^1_0(\Sigma)\times L^2(\Sigma)$ and $T_{\mathcal{D}_R}\mathcal{Q}=H^1(\Sigma)\times L^2(\Sigma)$.

\subsection{Dirichlet boundary conditions}

We will study the dynamics of a free scalar field defined on a bounded domain  $\Sigma\subset \mathbb{R}^n$ of the type specified above subject to boundary conditions of the Dirichlet type:
\begin{eqnarray}
&&\ddot{\Phi}-\Delta\Phi=0\quad\mathrm{in}\quad(t_1,t_2)\times\Sigma\label{waveequationD}\\
&&\Phi=0\quad\mathrm{in}\quad(t_1,t_2)\times\partial\Sigma\label{Dboundary}\\
&&\Phi(t_1)=Q_1\,,\,\Phi(t_2)=Q_2\,.\label{inicQ}
\end{eqnarray}
where $Q_1,Q_2\in H^2(\Sigma)\cap H_0^1(\Sigma)$.

\subsubsection{Variational approach}

The action is defined on the space of curves
\begin{eqnarray*}
& & \mathcal{C}_{\scriptscriptstyle{D}}(Q_1,Q_2,[t_1,t_2])\\
& & =\{\Phi\in C^0([t_1,t_2],H^2(\Sigma)\cap H_0^1(\Sigma))\cap C^1([t_1,t_2],H^1_0(\Sigma))\cap C^2([t_1,t_2],L^2(\Sigma)):\Phi(t_i)=Q_i,\, i=1,2\}
\end{eqnarray*}
with tangent spaces at $\Phi\in\mathcal{C}_{\scriptscriptstyle{D}}(Q_1,Q_2,[t_1,t_2])$ of the type
\begin{eqnarray*}
T_\Phi\mathcal{C}_{\scriptscriptstyle{D}}(Q_1,Q_2,[t_1,t_2])=\mathcal{C}_{\scriptscriptstyle{D}}(0,0,[t_1,t_2])\,.
\end{eqnarray*}
The tangent vectors $\delta\in T_\Phi\mathcal{C}_{\scriptscriptstyle{D}}$ are sometimes referred to as \textit{variations}.

The action $S_{\scriptscriptstyle{D}}:\mathcal{C}_{\scriptscriptstyle{D}}(Q_1,Q_2,[t_1,t_2])\rightarrow \mathbb{R}$ is given by
\begin{eqnarray*}
S_{\scriptscriptstyle{D}}(\Phi)=\int_{t_1}^{t_2}L_{\scriptscriptstyle{D}}(\Phi(t),\dot{\Phi}(t))\mathrm{d}t=\frac{1}{2}\int_{t_1}^{t_2}\!\!\mathrm{d}t\int_\Sigma (\dot{\Phi}^2-\vec{\nabla}\Phi\cdot\vec{\nabla}\Phi)\mathrm{vol}_\Sigma\,.
\end{eqnarray*}
In this domain the action $S_{\scriptscriptstyle{D}}$ is differentiable \cite{Eells}. The differential of $S_{\scriptscriptstyle{D}}$ can be computed in a straightforward way as
\begin{eqnarray}
\mathrm{d}S_{\scriptscriptstyle{D}}(\Phi)\cdot\delta
&=&\left.\frac{\mathrm{d}}{\mathrm{d}\lambda}\right|_{\lambda=0}S(\Phi+\lambda\delta)=
\int_{t_1}^{t_2}\!\!\mathrm{d}t\int_\Sigma(\dot{\Phi}\,\dot{\delta}
-\vec{\nabla}\Phi\cdot\vec{\nabla}\delta)\mathrm{vol}_\Sigma\label{direcderivative}\\
&=&\int_\Sigma\!\! \dot{\Phi}(t_2)\delta(t_2)\mathrm{vol}_\Sigma-\int_\Sigma\!\! \dot{\Phi}(t_1)\delta(t_1)\mathrm{vol}_\Sigma\nonumber\\
&+&\int_{t_1}^{t_2}\!\!\!\!\mathrm{d}t\int_\Sigma(-\ddot{\Phi}+\Delta\Phi)\delta\,\,\mathrm{vol}_\Sigma-
 \int_{t_1}^{t_2}\!\!\!\!\mathrm{d}t\int_{\partial\Sigma}(\vec{n}\cdot \vec{\nabla} \Phi \,\delta) |_{\partial\Sigma}\mathrm{vol}_{\partial\Sigma}\,,\nonumber
\end{eqnarray}
where $\vec{n}\cdot \vec{\nabla}\Phi$ denotes the (outward) normal derivative of $\Phi$ at the boundary of $\Sigma$.
In the space of curves that we are considering we have
$$
\mathrm{d}S(\Phi)\cdot\delta=\int_{t_1}^{t_2}\!\!\mathrm{d}t\int_\Sigma(-\ddot{\Phi}+\Delta\Phi)\delta\,\,\mathrm{vol}_\Sigma=
-\int_{t_1}^{t_2}\!\!\langle\ddot{\Phi}(t)-\Delta\Phi(t),\delta(t)\rangle_{L^2}\mathrm{d}t\,,
$$
where we have used $\left.\delta(t)\right|_{\partial\Sigma}=0$ and $\delta(t_1)=\delta(t_2)=0$. Hence, the condition $\mathrm{d}S(\Phi)\cdot \delta=0$ for all the vectors $\delta\in T_\Phi\mathcal{C}_{\scriptscriptstyle{D}}$ implies that
\begin{eqnarray*}
\ddot{\Phi}-\Delta\Phi=0\quad\mathrm{in}\quad(t_1,t_2)\times\Sigma\,.
\end{eqnarray*}
Notice that the Dirichlet boundary conditions have been incorporated by choosing the appropriate domain for the action, in this case $\mathcal{C}_{\scriptscriptstyle{D}}$. In fact, if we replace  $H^1_0(\Sigma)$ by $H^1(\Sigma)$ in the definition of $\mathcal{C}_{\scriptscriptstyle{D}}$ the critical points of the action would satisfy the wave equation with Neumann conditions originating from the surface integrals appearing in (\ref{direcderivative}).

\subsubsection{Hamiltonian approach}

We will show that the Hamiltonian dynamics of the scalar field with Dirichlet boundary conditions takes place in the
 second class (generalized) submanifold $$\mathcal{N}_{\scriptscriptstyle{D}}=(H^2(\Sigma)\cap H_0^1(\Sigma))\times  H_0^1(\Sigma)$$   of the weakly simplectic manifold  $(\mathcal{M}_{\scriptscriptstyle{D}},\omega_{\scriptscriptstyle{D}})$, where $$\mathcal{M}_{\scriptscriptstyle{D}}=\overline{\mathcal{N}}_{\scriptscriptstyle{D}}= H_0^1(\Sigma)\times L^2(\Sigma)$$
  and  $\omega_{\scriptscriptstyle{D}}$ is the pullback to $\mathcal{M}_{\scriptscriptstyle{D}}$ of the strong, canonical, symplectic form on $L^2(\Sigma)\times L^2(\Sigma)$. The (uniquely defined) Hamiltonian vector field that gives the dynamics of the system is
 $$
X_{\scriptscriptstyle{D}}:\mathcal{N}_{\scriptscriptstyle{D}}\rightarrow \underline{T\overline{\mathcal{N}}_{\scriptscriptstyle{D}}}=\mathcal{N}_{\scriptscriptstyle{D}}\times \mathcal{M}_{\scriptscriptstyle{D}}: \quad X(Q,P)=((Q,P),(P,\Delta_{\scriptscriptstyle{D}} Q))\,.
$$
Here $\Delta_{\scriptscriptstyle{D}}:H^2(\Sigma)\cap H_0^1(\Sigma)\rightarrow L^2(\Sigma)$ denotes the scalar Dirichlet Laplacian.

\bigskip

In the following we will use the GNH algorithm explained above. The first element that we need is the fiber derivative. In the present case, and taking into account that the domain of the Lagrangian is
$H_0^1(\Sigma)\times L^2(\Sigma)$, we have that $FL_{\scriptscriptstyle{D}}:H_0^1(\Sigma)\times L^2(\Sigma)\rightarrow L^2(\Sigma)\times L^2(\Sigma)^*$
$$
FL_{\scriptscriptstyle{D}}(Q,V)=(Q,\langle V,\cdot\rangle_{L^2})\in L^2(\Sigma)\times L^2(\Sigma)^*\,.
$$
As $L^2(\Sigma)^*$ is isomorphic, according to Riesz's theorem, to $L^2(\Sigma)$  we consider $FL_{\scriptscriptstyle{D}}:H_0^1(\Sigma)\times L^2(\Sigma)\rightarrow L^2(\Sigma)\times L^2(\Sigma)$ given by $FL_{\scriptscriptstyle{D}}(Q,V)=(Q,V)$, in other words $FL_{\scriptscriptstyle{D}}$ is simply the inclusion of $H_0^1(\Sigma)\times L^2(\Sigma)$ into $L^2(\Sigma)\times L^2(\Sigma)$. The primary constraint manifold is $\mathcal{M}_1:=H_0^1(\Sigma)\times L^2(\Sigma)$ understood as a \textit{generalized} submanifold of $L^2(\Sigma)\times L^2(\Sigma)$ in the sense that its topology is not the induced one but the natural one for $H_0^1(\Sigma)\times L^2(\Sigma)$.

The space $L^2(\Sigma)\times L^2(\Sigma)$ carries a canonical, strongly non-degenerate, symplectic form\footnote{This is so (see \cite{ChernoffMarsden}) because $L^2(\Sigma)$ is a Hilbert space and, hence, is reflexive.} given by
$$
\Omega(Q,P)((q_1,p_1),(q_2,p_2))=\langle q_1,p_2\rangle_{L^2}-\langle q_2,p_1\rangle_{L^2}
$$
where $Q,P,q_i,p_i\in L^2(\Sigma)$. The pullback $\omega:=FL_{\scriptscriptstyle{D}}^*\Omega$  of $\Omega$ to $\mathcal{M}_1$, that we must use in the GNH algorithm is weakly symplectic.
$$
\omega(Q,P)((q_1,p_1),(q_2,p_2))=\langle q_1,p_2\rangle_{L^2}-\langle q_2,p_1\rangle_{L^2}
$$
with $Q, q_i\in H_0^1(\Sigma)$ and $P,p_i\in L^2(\Sigma)$.

The Hamiltonian $H_{\scriptscriptstyle{D}}:\mathcal{M}_1\rightarrow \mathbb{R}$ is
\begin{eqnarray*}
H_{\scriptscriptstyle{D}}(Q,P)=
\frac{1}{2}\big(\langle P,P\rangle_{L^2}+\langle \vec{\nabla}Q,\vec{\nabla}Q\rangle_{\vec{L}^2}\big)\,
\end{eqnarray*}
and its differential $\mathrm{d}H_{\scriptscriptstyle{D}}:\mathcal{M}_1\rightarrow \mathcal{L}(\mathcal{M}_1,\mathbb{R})$ is given by
\begin{eqnarray*}
dH_{\scriptscriptstyle{D}}(Q,P)(q,p)=\langle P,p\rangle_{L^2}+\langle \vec{\nabla}Q,\vec{\nabla}q\rangle_{\vec{L}^2}
\end{eqnarray*}
for $q\in H_0^1(\Sigma)$ and $p\in L^2(\Sigma)$.

Vector fields on $\mathcal{M}_1$ are maps $$X:\mathcal{M}_1\rightarrow\mathcal{M}_1\times \mathcal{M}_1:(Q,P)\mapsto ((Q,P),(X_Q(Q,P),X_P(Q,P))).$$ It is immediate to get
$$
(i_X\omega)(Q,P)(q,p)=\langle X_Q,p\rangle_{L^2}-\langle q, X_{P}\rangle_{L^2}\,.
$$
We have to find now a submanifold $\mathcal{M}_2$ with smooth injective inmersion $\mathcal{M}_2\stackrel{\jmath_2}{\rightarrow} \mathcal{M}_1$ such that the equation
$$
(i_X\omega-dH)|_{\jmath_2(\mathcal{M}_2)}=0
$$
can be solved. This is equivalent to considering $(i_X\omega-dH)|_{\jmath_2(\mathcal{M}_2)}(q,p)=0$ for all $(q,p)\in \mathcal{M}_1$. This last condition is
\begin{equation}
\langle P,p\rangle_{L^2}+\langle \vec{\nabla}Q,\vec{\nabla}q\rangle_{\vec{L}^2}=\langle X_Q,p\rangle_{L^2}-\langle q, X_P\rangle_{L^2}\,,\quad\forall (q,p)\in H_0^1(\Sigma)\times L^2(\Sigma)\,.
\label{conditionscalarD}
\end{equation}

This is a linear, non-homogeneous equation for $(X_Q,X_P)$. At this point we have to find out the conditions that $(Q,P)\in H_0^1(\Sigma)\times L^2(\Sigma)$ must satisfy in order to guarantee that the equation can be solved and then obtain its most general solution.

The best way to proceed is to start by considering $q=0$, so that (\ref{conditionscalarD}) becomes $\langle P-X_Q,p\rangle_{L^2}=0$ for all $p\in L^2(\Sigma)$. We deduce two things from this last equation: the first is that $X_Q$ is fixed to be $X_Q(Q,P)=P$ as a consequence of the Hahn-Banach theorem. The second is that, as $X_Q$ is required to be an element of $H_0^1(\Sigma)$, $P$ itself must be restricted to be in $H_0^1(\Sigma)$.

By taking now $p=0$ we are led to solve the equation $\langle\vec{\nabla}Q,\vec{\nabla}q\rangle_{\vec{L}^2}=-\langle q, X_P\rangle_{L^2}$ for all $q\in H_0^1(\Sigma)$. As the right hand side is the scalar product $\langle q, X_P\rangle_{L^2}$ we need to find out the conditions that $Q\in H_0^1(\Sigma)$ must satisfy to guarantee that the left hand side can also be written as the $L^2(\Sigma)$ scalar product of an element of $L^2(\Sigma)$ and $q$ (remember that $X_P\in L^2(\Sigma)$). It is straightforward to see that the right condition is to require  that $\vec{\nabla}Q\in\vec{H}(\mathrm{div},\Sigma)$ or, equivalently, $Q\in H^1(\Delta,\Sigma)$ (see Appendix \ref{Sobolevspaces}), so that in order to guarantee the solvability of the equation we must take $Q\in H^1(\Delta,\Sigma)\cap H^1_0(\Sigma)$. This allows us to write $\langle \vec{\nabla}Q,\vec{\nabla}q\rangle_{\vec{L}^2}=-\langle \Delta Q,q\rangle_{\vec{L}^2}$ and the equation that we must solve becomes $\langle \Delta Q-X_P,q\rangle_{L^2}=0$ for all $q\in H_0^1(\Sigma)$. By using now that $H_0^1(\Sigma)$ is dense in $L^2(\Sigma)$, extending the previous condition by continuity to $L^2(\Sigma)$, and employing the Hahn-Banach theorem we conclude that $X_P(Q,P)=\Delta Q\in L^2(\Sigma)$. It is important to mention at this point that in the case of manifolds $\Sigma$ with smooth boundary $H^1(\Delta,\Sigma)\cap H^1_0(\Sigma)=H^2(\Sigma)\cap H^1_0(\Sigma)$, hence, in the following, we will take $Q\in  H^2(\Sigma)\cap H^1_0(\Sigma)$.

At this stage we have found that
\begin{eqnarray*}
&&\mathcal{M}_2:=\big(H^2(\Sigma)\cap H_0^1(\Sigma)\big)\times H_0^1(\Sigma)
\end{eqnarray*}
and a Hamiltonian vector field  given by
\begin{eqnarray*}
(X_Q,X_P):\mathcal{M}_2\rightarrow \mathcal{M}_1,\,  (Q,P)\mapsto (P,\Delta Q)\,.
\end{eqnarray*}
We have to obtain now $\displaystyle \mathcal{M}_3=\{m\in\mathcal{M}_2: X(m)\in T_m\overline{\mathcal{M}}_2\}\,.$
To this end we need to compute
$$
\overline{\mathcal{M}}_2=\mathrm{cl}(\jmath_2 \mathcal{M}_2)=\mathrm{cl}_{H^1_0\times L^2}\big((H^2\cap H_0^1)\times H_0^1\big)=\mathrm{cl}_{H^1_0}(H^2\cap H_0^1)\times\mathrm{cl}_{L^2}H_0^1=H^1_0\times L^2=\mathcal{M}_1\,.
$$
It is obvious that $\mathrm{cl}_{L^2}H_0^1=L^2$ because the smooth functions with compact support $C^\infty_0(\Sigma)$ are a subset of $H^1_0$ (it is actually dense by the definition of $H^1_0$) and $C^\infty_0(\Sigma)$ is dense in $L^2(\Sigma)$. Likewise, as $C^\infty_0(\Sigma)\subset H^2(\Sigma)\cap H^1_0(\Sigma)$ and $C^\infty_0(\Sigma)$ is dense in $H^1_0$, we conclude that $\mathrm{cl}_{H^1_0}(H^2\cap H_0^1)=H_0^1$ and, hence, $\overline{\mathcal{M}}_2=H_0^1\times L^2$. Now it is straightforward to see that for every $m\in\mathcal{M}_2$ we have $X(m)\in T_m\overline{\mathcal{M}}_2$,  that is, for $(Q,P)\in (H^2\cap H^1_0)\times H^1_0$ we have $(P,\Delta Q)\in H_0^1\times L^2$. We conclude, then, that $\displaystyle \mathcal{M}_3=\mathcal{M}_2$ and the GNH algorithm stops giving $\mathcal{N}:=\mathcal{M}_2$.

Several comments are in order now. The first one is about the condition of tangency of the Hamiltonian vector field $X(Q,P)=(P,\Delta Q)$ to the closure of $\mathcal{M}_2$ in the GNH algorithm. It is well known that,  in the absence of boundaries, the GNH algorithm with the requirement of \textit{strict tangency} does not stop because, in each step, one is forced to introduce Sobolev spaces of increasingly higher order of regularity \cite{Gotaythesis}. In the presence of boundaries one finds, in addition, a sequence of conditions on the boundary of the spatial manifold of the type
\begin{equation}
Q|_{\partial \Sigma}=0,\, P|_{\partial \Sigma}=0,\, \Delta Q|_{\partial \Sigma}=0,\, \Delta P|_{\partial \Sigma}=0,\ldots, \Delta^k Q|_{\partial \Sigma}=0,\, \Delta^k P|_{\partial \Sigma}=0,\ldots
\label{Frechetboundary}
\end{equation}
This means that the final description for the states $(Q,P)$ would require the introduction of the Fr\'echet manifold $C^\infty(\overline{\Sigma})\times C^\infty(\overline{\Sigma})$ (with the countable collection of semi-norms inherited by its definition as the intersection $\bigcap_{k=0}^\infty H^k(\Sigma)$) and, in addition, an infinite set (\ref{Frechetboundary}) of subsidiary conditions at the boundary  (this set of conditions  is discussed in \cite{Brezis} for the case of the wave and heat equation).

The second comment is that the integral curves of the field $X$ are solutions to the first order equations
\begin{eqnarray*}
&&\dot{Q}=P\\
&&\dot{P}=\Delta Q
\end{eqnarray*}
with initial data  $(Q_0,P_0)\in  (H^2(\Sigma)\cap H_0^1(\Sigma))\times  H_0^1(\Sigma)$: Notice that, in the previous equation, $\Delta=\Delta_{\scriptscriptstyle{D}}$ is just the Dirichlet Laplacian $\Delta_{\scriptscriptstyle{D}}:H^2(\Sigma)\cap H^1_0(\Sigma)\rightarrow L^2(\Sigma)$. These equations lead immediately to $\ddot{Q}-\Delta Q=0$ and are equivalent to (\ref{waveequationD})-(\ref{inicQ}) (see Appendix \ref{abstractwave} where, in this specific case,  $\mathcal{D}(-\Delta_{\scriptscriptstyle{D}})=H^2(\Sigma)\cap H^1_0(\Sigma)$ and $\mathcal{D}(\sqrt{-\Delta_{\scriptscriptstyle{D}}})=H^1_0(\Sigma)$). Notice also that the vector field $X$ has been completely fixed by the GNH algorithm, i.e. there is no arbitrariness in its determination, hence, in this case the boundary conditions do not imply the existence of gauge symmetries in the system, as expected. From a geometric point of view it is important to mention that the submanifold $\mathcal{M}_2\stackrel{\jmath_2}{\rightarrow} \mathcal{M}_1$ is \textit{second class} because $\mathrm{ker}(\flat)=\{0\}$. Finally, it is possible to show (see \cite{Brezis} and Appendix \ref{abstractwave}) that the vector field is integrable and defines a $C_0$-flux in $\mathcal{M}_2$, with the tangents to the integral curves belonging to $\overline{\mathcal{M}}_2=\mathcal{M}_1$.

\subsubsection{The symplectic-Lagrangian approach}

As the fiber derivative in this case is just the inclusion $H_0^1(\Sigma)\times L^2(\Sigma)\stackrel{\jmath}{\rightarrow}L^2(\Sigma)\times L^2(\Sigma)$, the pullback of the canonical symplectic form of $L^2(\Sigma)\times L^2(\Sigma)$ to the velocity phase space  $T_{\mathcal{D}_{\scriptscriptstyle{D}}}\mathcal{Q}=H_0^1(\Sigma)\times L^2(\Sigma)$ that defines $\Omega_L$ is precisely equivalent to the pullback to $\mathcal{M}_1$ that we used before. Actually, as we use $L^2(\Sigma)$ instead of its dual, the symplectic-Lagrangian approach in this case is identical to the Hamiltonian one described in the preceding section.

\subsection{Robin boundary conditions}

We study now the dynamics of a free scalar field defined on a bounded domain of $\Sigma\subset \mathbb{R}^n$ subject to boundary conditions of the Robin type:
\begin{eqnarray*}
&&\ddot{\Phi}-\Delta\Phi=0\quad\mathrm{in}\quad(t_1,t_2)\times\Sigma\\
&&\vec{n}\cdot\vec{\nabla}\Phi=B\Phi\quad\mathrm{in}\quad(t_1,t_2)\times\partial\Sigma\\
&&\Phi(t_1)=Q_1\,,\,\Phi(t_2)=Q_2\,.
\end{eqnarray*}
with $Q_1,Q_2\in H^2(\Sigma)$.

\subsubsection{Variational approach}

The action is defined on the space of curves
\begin{eqnarray*}
& & \mathcal{C}_{\scriptscriptstyle{R}}(Q_1,Q_2,[t_1,t_2])\\
& & =\{\Phi\in C^0([t_1,t_2],H^2(\Sigma))\cap C^1([t_1,t_2],H^1(\Sigma))\cap C^2([t_1,t_2],L^2(\Sigma)):\Phi(t_i)=Q_i,\, i=1,2\}
\nonumber
\end{eqnarray*}
with tangent spaces at $\Phi\in\mathcal{C}_{\scriptscriptstyle{R}}(Q_1,Q_2,[t_1,t_2])$ given by
\begin{eqnarray*}
T_\Phi\mathcal{C}_{\scriptscriptstyle{R}}(Q_1,Q_2,[t_1,t_2])=\mathcal{C}_{\scriptscriptstyle{R}}(0,0,[t_1,t_2])\,.
\end{eqnarray*}

The action $S_{\scriptscriptstyle{R}}:\mathcal{C}_{\scriptscriptstyle{R}}(Q_1,Q_2,[t_1,t_2])\rightarrow \mathbb{R}$ is
\begin{eqnarray*}
S_{\scriptscriptstyle{R}}(\Phi)=\int_{t_1}^{t_2}L_{\scriptscriptstyle{R}}(\Phi(t),\dot{\Phi}(t))\mathrm{d}t
=\frac{1}{2}\int_{t_1}^{t_2}\!\!\mathrm{d}t\int_\Sigma (\dot{\Phi}^2-\vec{\nabla}\Phi\cdot\vec{\nabla}\Phi)\mathrm{vol}_\Sigma+
\frac{1}{2}\int_{t_1}^{t_2}\!\!\mathrm{d}t\int_{\delta\Sigma}B\Phi^2|_{\partial\Sigma}\,\mathrm{vol}_{\partial\Sigma}\,.
\end{eqnarray*}
In this domain the action $S_{\scriptscriptstyle{R}}$ is differentiable. The differential of $S_{\scriptscriptstyle{R}}$ at such $\Phi$ acting on a vector $\delta\in T_\Phi\mathcal{C}_{\scriptscriptstyle{R}}(Q_1,Q_2,[t_1,t_2])$ can be computed in a straightforward way as
\begin{eqnarray*}
\mathrm{d}S_{\scriptscriptstyle{R}}(\Phi)\cdot\delta&=&
\int_{t_1}^{t_2}\!\!\mathrm{d}t\int_\Sigma(\dot{\Phi}\,\dot{\delta}-
\vec{\nabla}\Phi\cdot\vec{\nabla}\delta)\mathrm{vol}_\Sigma+
\int_{t_1}^{t_2}\!\!\mathrm{d}t\int_{\delta\Sigma}B\Phi \delta|_{\partial\Sigma}\,\mathrm{vol}_{\partial\Sigma}\nonumber\\
&=&\int_{t_1}^{t_2}\!\!\mathrm{d}t\int_\Sigma(-\ddot{\Phi}+\Delta\Phi)\delta\,\,\mathrm{vol}_\Sigma
+\int_{t_1}^{t_2}\!\!\mathrm{d}t\int_{\partial\Sigma}\left.(B\Phi-\vec{n}\cdot\vec{\nabla} \Phi) \delta\right|_{\partial\Sigma}\mathrm{vol}_{\partial\Sigma}\\
&=&\int_{t_1}^{t_2}\!\!\langle -\ddot{\Phi}(t)+\Delta\Phi(t),\delta(t)\rangle_{L^2(\Sigma)}\,\,\mathrm{d}t
+\int_{t_1}^{t_2}\!\! \langle B\Phi(t)-\vec{n}\cdot\vec{\nabla} \Phi(t), \delta(t)\rangle_{L^2(\partial\Sigma)}\mathrm{d}t\,,\nonumber
\end{eqnarray*}
where $\vec{n}\cdot\vec{\nabla}\Phi$ denotes the (outward) normal derivative of $\Phi$ at the boundary of $\Sigma$ and we have used $\delta(t_1)=\delta(t_2)=0$. Hence, the condition $\mathrm{d}S(\Phi)\cdot \delta=0$ for all the vectors $\delta$ implies that
\begin{eqnarray}
&&\ddot{\Phi}-\Delta\Phi=0\quad\mathrm{in}\quad(t_1,t_2)\times\Sigma\,,\label{bulkR}\\
&&\vec{n}\cdot\vec{\nabla}\Phi(t)|_{\partial\Sigma}=B\Phi(t)|_{\partial\Sigma}\,.\label{boundaryR}
\end{eqnarray}
The first condition (\ref{bulkR}) is obtained by considering variations $\delta(t)\in H^1_0(\Sigma)\subset H^1(\Sigma)$. Once this necessary condition is obtained the second set of equations (the boundary conditions \ref{boundaryR}) come from variations with $\delta(t)\in H^1(\Sigma)$ (that may not vanish at $\partial\Sigma$). Notice that the Robin boundary conditions appear now as conditions on the critical points of the action (variational equations) and are not incorporated in the functional space $\mathcal{C}_{\scriptscriptstyle{R}}$ (as happened in the Dirichlet case). It is also worth pointing out that the Neumann boundary conditions correspond to the choice $B=0$.

\subsubsection{Hamiltonian approach}

We will show now that the Hamiltonian dynamics of the scalar field with Robin boundary conditions takes place in the
 second class (generalized) submanifold
 $$\mathcal{N}_{\scriptscriptstyle{R}}=H_\partial^2(\Sigma)\times  H^1(\Sigma)\,,\quad H_\partial^2(\Sigma):=\{Q\in H^2(\Sigma):( B Q-\vec{n}\cdot\vec{\nabla} Q)|_{\partial\Sigma}=0\}\,,$$
 of the weakly simplectic manifold  $(\mathcal{M}_{\scriptscriptstyle{R}},\omega_{\scriptscriptstyle{R}})$, where
 $$\mathcal{M}_{\scriptscriptstyle{R}}=\overline{\mathcal{N}}_{\scriptscriptstyle{R}}= H^1(\Sigma)\times L^2(\Sigma)$$
  and  $\omega_{\scriptscriptstyle{R}}$ is the pullback to $\mathcal{M}_{\scriptscriptstyle{R}}$ of the strong, canonical, symplectic form on $L^2(\Sigma)\times L^2(\Sigma)$. The Hamiltonian vector field
 $$
X_{\scriptscriptstyle{R}}:\mathcal{N}_{\scriptscriptstyle{R}}\rightarrow \underline{T\overline{\mathcal{N}}_{\scriptscriptstyle{R}}}=\mathcal{N}_{\scriptscriptstyle{R}}\times \mathcal{M}_{\scriptscriptstyle{R}}: \quad X(Q,P)=((Q,P),(P,\Delta_{\scriptscriptstyle{R}} Q))\,.
$$
is defined in terms of the scalar Robin Laplacian $\Delta_{\scriptscriptstyle{R}}:H_\partial^2(\Sigma)\rightarrow L^2(\Sigma)$.

\bigskip

We study the Hamiltonian formulation by using the GNH algorithm. The fiber derivative is now $FL_{\scriptscriptstyle{R}}:H^1(\Sigma)\times L^2(\Sigma)\rightarrow L^2(\Sigma)\times L^2(\Sigma)^*$
$$
FL_{\scriptscriptstyle{R}}(Q,V)=(Q,\langle V,\cdot\rangle_{L^2})\in L^2(\Sigma)\times L^2(\Sigma)^*\,.
$$
As we did before, we identify $L^2(\Sigma)^*$ with $L^2(\Sigma)$ and consider $FL_{\scriptscriptstyle{R}}:H^1(\Sigma)\times L^2(\Sigma)\rightarrow L^2(\Sigma)\times L^2(\Sigma)$. By doing this $FL_{\scriptscriptstyle{R}}$ is the inclusion of $H^1(\Sigma)\times L^2(\Sigma)$ into $L^2(\Sigma)\times L^2(\Sigma)$. The primary constraint manifold is $\mathcal{M}_1:=H^1(\Sigma)\times L^2(\Sigma)$ understood as a \textit{generalized} submanifold of $L^2(\Sigma)\times L^2(\Sigma)$.

The pullback of the canonical symplectic form of $L^2(\Sigma)\times L^2(\Sigma)$ to $\mathcal{M}_1$ is the weakly symplectic given by
$$
\omega(Q,P)((q_1,p_1),(q_2,p_2))=\langle q_1,p_2\rangle_{L^2}-\langle q_2,p_1\rangle_{L^2}\,,
$$
with $Q, q_i\in H^1(\Sigma)$ and $P,p_i\in L^2(\Sigma)$. The Hamiltonian $H_{\scriptscriptstyle{R}}:\mathcal{M}_1\rightarrow \mathbb{R}$ is
\begin{eqnarray*}
H_{\scriptscriptstyle{R}}(Q,P)=
\frac{1}{2}\big(\langle P,P\rangle_{L^2(\Sigma)}+\langle \vec{\nabla}Q,\vec{\nabla}Q\rangle_{\vec{L}^2(\Sigma)}+\langle b Q,b Q\rangle_{L^2(\partial\Sigma)}\big)\,
\end{eqnarray*}
where we have made use of the fact that $B\leq0$ to write $-B=b^2$ with $b\geq0$. It is important to notice that this condition on $B$ guarantees the non-negativity of the Hamiltonian.
 Also, as we are considering $B\in C^\infty(\partial\Sigma)$ we also have $b\in C^0(\partial\Sigma)$.

The differential $\mathrm{d}H_{\scriptscriptstyle{R}}:\mathcal{M}_1\rightarrow \mathcal{L}(\mathcal{M}_1,\mathbb{R})$ is given by
\begin{eqnarray*}
dH_{\scriptscriptstyle{R}}(Q,P)(q,p)=\langle P,p\rangle_{L^2(\Sigma)}+\langle \vec{\nabla}Q,\vec{\nabla}q\rangle_{\vec{L}^2(\Sigma)}-\langle B Q, q\rangle_{L^2(\partial\Sigma)}\,.
\end{eqnarray*}
for $q\in H^1(\Sigma)$ and $p\in L^2(\Sigma)$. As before, if we denote the vector fields on $\mathcal{M}_1$ by
$$X:\mathcal{M}_1\rightarrow\mathcal{M}_1\times \mathcal{M}_1:(Q,P)\mapsto ((Q,P),(X_Q(Q,P),X_P(Q,P))),$$
it is immediate to get $i_X\omega$
$$
(i_X\omega)(Q,P)(q,p)=\langle X_Q,p\rangle_{L^2}-\langle q, X_{P}\rangle_{L^2}\,.
$$
We have to find now a submanifold $\mathcal{M}_2$ with smooth injective inmersion $\jmath_2:\mathcal{M}_2\rightarrow \mathcal{M}_1$ such that the equation
$$
(i_X\omega-dH)|_{\jmath_2(\mathcal{M}_2)}=0
$$
can be solved. This is equivalent to considering $(i_X\omega-dH)|_{\jmath_2(\mathcal{M}_2)}(q,p)=0$ for all $(q,p)\in \mathcal{M}_1$. This last condition is
\begin{eqnarray*}
\langle P,p\rangle_{L^2(\Sigma)}+\langle \vec{\nabla}Q,\vec{\nabla}q\rangle_{\vec{L}^2(\Sigma)}-\langle B Q, q\rangle_{L^2(\partial\Sigma)}=\langle X_Q,p\rangle_{L^2(\Sigma)}-\langle q, X_P\rangle_{L^2(\Sigma)}\,,
\end{eqnarray*}
for all $(q,p)\in H^1(\Sigma)\times L^2(\Sigma)$. This equation cannot be solved in general for arbitrary values of $(Q,P)\in H^1(\Sigma)\times L^2(\Sigma)$. A direct reasoning, that parallels the one used in the case of the Dirichlet boundary conditions, tells us that $X_Q=P$ with $P\in H^1(\Sigma)$. Furthermore we must require $Q\in H^2(\Sigma)$ where, as before, we have made use of the regularity of the boundary $\partial\Sigma$ to trade the Sobolev space $H^1(\Delta,\Sigma)$ for $H^2(\Sigma)$ (see chapter 5 of \cite{Taylor}). The condition that remains to be solved is
$$
\langle -\Delta Q,q\rangle_{L^2(\Sigma)}-\langle B Q-\vec{n}\cdot\vec{\nabla} Q, q\rangle_{L^2(\partial\Sigma)}=-\langle q, X_P\rangle_{L^2(\Sigma)}\,\quad\forall q\in H^1(\Sigma)\,.
$$
From this it is obvious that $X_P=\Delta Q$ and also that $(B Q-\vec{n}\cdot\vec{\nabla} Q)|_{\partial\Sigma}=0$. Hence we have that $\mathcal{M}_2:=H_\partial^2(\Sigma)\times H^1(\Sigma)$ where
$$
H_\partial^2(\Sigma)=\{Q\in H^2(\Sigma)\, :\, ( B Q-\vec{n}\cdot\vec{\nabla} Q)|_{\partial\Sigma}=0\}\,.
$$
With the induced topology, $H_\partial^2(\Sigma)$ is a closed linear subspace of $H^2(\Sigma)$ and, hence, a Hilbert space. At this stage we have found that
\begin{eqnarray*}
&&\mathcal{M}_2=H^2_\partial(\Sigma)\times H^1(\Sigma)\,,\\
&&(X_Q,X_P):\mathcal{M}_2\rightarrow  \mathcal{M}_1:(Q,P)\mapsto(P,\Delta Q)\,.
\end{eqnarray*}

We have to obtain now $\displaystyle \mathcal{M}_3=\{m\in\mathcal{M}_2:X(m)\in T_m\overline{\mathcal{M}}_2\}\,.$
To this end we need to compute
$$
\overline{ \mathcal{M}}_2=\mathrm{cl}(\jmath_2 \mathcal{M}_2)=\mathrm{cl}_{H^1\times L^2}\big(H^2_\partial\times H^1\big)=\mathrm{cl}_{H^1}(H^2_\partial)\times\mathrm{cl}_{L^2}H^1=H^1\times L^2=\mathcal{M}_1\,.
$$
It is obvious now that $\mathrm{cl}_{L^2}H^1=L^2$ because $C^\infty_0(\Sigma)\subset H^1$ and $C^\infty_0(\Sigma)$ is dense in $L^2(\Sigma)$.

The argument to prove that $\mathrm{cl}_{H^1}H^2_\partial=H^1$ is slightly more subtle and goes as follows. Let us consider the following scalar product in $H^1(\Sigma)$
$$
\langle\!\langle u,v\rangle\!\rangle_{H^1(\Sigma)}=\langle u,v\rangle_{L^2(\Sigma)}+\langle b u , b v\rangle_{L^2(\partial\Sigma)}+\langle \vec{\nabla}u,\vec{\nabla}v\rangle_{\vec{L}^2(\Sigma)}
$$
with associated norm denoted as $|||v|||_{H^1(\Sigma)}$. It is straightforward to prove the equivalence of the norms $|||\cdot|||_{H^1(\Sigma)}$ and $||\cdot||_{H^1(\Sigma)}$ because $\|v\|^2_{H^1(\Sigma)}\leq|||v|||^2_{H^1(\Sigma)}$, and
\begin{eqnarray*}
|||v|||^2_{H^1(\Sigma)}&=&\|v\|^2_{H^1(\Sigma)}+\|b v|_{\partial\Sigma}\|^2_{L^2(\partial\Sigma)}\leq\|v\|^2_{H^1(\Sigma)}+
(\max_{\partial\Sigma}b)^2\|v|_{\partial\Sigma}\|^2_{L^2(\partial\Sigma)}\\
&\leq& \|v\|^2_{H^1(\Sigma)}+ (\max_{\partial\Sigma}b)^2\|\gamma\|^2\|v\|^2_{H^1(\Sigma)}=
\big(1+(\max_{\partial\Sigma}b)^2\|\gamma\|^2\big)\|v\|^2_{H^1(\Sigma)}\,,
\end{eqnarray*}
as a consequence of the compactness of $\partial\Sigma$ and the continuity of the trace operator. Now the subspace $H^2_\partial(\Sigma)$ will be dense in $H^1(\Sigma)$ iff $(H^2_\partial(\Sigma))^\perp=\{0\}$ (with respect to the scalar product $\langle\!\langle \cdot,\cdot\rangle\!\rangle_{H^1(\Sigma)}$). In order to compute $(H^2_\partial(\Sigma))^\perp$ we take an orthonormal basis of eigenstates $u_k$, $k\in \mathbb{N}$,  of the Laplace operator with Robin boundary conditions and eigenvalues $-\lambda_k^2$, and make use of the known fact that for sufficiently regular $\partial\Sigma$ these eigenfunctions are smooth, i.e. $u_k\in C^\infty(\overline{\Sigma})$, $\forall k\in \mathbb{N}$. The condition
\begin{eqnarray*}
0&=&\langle\!\langle u_k,v\rangle\!\rangle_{H^1(\Sigma)}=\langle u_k,v\rangle_{L^2(\Sigma)}+\langle b u_k , b v\rangle_{L^2(\partial\Sigma)}+\langle \vec{\nabla}u_k,\vec{\nabla}v\rangle_{\vec{L}^2(\Sigma)}\\
&=&(1+\lambda^2_k)\langle u_k,v\rangle_{L^2(\Sigma)}+\langle b u_k , b v\rangle_{L^2(\partial\Sigma)}+\int_{\partial\Sigma} v \vec{n}\cdot\vec{\nabla} u_k\\
&=&(1+\lambda^2_k)\langle u_k,v\rangle_{L^2(\Sigma)}+\langle b u_k , b v\rangle_{L^2(\partial\Sigma)}-\int_{\partial\Sigma} b^2 v u_k\\
&=&(1+\lambda^2_k)\langle u_k,v\rangle_{L^2(\Sigma)}\,,
\end{eqnarray*}
for all $k\in \mathbb{N}$ implies $v=0$ and we conclude that $(H^2_\partial(\Sigma))^\perp=\{0\}$ so that $\mathrm{cl}_{H^1}H^2_\partial=H^1$. We have then $\mathcal{M}_3=\mathcal{M}_2\,,$ and the GNH algorithm stops giving $\mathcal{N}:=\mathcal{M}_2$.

As before,  the integral curves of the Hamiltonian vector field $X(Q,P)=(P,\Delta_{\scriptscriptstyle{R}}Q)$ reproduce the evolution given by the wave equation with Robin boundary conditions. In particular, the conditions discussed for the abstract wave equation in Appendix \ref{abstractwave} are satisfied in this specific case where $\mathcal{D}(-\Delta_{\scriptscriptstyle{R}})=H^2_\partial(\Sigma)$ and $\mathcal{D}(\sqrt{-\Delta_{\scriptscriptstyle{R}}})=H^1(\Sigma)$. Finally, the submanifold $\mathcal{M}_2\stackrel{\jmath_2}{\rightarrow} \mathcal{M}_1$ is, again, a \textit{second class} submanifold of $(\mathcal{M}_1,\omega)$.

\subsubsection{The symplectic-Lagrangian approach}

As the fiber derivative is the inclusion $H^1(\Sigma)\times L^2(\Sigma)\stackrel{\jmath}{\rightarrow}L^2(\Sigma)\times L^2(\Sigma)$, as in the case of Dirichlet boundary conditions, the symplectic-Lagrangian approach is exactly the same as the Hamiltonian approach discussed above.

\section{The electromagnetic field in the presence of boundaries}\label{EM}

We study now the electromagnetic field defined on a bounded region. The main difference between this example and the case of the scalar field discussed in the preceding section is the presence of a gauge symmetry. We will explore here how the boundary changes the constraint analysis for the system. We will follow the same scheme used in the discussion of the scalar field and consider different types of boundary conditions that are the natural generalizations of the Dirichlet and Neumann ones. In particular the Dirichlet case has a clear physical interpretation as it corresponds to the perfect conductor boundary conditions. This is dealt with by a suitable choice of domain for the Lagrangian.  The Neumann boundary conditions, on the other hand, behave as those of the scalar Robin case in the sense that the Hamiltonian GNH analysis provides additional conditions on the boundary values of the fields. Of course we will find also the expected constraints associated with the usual gauge symmetry of electromagnetism.

The electromagnetic field will be represented by a $U(1)$ connection on the 4-manifold $[t_1,t_2]\times \Sigma$, we will restrict ourselves to connected manifolds $\Sigma$ with boundary and, hence, represent them as a 1-form field $A:[t_1,t_2]\times \Sigma\rightarrow\Lambda^1([t_1,t_2]\times \Sigma)$. For ease of comparison we will use the metric to transform 1-forms into vectors fields and use a 3-vector notation in the following.

We will consider two types of Maxwell  Lagrangians, $L_{\scriptscriptstyle{D}}: \mathbf{H}^1_\partial(\Sigma)\times \mathbf{L}^2(\Sigma)\rightarrow\mathbb{R}$ and $L_{\scriptscriptstyle{N}}: \mathbf{H}^1(\Sigma)\times \mathbf{L}^2(\Sigma)\rightarrow\mathbb{R}$
given by
\begin{eqnarray*}
\!\!\!\! L_{\scriptscriptstyle{D,N}}(Q,V)&=&\frac{1}{2}\langle\vec{V},\vec{V}\rangle_{\vec{L}^2}+\langle\vec{\nabla}Q_\perp,\vec{V}\rangle_{\vec{L}^2}
+\frac{1}{2}\langle\vec{\nabla}Q_\perp,\vec{\nabla}Q_\perp\rangle_{\vec{L}^2}-\frac{1}{2}\langle \vec{\nabla}\times \vec{Q}, \vec{\nabla}\times \vec{Q} \rangle_{\vec{L}^2}\,.
\end{eqnarray*}
The preceding Lagrangians are defined in an fixed inertial frame of the Minkowski space-time and we are using the notation
\begin{eqnarray*}
&&\mathbf{L}^2(\Sigma):=L^2_{\perp}(\Sigma)\times \vec{L}^2(\Sigma)\,,\,\,\, \mathbf{H}^1(\Sigma):=H^1_{\perp}(\Sigma)\times \vec{H}(\mathrm{curl},\Sigma)\,,\,\,\,\mathbf{H}^1_\partial(\Sigma):=H^1_{0\perp}(\Sigma)\times \vec{H}_0(\mathrm{curl},\Sigma)
\end{eqnarray*}
The different functional spaces that we use in this section are described in Appendix \ref{Sobolevspaces}.

As in the case of the scalar field, the configuration space is taken to be $\mathcal{Q}:=\mathbf{L}^2(\Sigma)$. The presence of derivatives in some terms of the Lagrangian forces us to consider the manifold domains  $\mathcal{D}_{\scriptscriptstyle{D}}:=\mathbf{H}^1_\partial(\Sigma)$ and  $\mathcal{D}_{\scriptscriptstyle{N}}:=\mathbf{H}^1(\Sigma)$ and, hence,
$L_{\scriptscriptstyle{D,N}}: T_{\mathcal{D}_{\scriptscriptstyle{D,N}}}\mathcal{Q}\rightarrow\mathbb{R}$.

\subsection{The perfect conductor boundary conditions}

We will study here the dynamics of the electromagnatic field on a bounded domain $\Sigma\subset \mathbb{R}^3$, with a smooth boundary, subject to the perfect conductor boundary conditions. In terms of $A=(A_perp,\vec{A})$, the Maxwell equations are
\begin{eqnarray*}
&\ddot{\vec{A}}+\vec{\nabla}\dot{A}_\perp-\Delta\vec{A}+\vec{\nabla}(\vec{\nabla}\cdot \vec{A})=\vec{0}&\hspace*{4mm}\mathrm{in}\quad(t_1,t_2)\times\Sigma
\\
&\vec{\nabla}\cdot(\dot{\vec{A}}+\vec{\nabla}A_\perp)=0&\hspace*{4mm}\mathrm{in}\quad(t_1,t_2)\times\Sigma\\
&\vec{n}\times \vec{A}=\vec{0}&\hspace*{4mm}\mathrm{in}\quad(t_1,t_2)\times\partial\Sigma\\
&A_\perp=0&\hspace*{4mm}\mathrm{in}\quad(t_1,t_2)\times\partial\Sigma\\
&\vec{A}(t_1)=\vec{Q}_1\,,\, A_\perp(t_1)=Q_{\perp1}\,,& \hspace*{2mm}\vec{A}(t_2)=\vec{Q}_2\,,\, A_\perp(t_2)=Q_{\perp2}
\end{eqnarray*}
with $\vec{Q}_i\in \vec{H}^2_\partial(\mathrm{curl},\Sigma)$ and $Q_{\perp i}\in H^2_\perp(\Sigma)\cap H^1_{0\perp}(\Sigma)$.

\subsubsection{Variational approach}

The action is defined on the space of curves
\begin{eqnarray*}
& & \mathcal{C}_{\scriptscriptstyle{D}}(Q_1,Q_2,[t_1,t_2])=\{A\in \mathcal{C}_{\scriptscriptstyle{D}}^0([t_1,t_2],\Sigma) \cap \mathcal{C}_{\scriptscriptstyle{D}}^1([t_1,t_2],\Sigma) \cap\mathcal{C}_{\scriptscriptstyle{D}}^2([t_1,t_2],\Sigma) :A(t_i)\!=\!Q_i,\, i\!=\!1,2\}\}
\end{eqnarray*}
where
\begin{eqnarray*}
\mathcal{C}_{\scriptscriptstyle{D}}^0([t_1,t_2],\Sigma) &:=&C^0\big([t_1,t_2],(H^2_\perp(\Sigma)\cap H^1_{0\perp }(\Sigma))\times\vec{H}^2_\partial(\mathrm{curl},\Sigma)\big)\,,\\
\mathcal{C}_{\scriptscriptstyle{D}}^1([t_1,t_2],\Sigma)&:=&C^1\big([t_1,t_2],L^2_\perp(\Sigma)\times
(\vec{H}_0(\mathrm{curl},\Sigma)\cap\vec{H}(\mathrm{div},\Sigma))\big)\,,\\
\mathcal{C}_{\scriptscriptstyle{D}}^2([t_1,t_2],\Sigma)&:=& C^2\big([t_1,t_2],\vec{L}^2(\Sigma)\big)\,.
\nonumber
\end{eqnarray*}
The tangent spaces at $A\in\mathcal{C}_{\scriptscriptstyle{D}}(Q_1,Q_2,[t_1,t_2])$ are $T_A\mathcal{C}_{\scriptscriptstyle{D}}(Q_1,Q_2,[t_1,t_2])=\mathcal{C}_{\scriptscriptstyle{D}}(0,0,[t_1,t_2])$.

The action $S_{\scriptscriptstyle{D}}:\mathcal{C}_{\scriptscriptstyle{D}}(Q_1,Q_2,[t_1,t_2])\rightarrow \mathbb{R}$ is given by
\begin{eqnarray*}
S_{\scriptscriptstyle{D}}(A)&=&\int_{t_1}^{t_2}L_{\scriptscriptstyle{D}}(A(t),\dot{A}(t))\mathrm{d}t\\
&=&\int_{t_1}^{t_2}\left(\frac{1}{2}\langle\dot{\vec{A}},\dot{\vec{A}}\rangle_{\vec{L}^2}
+\langle\vec{\nabla}A_\perp,\dot{\vec{A}}\rangle_{\vec{L}^2}
+\frac{1}{2}\langle\vec{\nabla}A_\perp,\vec{\nabla}A_\perp\rangle_{\vec{L}^2}-\frac{1}{2}\langle \vec{\nabla}\times \vec{A}, \vec{\nabla}\times \vec{A} \rangle_{\vec{L}^2}\right)\mathrm{d}t\,.\nonumber
\end{eqnarray*}
The differential of $S_{\scriptscriptstyle{D}}$ at $A$ acting on a vector $\delta\in T_A\mathcal{C}_{\scriptscriptstyle{D}}(Q_1,Q_2,[t_1,t_2])$ can be computed in a straightforward way as
\begin{eqnarray*}
\mathrm{d}S_{\scriptscriptstyle{D}}(A)\cdot\delta
&=&\int_{t_1}^{t_2}\!\!\mathrm{d}t\int_\Sigma \big(-\delta_\perp(\Delta A_\perp+\vec{\nabla}\cdot \dot{\vec{A}})+\vec{\delta}\cdot(\Delta \vec{A}-\ddot{\vec{A}}-\vec{\nabla}(\vec{\nabla}\cdot \vec{A})-\vec{\nabla}\dot{A}_\perp)\big)\mathrm{vol}_\Sigma\,,
\nonumber
\end{eqnarray*}
where we have used that $\delta_\perp(t)\in H_{0\perp}^1(\Sigma)$ and  $\vec{\delta}(t)\in\vec{H}_0(\mathrm{curl},\Sigma)$  for each $t\in[t_1,t_2]$.

Hence, the condition $\mathrm{d}S_{\scriptscriptstyle{D}}(A)\cdot \delta=0$ for all the vectors $\delta\in T_A\mathcal{C}_{\scriptscriptstyle{D}}$ implies
\begin{eqnarray*}
&\ddot{\vec{A}}+\vec{\nabla}\dot{A}_\perp-\Delta\vec{A}+\vec{\nabla}(\vec{\nabla}\cdot \vec{A})=0&\quad\mathrm{in}\quad(t_1,t_2)\times\Sigma\,\\
&\vec{\nabla}\cdot(\dot{\vec{A}}+\vec{\nabla}A_\perp)=0&\quad\mathrm{in}\quad(t_1,t_2)\times\Sigma\,
\end{eqnarray*}
These are just the Maxwell equations in $\Sigma$ subject to the perfect conductor boundary conditions in $\partial\Sigma$ introduced in the definition of the domain for the action (and the Lagrangian).

\subsubsection{Hamiltonian approach}

We will show that the Hamiltonian dynamics of the electromagnetic field with perfect conductor boundary conditions takes place in the
 first class (generalized) submanifold $$\mathcal{N}_{\scriptscriptstyle{D}}:=\{(Q,\vec{P})\,:\, Q_\perp\in H^1_{0\perp}(\Sigma),\,\vec{Q}\in \vec{H}^2_\partial(\mathrm{curl},\Sigma),\, \vec{P}\in\vec{H}_0(\mathrm{curl},\Sigma)\cap  \vec{H}(\mathrm{div},\Sigma),\, \vec{\nabla}\cdot\vec{P}=0\}$$
 of the presimplectic manifold  $(\mathcal{M}_{\scriptscriptstyle{D}},\omega_{\scriptscriptstyle{D}})$, where $$\mathcal{M}_{\scriptscriptstyle{D}}=\mathbf{H}_\partial^1(\Sigma)\times\vec{L}^2(\Sigma)$$
  and  $\omega_{\scriptscriptstyle{D}}$ is the pullback to $\mathcal{M}_{\scriptscriptstyle{D}}$ of the strong, canonical, symplectic form on $\mathbf{L}^2(\Sigma)\times \mathbf{L}^2(\Sigma)$. The class of Hamiltonian vector fields that defines the dynamics of the system is given by
 $$
X_{\scriptscriptstyle{D}}:\mathcal{N}_{\scriptscriptstyle{D}}\rightarrow \underline{T\overline{\mathcal{N}}_{\scriptscriptstyle{D}}}=\mathcal{N}_{\scriptscriptstyle{D}}\times \mathcal{M}_{\scriptscriptstyle{D}}: \quad (Q,\vec{P})\mapsto \big((Q,\vec{P}),((X_{Q\perp}(Q,\vec{P}),\vec{X}_{Q}(Q,\vec{P})), \vec{X}_{\vec{P}}(Q,\vec{P}))\big)
$$
where
\begin{eqnarray*}
\vec{X}_{Q}(Q,\vec{P})=\vec{P}-\vec{\nabla}Q_\perp\,,\quad
 \vec{X}_{\vec{P}}(Q,\vec{P})=-\vec{\nabla}\times \vec{\nabla}\times \vec{Q}\,,
\end{eqnarray*}
and $X_{Q\perp}(Q,\vec{P})$ is any (continuous) function.

\bigskip

As in the previous  cases, we study the Hamiltonian formulation with the help of the GNH algorithm. The fiber derivative
$FL_{\scriptscriptstyle{D}}:\mathbf{H}_\partial^1(\Sigma)\times \mathbf{L}^2(\Sigma)\rightarrow \mathbf{L}^2(\Sigma)\times \mathbf{L}^2(\Sigma)^*$  is given by the expression
$$
FL_{\scriptscriptstyle{D}}(Q,V)=(Q,\langle \vec{V}+\vec{\nabla}Q_\perp, \mathrm{proj}(\cdot) \rangle_{\vec{L}^2})\in \mathbf{H}_\partial^1(\Sigma)\times \mathbf{L}^2(\Sigma)^*\,,
$$
where $\mathrm{proj}(a,\vec{a})=\vec{a}$. As in the case of the scalar field, we will use the Riesz representation theorem to swap $\mathbf{L}^2(\Sigma)^*$ for $\mathbf{L}^2(\Sigma)$ in which case the fiber derivative becomes
$$
FL_{\scriptscriptstyle{D}}(Q,V)=(Q,(0,\vec{V}+\vec{\nabla}Q_\perp))\in \mathbf{H}_\partial^1(\Sigma)\times \mathbf{L}^2(\Sigma)\,.
$$
The image of $\mathbf{H}_\partial^1(\Sigma)\times \mathbf{L}^2(\Sigma)$ under the fiber derivative is
$$
\mathcal{M}_1:=\mathbf{H}_\partial^1(\Sigma)\times(\{0\}\times \vec{L}^2(\Sigma))\cong \mathbf{H}_\partial^1(\Sigma)\times\vec{L}^2(\Sigma)\,.
$$
The pull-back to $\mathcal{M}_1$ of the canonical symplectic form in $\mathbf{L}^2(\Sigma)\times\mathbf{L}^2(\Sigma)$ is
$$
\omega(Q,\vec{P})((q_1,\vec{p}_1),(q_2,\vec{p}_2))=
\langle\vec{q}_1,\vec{p}_2\rangle_{\vec{L}^2}-\langle\vec{q}_2,\vec{p}_1\rangle_{\vec{L}^2}\,
$$
with $(Q,\vec{P}), (q_i,\vec{p}_i)\in\mathbf{H}_\partial^1(\Sigma)\times\vec{L}^2(\Sigma)$. Notice that, at variance with the scalar field case, this symplectic form is degenerate on the primary constraint submanifold $\mathcal{M}_1$.

The energy is now
\begin{eqnarray*}
H_{\scriptscriptstyle{D}}\circ FL_{\scriptscriptstyle{D}}(Q,V)=\frac{1}{2}\langle\vec{V},\vec{V}\rangle_{\vec{L}^2}+\frac{1}{2}\langle \vec{\nabla}\times \vec{Q}, \vec{\nabla}\times \vec{Q} \rangle_{\vec{L}^2}-\frac{1}{2}\langle\vec{\nabla}Q_\perp,\vec{\nabla}Q_\perp\rangle_{\vec{L}^2}
\,,
\end{eqnarray*}
and the Hamiltonian $H_{\scriptscriptstyle{D}}:\mathcal{M}_1\rightarrow \mathbb{R}$  and its differential are given by
\begin{eqnarray*}
H_{\scriptscriptstyle{D}}(Q,\vec{P})&=&\frac{1}{2}\langle\vec{P},\vec{P}\rangle_{\vec{L}^2}-\langle\vec{P},\vec{\nabla}Q_\perp\rangle_{\vec{L}^2}
+
\frac{1}{2}\langle\vec{\nabla}\times\vec{Q},\vec{\nabla}\times\vec{Q}\rangle_{\vec{L}^2}\,.
\\
\mathrm{d}H_{\scriptscriptstyle{D}}(Q,\vec{P})(q,\vec{p})&=&\langle\vec{P},\vec{p}\rangle_{\vec{L}^2}
-\langle\vec{\nabla}Q_\perp,\vec{p}\rangle_{\vec{L}^2}
-\langle\vec{P},\vec{\nabla}q_\perp\rangle_{\vec{L}^2}
+\langle\vec{\nabla}\times\vec{Q},\vec{\nabla}\times\vec{q}\rangle_{\vec{L}^2}\,.
\nonumber
\end{eqnarray*}
Vector fields on $\mathcal{M}_1$ are maps $X:\mathcal{M}_1\rightarrow \mathcal{M}_1\times \mathcal{M}_1:(Q,\vec{P})\mapsto((Q,\vec{P}),(X_Q(Q,\vec{P}),\vec{X}_{\vec{P}}(Q,\vec{P})))$ and, hence,
$$
i_X\omega(Q,\vec{P})(q,\vec{p})=\langle\vec{X}_Q(Q,\vec{P}),\vec{p}\rangle_{\vec{L}^2}-
\langle\vec{X}_{\vec{P}}(Q,\vec{P}),\vec{q}\rangle_{\vec{L}^2}\,.
$$We have to find now a submanifold $\mathcal{M}_2$ with smooth injective immersion $\jmath_2:\mathcal{M}_2\rightarrow \mathcal{M}_1$ such that the equation
$$
(i_X\omega-dH)|_{\jmath_2(\mathcal{M}_2)}=0
$$
can be solved. In order to do this we have to find the general solution to the equation
\begin{eqnarray*}
&&\langle\vec{X}_Q,\vec{p}\rangle_{\vec{L}^2}-
\langle\vec{X}_{\vec{P}},\vec{q}\rangle_{\vec{L}^2}=\langle\vec{P}-\vec{\nabla}Q_\perp,\vec{p}\rangle_{\vec{L}^2}
-\langle\vec{P},\vec{\nabla} q_\perp\rangle_{\vec{L}^2}+
\langle\vec{\nabla}\times\vec{Q},\vec{\nabla}\times\vec{q}\rangle_{\vec{L}^2}\nonumber
\end{eqnarray*}
for all $((q_\perp,\vec{q}),\vec{p})\in \mathbf{H}^1_\partial\times \vec{L}^2$. We have to find first the conditions on $(Q,\vec{P})$ that guarantee that the previous equation can be solved and then get the solutions for the vector field $(X_Q,\vec{X}_{\vec{P}})$. This defines the submanifold
$$
\mathcal{M}_2:=\{(Q,\vec{P})\,:\, Q_\perp\in H^1_{0\perp}(\Sigma),\,\vec{Q}\in \vec{H}^2_\partial(\mathrm{curl},\Sigma),\, \vec{P}\in\vec{H}_0(\mathrm{curl},\Sigma)\cap  \vec{H}(\mathrm{div},\Sigma),\, \vec{\nabla}\cdot\vec{P}=0\}
$$
that is obtained as follows:
\begin{itemize}
\item Consider first the case
 $q_\perp=0$ and $\vec{p}=\vec{0}$. This forces us to take $\vec{Q}\in \vec{H}^2_\partial(\mathrm{curl},\Sigma)$ and then solve (using the boundary conditions and Green's theorem in the form (\ref{Green2}))
$$
-\langle\vec{X}_{\vec{P}},\vec{q}\rangle_{\vec{L}^2}=\langle\vec{\nabla}\times\vec{Q},\vec{\nabla}\times\vec{q}\rangle_{\vec{L}^2}
=\langle\vec{\nabla}\times\vec{\nabla}\times \vec{Q},\vec{q} \rangle_{\vec{L}^2}\,,\quad \forall \vec{q}\in \vec{H}_0(\mathrm{curl},\Sigma)\,,
$$
to get $\vec{X}_{\vec{P}}=-\vec{\nabla}\times\vec{\nabla}\times\vec{Q}$, which is always in $\vec{L}^2$.

\item We consider now the situation where $q_\perp=0$ and $\vec{q}=\vec{0}$ and find $\vec{X}_Q=\vec{P}-\vec{\nabla}Q_\perp$. As $\vec{X}_Q\in \vec{H}_0(\textrm{curl},\Sigma)$ we require that
$\vec{P}-\vec{\nabla}Q_\perp \in \vec{H}(\mathrm{curl},\Sigma)$ and  $\vec{n}\times(\vec{P}-\vec{\nabla}Q_\perp)|_{\partial\Sigma}=0$. However, $Q_\perp\in H^1_{0\perp}(\Sigma)$ implies that $\vec{\nabla}Q_\perp\in \vec{H}(\mathrm{curl},\Sigma)$ and the condition $Q_\perp|_{\partial\Sigma}=0$ implies $\vec{n}\times \vec{\nabla}Q_\perp|_{\partial\Sigma}=0$. This means that the constraint  $\vec{n}\times(\vec{P}-\vec{\nabla}Q_\perp)|_{\partial\Sigma}=0$ is equivalent to $\vec{n}\times\vec{P}|_{\partial\Sigma}=0$ and, hence, we will require that $\vec{P}\in \vec{H}_0(\mathrm{curl},\Sigma)$.

\item Finally, if $\vec{q}=\vec{p}=\vec{0}$ the condition $\langle\vec{P},\vec{\nabla} q_\perp\rangle_{\vec{L}^2}=0$  implies that $0=\vec{\nabla}\cdot \vec{P}\in L^2(\Sigma)$ and, hence $\vec{P}\in \vec{H}_0(\mathrm{curl},\Sigma)\cap  \vec{H}(\mathrm{div},\Sigma)$ satisfying the additional condition  $\vec{\nabla}\cdot \vec{P}=0$. In this process the component $X_{Q\perp}$ is left arbitrary.
\end{itemize}

We have then found the continuous vector field $X:\mathcal{M}_2\rightarrow \mathcal{M}_2\times \mathcal{M}_1$ given by
\begin{eqnarray*}
& &\vec{X}_Q(Q,\vec{P})=\vec{P}-\vec{\nabla}Q_\perp\,,\\
& & \vec{X}_{\vec{P}}(Q,\vec{P})=-\vec{\nabla}\times\vec{\nabla}\times\vec{Q}\,.
\end{eqnarray*}
and $X_{Q\perp}:\mathcal{M}_2\rightarrow H^1_{0\perp}$ any  arbitrary continuous function.

We have to check now if the vector fields obtained in the previous step are tangent to the closure of $\mathcal{M}_2$ in $\mathcal{M}_1$. If this is so the GNH algorithm terminates because we would have $\mathcal{M}_3=\mathcal{M}_2$. We will see that this is the case. First, notice that (see Appendix \ref{functionalspaces})
\begin{eqnarray*}
\overline{\mathcal{M}}_2&=&\mathrm{cl}_{\mathcal{M}_1}(\mathcal{M}_2)\\
&=&\mathrm{cl}_{\mathbf{H}^1_\partial}\big(H^1_0(\Sigma)\times \vec{H}^2_\partial(\mathrm{curl},\Sigma)\big)\times \mathrm{cl}_{\vec{L}^2}\big(\{\vec{P}\in\vec{H}_0(\mathrm{curl},\Sigma)\cap  \vec{H}(\mathrm{div},\Sigma)):\vec{\nabla}\cdot\vec{P}=0\}\big)\\
&=&\mathbf{H}^1_\partial(\Sigma)\times (\vec{L}^2_{h_{\scriptscriptstyle{D}}}(\Sigma)\oplus\vec{L}^2_{T_{\scriptscriptstyle{D}}}(\Sigma))\,.
\end{eqnarray*}

We have to find those points  $(Q,\vec{P})\in \mathcal{M}_2$ for which the vector field $X$ is tangent to $\overline{\mathcal{M}}_2$, i.e. such that
\begin{eqnarray*}
&&X_{Q\perp}(Q,\vec{P})\in H^1_{0\perp}(\Sigma)\,,\\
&&\vec{X}_{Q}(Q,\vec{P})=\vec{P}-\vec{\nabla}Q_\perp\in \vec{H}_0(\mathrm{curl},\Sigma)\,,\\
&& \vec{X}_{\vec{P}}(Q,\vec{P})=-\vec{\nabla}\times \vec{\nabla}\times \vec{Q}\in \vec{L}^2_{h_{\scriptscriptstyle{D}}}(\Sigma)\oplus\vec{L}^2_{T_{\scriptscriptstyle{D}}}(\Sigma)\,.
\end{eqnarray*}

These conditions are satisfied in the whole of $\mathcal{M}_2$. The first two can be trivially checked whereas the last one is proved in Appendix \ref{functionalspaces}. We see then that the GNH algorithm stops. We have identified the submanifold $\mathcal{M}_2$ where the dynamics is well defined as well as the form of the Hamiltonian vector fields whose integral curves give the dynamics of the system.

The presence of arbitrary functions in $X$ signals the existence of gauge symmetries. In fact, the (generalized) submanifold $\mathcal{M}_2\stackrel{\jmath_2}{\rightarrow} \mathcal{M}_1$ of $(\mathcal{M}_1,\omega)$ is a \emph{first class} submanifold. In order to see this we have to check that $T\mathcal{M}_2^\perp\subset j_{2*}(T\mathcal{M}_2)$, where
$$
T\mathcal{M}_2^\perp:=\{Z\in T\mathcal{M}_1|_{\mathcal{M}_2} \,:\, \omega|_{\mathcal{M}_2}(Z,Y)=0\,,\, \forall Y\in j_{2*}(T\mathcal{M}_2)\}\,.
$$
To this end we show that
$$
Z\in T\mathcal{M}_2^\perp \Leftrightarrow Z_{Q\perp}\in H^1_{0\perp}(\Sigma)\,,\, \vec{Z}_{\vec{Q}}\in \vec{\nabla}H^1_0(\Sigma)\subset \vec{H}^2_\partial(\mathrm{curl},\Sigma)\,,\, \vec{Z}_{\vec{P}}=\vec{0}\in \vec{C}_{\scriptscriptstyle{D}}(\Sigma)
$$
where
$$
\vec{C}_{\scriptscriptstyle{D}}(\Sigma):=\{\vec{P}\in \vec{H}_0(\mathrm{curl},\Sigma)\cap\vec{H}(\mathrm{div},\Sigma) :\vec{\nabla}\cdot\vec{P}=0\}\,.
$$
The vector fields $Y\in\jmath_{2*}(T\mathcal{M}_2)$ can be thought of as maps
$$
Y:\mathcal{M}_2\rightarrow \mathcal{M}_2\times \mathcal{M}_2:(Q,\vec{P})\mapsto((Q,\vec{P}),(Y_Q(Q,\vec{P}),\vec{Y}_{\vec{P}}(Q,\vec{P})))\,,
$$
that is, $(Q,\vec{P})\in\mathcal{M}_2$, $Y_{Q\perp}(Q,\vec{P})\in H^1_{0\perp}$, $\vec{Y}_{\vec{Q}}(Q,\vec{P})\in \vec{H}^2_\partial(\mathrm{curl},\Sigma)$ and $\vec{Y}_{\vec{P}}(Q,\vec{P})\in \vec{C}_{\scriptscriptstyle{D}}(\Sigma)$, whereas vector fields $\left.Z\in T\mathcal{M}_1\right|_{\mathcal{M}_2}$ are now maps
$$
Z:\mathcal{M}_2\rightarrow \mathcal{M}_2\times \mathcal{M}_1:(Q,\vec{P})\mapsto((Q,\vec{P}),(Z_Q(Q,\vec{P}),\vec{Z}_{\vec{P}}(Q,\vec{P})))
$$
with $(Q,\vec{P})\in\mathcal{M}_2$, $Z_{Q\perp}(Q,\vec{P})\in H^1_{0\perp}$, $\vec{Z}_{\vec{Q}}(Q,\vec{P})\in \vec{H}(\mathrm{curl},\Sigma)$ and $\vec{Z}_{\vec{P}}(Q,\vec{P})\in \vec{L}^2(\Sigma)$. The condition that defines $T\mathcal{M}_2^\perp$ is
$$
\omega(Q,\vec{P})((Z_Q,\vec{Z}_{\vec{P}}),(Y_Q,\vec{Y}_{\vec{P}}))=\langle \vec{Z}_{\vec{Q}},\vec{Y}_{\vec{P}}\rangle_{\vec{L}^2}-\langle \vec{Z}_{\vec{P}},\vec{Y}_{\vec{Q}}\rangle_{\vec{L}^2}=0\,,
$$
for all the possible values of the fields $Y$ written above. This leads to the following two conditions
\begin{eqnarray*}
&&\langle\vec{Z}_{\vec{P}},\vec{Y}_{\vec{Q}}\rangle_{\vec{L}^2}=0\,,\quad \forall \vec{Y}_{\vec{Q}}\in\vec{H}^2_\partial(\mathrm{curl},\Sigma)\,,\\
&&\langle\vec{Z}_{\vec{Q}},\vec{Y}_{\vec{P}}\rangle_{\vec{L}^2}=0\,,\quad \forall \vec{Y}_{\vec{P}}\in \vec{C}_{\scriptscriptstyle{D}}(\Sigma)\,.
\end{eqnarray*}
The first one implies that $\vec{Z}_{\vec{P}}=\vec{0}$ because $\vec{C}^\infty_0(\Sigma)$ is dense in $\vec{L}^2(\Sigma)$ and $\vec{C}^\infty_0(\Sigma)\subset \vec{H}^2_\partial(\mathrm{curl},\Sigma)$. The second condition implies that $\vec{Z}_{\vec{Q}}\in \vec{C}^\perp_{\scriptscriptstyle{D}}=\vec{L}^2_{L_{\scriptscriptstyle{D}}}=\vec{\nabla}H_0^1$. Finally there is no condition on $Z_{Q\perp}$. As $\vec{0}\in\vec{C}_{\scriptscriptstyle{D}}(\Sigma)$ it only remains to check that $\vec{\nabla}H_0^1\subset \vec{H}^2_\partial(\mathrm{curl},\Sigma)$. This amounts to showing that for every $\varphi \in H^1_0(\Sigma)$, $\vec{\nabla}\varphi\in\vec{H}^2(\mathrm{curl},\Sigma)$, and the trace $(\vec{n}\times\vec{\nabla})|_{\partial\Sigma}$ is defined and it is zero. It is straightforward to prove that this is indeed the case.

\subsubsection{Symplectic Lagrangian approach}

In the symplectic Lagrangian approach there is no primary constraint surface. To begin with we must pull-back the canonical symplectic form from $\mathbf{L}^2(\Sigma)\times\mathbf{L}^2(\Sigma)^*$ to the Hilbert space $\mathcal{M}_1:=\mathbf{H}^1_\partial(\Sigma)\times\mathbf{L}^2(\Sigma)$. By doing this we obtain
\begin{eqnarray*}
&&\Omega_{L_{\scriptscriptstyle{D}}}(Q,V)((q_1,v_1),(q_2,v_2))=\langle\vec{\nabla}q_{2\perp}
+\vec{v}_2,\vec{q}_1\rangle_{\vec{L}^2}-\langle\vec{\nabla}q_{1\perp}+\vec{v}_1,\vec{q}_2\rangle_{\vec{L}^2}\,.
\end{eqnarray*}
The energy and its differential are given by
\begin{eqnarray*}
&&E_{\scriptscriptstyle{D}}(Q,V)=\frac{1}{2}\langle\vec{V},\vec{V}\rangle_{\vec{L}^2}+\frac{1}{2}\langle \vec{\nabla}\times \vec{Q}, \vec{\nabla}\times \vec{Q} \rangle_{\vec{L}^2}-\frac{1}{2}\langle\vec{\nabla}Q_\perp,\vec{\nabla}Q_\perp\rangle_{\vec{L}^2}\nonumber\\
&&\mathrm{d}E_{\scriptscriptstyle{D}}(Q,V)(q,v)=\langle\vec{V},\vec{v}\rangle_{\vec{L}^2}+\langle \vec{\nabla}\times \vec{Q}, \vec{\nabla}\times \vec{q}\rangle_{\vec{L}^2}-\langle\vec{\nabla}Q_\perp,\vec{\nabla}q_\perp\rangle_{\vec{L}^2}\,.\nonumber
\end{eqnarray*}
For a vector field
$$ Y:\mathbf{H}_\partial^1\times\mathbf{L}^2\rightarrow(\mathbf{H}_\partial^1\times\mathbf{L}^2)
\times(\mathbf{H}^1_\partial\times\mathbf{L}^2):(Q,V)\mapsto((Q,V),(Y_Q(Q,V),Y_V(Q,V)))$$
we have
$$
i_Y\Omega_{L_{\scriptscriptstyle{D}}}(Q,V)(q,v)=\langle\vec{Y}_Q,\vec{\nabla}q_{\perp}+\vec{v}\rangle_{\vec{L}^2}
-\langle\vec{\nabla}Y_{Q\perp}+\vec{Y}_V,\vec{q}\rangle_{\vec{L}^2}\,.
$$

We must find now a Banach manifold $\mathcal{M}_2$, with smooth injective immersion $\mathcal{M}_2\stackrel{\jmath_2}{\rightarrow} \mathbf{H}^1_\partial(\Sigma)\times \mathbf{L}^2(\Sigma)$, such that we can solve the equation
$$
(i_Y\Omega_{L_{\scriptscriptstyle{D}}}-\mathrm{d}E_{\scriptscriptstyle{D}})|_{\jmath_2(\mathcal{M}_2)}=0\,.
$$
This amounts to finding the general solution of the equation
\begin{equation}
\langle\vec{Y}_Q,\vec{\nabla}q_{\perp}+\vec{v}\rangle_{\vec{L}^2}
-\langle\vec{\nabla}Y_{Q\perp}+\vec{Y}_V,\vec{q}\rangle_{\vec{L}^2}=
\langle\vec{V},\vec{v}\rangle_{\vec{L}^2}+\langle\vec{\nabla}\times \vec{Q},\vec{\nabla}\times \vec{q}\rangle_{\vec{L}^2}-\langle\vec{\nabla}Q_\perp,\vec{\nabla}q_\perp\rangle_{\vec{L}^2}\,,
\label{ecLagSymD}
\end{equation}
for all $q_\perp\in H^1_{0\perp}$, $\vec{q}\in\vec{H}_0(\mathrm{curl},\Sigma)$, $v_\perp\in L^2_\perp(\Sigma)$ and $\vec{v}\in\vec{L}^2(\Sigma)$. This is solved sequentially as in the case of the Hamiltonian formulation to get the submanifold\footnote{As in the case of the scalar field we make use of the fact that under the regularity conditions that we are imposing on the boundary $\partial\Sigma$ we have $H^1_0(\Sigma)\cap H^1(\Delta,\Sigma)=H^1_0(\Sigma)\cap H^2(\Sigma)$.}
\begin{eqnarray*}
\mathcal{M}_2&=&\{Q_\perp \in H^2_\perp(\Sigma)\cap H^1_{0\perp}(\Sigma),\, \vec{Q}\in\vec{H}^2_\partial(\mathrm{curl},\Sigma),\hspace*{9cm}\\
&&\hspace*{4.1cm}V_\perp\in L^2_\perp(\Sigma),\, \vec{V}\in \vec{H}_0(\mathrm{curl},\Sigma)\cap\vec{H}(\mathrm{div},\Sigma), \, \vec{\nabla}\cdot(\vec{V}+\vec{\nabla}Q_\perp)=0\}
\end{eqnarray*}
and
$\vec{Y}_Q=\vec{V}$, $\vec{Y}_V=-\vec{\nabla}\times\vec{\nabla}\times \vec{Q}-\vec{\nabla}Y_{Q\perp}$; with $Y_{Q\perp}$ and $Y_{V\perp}$ arbitrary within the spaces where they are defined ($H^1_{0\perp}(\Sigma)$ and $L^2(\Sigma)$ respectively).

The condition $\vec{\nabla}\cdot(\vec{V}+\vec{\nabla}Q_\perp)=0$ involves both $\vec{V}$ and $Q_\perp$. This is mildly inconvenient when checking that the GNH algorithm stops at this stage. However, it suggests a simple way to proceed. Let us introduce
\begin{eqnarray*}
\widetilde{\mathcal{M}}_2&=&\{Q_\perp \in H^2_\perp(\Sigma)\cap H^1_{0\perp}(\Sigma),\,\vec{Q}\in\vec{H}_\partial^2(\mathrm{curl},\Sigma),\hspace*{5cm}\\
&&\hspace*{5.5cm}P_\perp\in L^2_\perp,\,\vec{P}\in\vec{H}_0(\mathrm{curl},\Sigma)\cap\vec{H}(\mathrm{div},\Sigma),\, \vec{\nabla}\cdot\vec{P}=0\}
\end{eqnarray*}
and the linear isomorphism $F:\mathcal{M}_2\rightarrow\widetilde{\mathcal{M}}_2:(Q,V)\mapsto(Q,(V_\perp,\vec{V}+\vec{\nabla}Q_\perp))$. Notice that this is well defined because $Q_\perp|_{\partial\Sigma}=0$ implies $\vec{n}\times\vec{\nabla}Q_\perp|_{\partial\Sigma}=\vec{0}$. Also, under this map the field $Y:\mathcal{M}_2\rightarrow\mathcal{M}_1$ transforms into $X:\widetilde{\mathcal{M}}_2\rightarrow \mathcal{M}_1$ given by $\vec{X}_Q=\vec{P}-\vec{\nabla}Q_\perp$, $\vec{X}_{\vec{P}}=-\vec{\nabla}\times\vec{\nabla}\times\vec{Q}$ with $X_{Q\perp}$ and $X_{P\perp}$ arbitrary within the spaces where they are defined. At this point we can use the results that we obtained in the study of the Hamiltonian formulation to guarantee that the GNH algorithm stops and also that the generalized submanifold $\mathcal{M}_2$ is first class\footnote{See, however, the footnote at the end of Subsection \ref{subH}.}.

As we can see the Hamiltonian and the symplectic-Lagrangian formalisms are very similar. The primary constraint submanifold in the Hamiltonian formulation corresponds to the points in $\widetilde{\mathcal{M}}_2$ with $P_\perp=0$. Within this submanifold the only difference is that the $Q_\perp$ fields live in different spaces\footnote{Notice, however, that the $Q_\perp$ are physically irrelevant.} and the presence of an extra arbitrary vector field component $X_{Q\perp}$ due to so called \emph{second order problem} \cite{Gotaythesis}. Notice, however, that in order to avoid the second order problem \cite{Gotaythesis} it suffices to fix its value to $X_{Q\perp}=P_\perp$ (or, equivalently, $Y_{Q\perp}=V_\perp$).

\subsection{Neumann  boundary conditions}

We will study here the dynamics of the electromagnatic field on a bounded domain $\Sigma \subset\mathbb{R}^3$  subject to boundary conditions that generalize the Neumann boundary conditions for the scalar field:
\begin{eqnarray}
&\ddot{\vec{A}}+\vec{\nabla}\dot{A}_\perp-\Delta\vec{A}+\vec{\nabla}(\vec{\nabla}\cdot \vec{A})=\vec{0}&\hspace*{4mm}\mathrm{in}\quad(t_1,t_2)\times\Sigma
\label{MN1}\\
&\vec{\nabla}\cdot(\dot{\vec{A}}+\vec{\nabla}A_\perp)=0&\hspace*{4mm}\mathrm{in}\quad(t_1,t_2)\times\Sigma\label{MN2}\\
&\vec{n}\cdot (\dot{\vec{A}}+\vec{\nabla}A_\perp)=0&\hspace*{4mm}\mathrm{in}\quad(t_1,t_2)\times\partial\Sigma\label{BN1}\\
&\vec{n}\times (\vec{\nabla}\times \vec{A})=\vec{0}&\hspace*{4mm}\mathrm{in}\quad(t_1,t_2)\times\partial\Sigma\label{BN2}\\
&\vec{A}(t_1)=\vec{Q}_1\,,\, A_\perp(t_1)=Q_{\perp1}\,,& \hspace*{2mm}\vec{A}(t_2)=\vec{Q}_2\,,\, A_\perp(t_2)=Q_{\perp2}\label{inicMN}
\end{eqnarray}
with $\vec{Q}_i\in \vec{H}^2(\mathrm{curl},\Sigma)$ and $Q_{\perp i}\in H^1_\perp(\Sigma)$.

\subsubsection{Variational approach}

The action is defined on the space of curves
\begin{eqnarray*}
& & \mathcal{C}_{\scriptscriptstyle{N}}(Q_1,Q_2,[t_1,t_2])=\{A\in \mathcal{C}_{\scriptscriptstyle{N}}^0([t_1,t_2],\Sigma) \cap \mathcal{C}_{\scriptscriptstyle{N}}^1([t_1,t_2],\Sigma) \cap\mathcal{C}_{\scriptscriptstyle{N}}^2([t_1,t_2],\Sigma) :A(t_i)\!=\!Q_i,\, i\!=\!1,2\}\}
\end{eqnarray*}
where
\begin{eqnarray*}
\mathcal{C}_{\scriptscriptstyle{N}}^0([t_1,t_2],\Sigma) &:=&C^0\big([t_1,t_2],\mathbf{H}^2(\Sigma)\big)\,,\\
\mathcal{C}_{\scriptscriptstyle{N}}^1([t_1,t_2],\Sigma)&:=&C^1\big([t_1,t_2],L^2_\perp(\Sigma)\times(\vec{H}(\mathrm{curl},\Sigma)\cap\vec{H}(\mathrm{div},\Sigma))\big)\,,\\
\mathcal{C}_{\scriptscriptstyle{N}}^2([t_1,t_2],\Sigma)&:=& C^2\big([t_1,t_2],\vec{L}^2(\Sigma)\big)\,.
\nonumber
\end{eqnarray*}
and we have used the notation $\mathbf{H}^2(\Sigma):=H^2_\perp(\Sigma)\times\vec{H}^2(\mathrm{curl},\Sigma)$.
The tangent spaces at $A\in\mathcal{C}_{\scriptscriptstyle{N}}(Q_1,Q_2,[t_1,t_2])$ are $T_A\mathcal{C}_{\scriptscriptstyle{N}}(Q_1,Q_2,[t_1,t_2])=\mathcal{C}_{\scriptscriptstyle{N}}(0,0,[t_1,t_2])$.

The action $S_{\scriptscriptstyle{N}}:\mathcal{C}_{\scriptscriptstyle{N}}(Q_1,Q_2,[t_1,t_2])\rightarrow \mathbb{R}$ is given by
\begin{eqnarray*}
S_{\scriptscriptstyle{N}}(A)&=&\int_{t_1}^{t_2}L_{\scriptscriptstyle{N}}(A(t),\dot{A}(t))\mathrm{d}t\\
&=&\int_{t_1}^{t_2}\left(\frac{1}{2}\langle\dot{\vec{A}},\dot{\vec{A}}\rangle_{\vec{L}^2}
+\langle\vec{\nabla}A_\perp,\dot{\vec{A}}\rangle_{\vec{L}^2}
+\frac{1}{2}\langle\vec{\nabla}A_\perp,\vec{\nabla}A_\perp\rangle_{\vec{L}^2}-\frac{1}{2}\langle \vec{\nabla}\times \vec{A}, \vec{\nabla}\times \vec{A} \rangle_{\vec{L}^2}\right)\mathrm{d}t\,,\nonumber
\end{eqnarray*}
and is differentiable in its domain. The differential of $S_{\scriptscriptstyle{N}}$ at $A$ acting on a vector $\delta\in T_A\mathcal{C}_{\scriptscriptstyle{N}}(Q_1,Q_2,[t_1,t_2])$ is
\begin{eqnarray*}
\mathrm{d}S_{\scriptscriptstyle{N}}(A)\cdot\delta
&=&\int_{t_1}^{t_2}\!\!\mathrm{d}t\int_\Sigma \big(-\delta_\perp(\Delta A_\perp+\vec{\nabla}\cdot \dot{\vec{A}})+\vec{\delta}\cdot(\Delta \vec{A}-\ddot{\vec{A}}-\vec{\nabla}(\vec{\nabla}\cdot \vec{A})-\vec{\nabla}\dot{A}_\perp)\big)\mathrm{vol}_\Sigma\,,
\nonumber\\
&+&\int_{t_1}^{t_2}\!\!\mathrm{d}t\int_{\partial\Sigma} \delta_{\perp} (\vec{n}\cdot(\dot{\vec{A}}+\vec{\nabla}A_\perp))|_{\partial\Sigma}\,\mathrm{vol}_{\partial\Sigma}
+\int_{t_1}^{t_2}\!\!\mathrm{d}t\int_{\partial\Sigma} \vec{\delta}\cdot (\vec{n}\times \vec{\nabla}\times \vec{A})|_{\partial\Sigma}\,\mathrm{vol}_{\partial\Sigma}
\end{eqnarray*}
Hence, the condition $\mathrm{d}S_{\scriptscriptstyle{N}}(A)\cdot \delta=0$ for all the vectors $\delta\in T_A\mathcal{C}_{\scriptscriptstyle{N}}$ implies equations (\ref{MN1})-(\ref{inicMN})

\subsubsection{Hamiltonian approach}\label{subH}

We will show that the Hamiltonian dynamics of the electromagnetic field with Neumann  conditions takes place in the
 first class (generalized) submanifold $$\mathcal{N}_{\scriptscriptstyle{N}}:=\{(Q,\vec{P})\in \mathbf{H}^2(\Sigma)\times(\vec{H}(\mathrm{curl},\Sigma)\cap  \vec{H}_0(\mathrm{div},\Sigma))\,:\,\vec{\nabla}\cdot\vec{P}=0\,,\quad \vec{n}\times (\vec{\nabla}\times \vec{Q})|_{\partial\Sigma}=\vec{0}\}$$   of the presimplectic manifold  $(\mathcal{M}_{\scriptscriptstyle{N}},\omega_{\scriptscriptstyle{N}})$, where $$\mathcal{M}_{\scriptscriptstyle{N}}=\mathbf{H}^1(\Sigma)\times\vec{L}^2(\Sigma)$$
  and  $\omega_{\scriptscriptstyle{N}}$ is the pullback to $\mathcal{M}_{\scriptscriptstyle{N}}$ of the strong, canonical, symplectic form on $\mathbf{L}^2(\Sigma)\times \mathbf{L}^2(\Sigma)$. The class of Hamiltonian vector fields that defines the dynamics of the system is given by
 $$
X_{\scriptscriptstyle{N}}:\mathcal{N}_{\scriptscriptstyle{N}}\rightarrow \underline{T\overline{\mathcal{N}}_{\scriptscriptstyle{N}}}=\mathcal{N}_{\scriptscriptstyle{N}}\times \mathcal{M}_{\scriptscriptstyle{N}}: \quad (Q,\vec{P})\mapsto \big((Q,\vec{P}),((X_{Q\perp}(Q,\vec{P}),\vec{X}_{Q}(Q,\vec{P})), \vec{X}_{\vec{P}}(Q,\vec{P}))\big)
$$
where
\begin{eqnarray*}
\vec{X}_{Q}(Q,\vec{P})=\vec{P}-\vec{\nabla}Q_\perp\,,\quad
 \vec{X}_{\vec{P}}(Q,\vec{P})=-\vec{\nabla}\times \vec{\nabla}\times \vec{Q}\,,
\end{eqnarray*}
and $X_{Q\perp}(Q,\vec{P})$ is any (continuous) function.

\bigskip

In order to make use of the GNH algorithm, let us consider first the fiber derivative
$FL_{\scriptscriptstyle{N}}:\mathbf{H}^1(\Sigma)\times \mathbf{L}^2(\Sigma)\rightarrow \mathbf{L}^2(\Sigma)\times \mathbf{L}^2(\Sigma)^*$ given by the expression
$$
FL_{\scriptscriptstyle{N}}(Q,V)=(Q,(0,\vec{V}+\vec{\nabla}Q_\perp))\in \mathbf{H}^1(\Sigma)\times \mathbf{L}^2(\Sigma)\,,
$$
where we have used the Riesz representation theorem to identify $\mathbf{L}^{2}(\Sigma)^*$ and $\mathbf{L}^{2}(\Sigma)$.
The image of $\mathbf{H}^1(\Sigma)\times \mathbf{L}^2(\Sigma)$ under the fiber derivative is
$$
\mathcal{M}_1:=\mathbf{H}^1(\Sigma)\times(\{0\}\times \vec{L}^2(\Sigma))\cong \mathbf{H}^1(\Sigma)\times\vec{L}^2(\Sigma)\,.
$$
The pull-back of the canonical symplectic form in $\mathbf{L}^2(\Sigma)\times\mathbf{L}^2(\Sigma)$ to $\mathcal{M}_1$ is
$$
\omega(Q,\vec{P})((q_1,\vec{p}_1),(q_2,\vec{p}_2))=
\langle\vec{q}_1,\vec{p}_2\rangle_{\vec{L}^2}-\langle\vec{q}_2,\vec{p}_2\rangle_{\vec{L}^2}\,
$$
with $(Q,\vec{P}), (q_i,\vec{p}_i)\in\mathbf{H}^1(\Sigma)\times\vec{L}^2(\Sigma)$.
The energy is now
\begin{eqnarray*}
H_{\scriptscriptstyle{N}}\circ FL(Q,V)=\frac{1}{2}\langle\vec{V},\vec{V}\rangle_{\vec{L}^2}+\frac{1}{2}\langle \vec{\nabla}\times \vec{Q}, \vec{\nabla}\times \vec{Q} \rangle_{\vec{L}^2}-\frac{1}{2}\langle\vec{\nabla}Q_\perp,\vec{\nabla}Q_\perp\rangle_{\vec{L}^2}\,,
\end{eqnarray*}
and the Hamiltonian $H_{\scriptscriptstyle{N}}:\mathcal{M}_1\rightarrow \mathbb{R}$  and its differential are given by
\begin{eqnarray*}
H_{\scriptscriptstyle{N}}(Q,\vec{P})&=&\frac{1}{2}\langle\vec{P},\vec{P}\rangle_{\vec{L}^2}
-\langle\vec{P},\vec{\nabla}Q_\perp\rangle_{\vec{L}^2}
+
\frac{1}{2}\langle\vec{\nabla}\times\vec{Q},\vec{\nabla}\times\vec{Q}\rangle_{\vec{L}^2}\,.
\\
\mathrm{d}H_{\scriptscriptstyle{N}}(Q,\vec{P})(q,\vec{p})&=&\langle\vec{P},\vec{p}\rangle_{\vec{L}^2}
-\langle\vec{\nabla}Q_\perp,\vec{p}\rangle_{\vec{L}^2}
-\langle\vec{P},\vec{\nabla}q_\perp\rangle_{\vec{L}^2}
+\langle\vec{\nabla}\times\vec{Q},\vec{\nabla}\times\vec{q}\rangle_{\vec{L}^2}\,.
\end{eqnarray*}
Vector fields on $\mathcal{M}_1$ are maps $X:\mathcal{M}_1\rightarrow \mathcal{M}_1\times \mathcal{M}_1:(Q,\vec{P})\mapsto((Q,\vec{P}),(X_Q(Q,\vec{P}),\vec{X}_{\vec{P}}(Q,\vec{P})))$ and, hence,
$$
i_X\omega(Q,\vec{P})(q,\vec{p})=\langle\vec{X}_Q(Q,\vec{P}),\vec{p}\rangle_{\vec{L}^2}-
\langle\vec{X}_{\vec{P}}(Q,\vec{P}),\vec{q}\rangle_{\vec{L}^2}\,.
$$
We have to find now a submanifold $\mathcal{M}_2$ with smooth injective immersion $\jmath_2:\mathcal{M}_2\rightarrow \mathcal{M}_1$ such that the equation
$$
(i_X\omega-dH_{\scriptscriptstyle{N}})|_{\jmath_2(\mathcal{M}_2)}=0
$$
can be solved. In order to do this we have to find the general solution to the equation
\begin{eqnarray*}
&&\langle\vec{X}_Q,\vec{p}\rangle_{\vec{L}^2}-
\langle\vec{X}_{\vec{P}},\vec{q}\rangle_{\vec{L}^2}=\langle\vec{P}-\vec{\nabla}Q_\perp,\vec{p}\rangle_{\vec{L}^2}
-\langle\vec{P},\vec{\nabla} q_\perp\rangle_{\vec{L}^2}+
\langle\vec{\nabla}\times\vec{Q},\vec{\nabla}\times\vec{q}\rangle_{\vec{L}^2}\,.\nonumber
\end{eqnarray*}
for all $((q_\perp,\vec{q}),\vec{p})\in \mathbf{H}^1(\Sigma)\times \vec{L}^2(\Sigma)$. The conditions for $(Q,\vec{P})$ that guarantee that the previous equation can be solved and the solutions for the vector field $(X_Q,\vec{X}_{\vec{P}})$ are obtained by following the same steps that we have detailed in the case of the Dirichlet boundary conditions. We get the submanifold
$$
\mathcal{M}_2:=\{(Q,\vec{P})\in \mathbf{H}^2(\Sigma)\times(\vec{H}(\mathrm{curl},\Sigma)\cap  \vec{H}(\mathrm{div},\Sigma)):\vec{\nabla}\cdot\vec{P}=0\,, \vec{n}\cdot \vec{P}|_{\partial\Sigma}=0\,, \vec{n}\times(\vec{\nabla}\times\vec{Q})|_{\partial\Sigma}=\vec{0}\}\,,
$$
where $\mathbf{H}^2(\Sigma)=H^1_\perp(\Sigma)\times \vec{H}^2(\mathrm{curl},\Sigma)$,  and  the continuous vector field $X:\mathcal{M}_2\rightarrow \mathcal{M}_2\times \mathcal{M}_1$ is given by
\begin{eqnarray*}
& &\vec{X}_Q(Q,\vec{P})=\vec{P}-\vec{\nabla}Q_\perp\,,\\
& & \vec{X}_{\vec{P}}(Q,\vec{P})=-\vec{\nabla}\times\vec{\nabla}\times\vec{Q}\,.
\end{eqnarray*}
with $X_{Q\perp}:\mathcal{M}_2\rightarrow H^1_{\perp}$ an arbitrary continuous function.

We check now that the vector fields obtained in the previous step are tangent to the closure of $\mathcal{M}_2$ in $\mathcal{M}_1$. To this end we must first compute  (see Appendix \ref{functionalspaces})
\begin{eqnarray*}
\overline{\mathcal{M}}_2&=&\mathrm{cl}_{\mathcal{M}_1}(\mathcal{M}_2)=\mathbf{H}^1(\Sigma)\times (\vec{L}^2_{h_{\scriptscriptstyle{N}}}(\Sigma)\oplus\vec{L}^2_{T_{\scriptscriptstyle{N}}}(\Sigma))\,,
\end{eqnarray*}
and then find the points $(Q,\vec{P})\in \mathcal{M}_2$ such that the field $X$ is tangent to  $\overline{\mathcal{M}}_2$. i.e. such that
\begin{eqnarray*}
&&X_{Q\perp}(Q,\vec{P})\in H^1_{\perp}(\Sigma)\\
&&\vec{X}_{Q}(Q,\vec{P})=\vec{P}-\vec{\nabla}Q_\perp\in \vec{H}(\mathrm{curl},\Sigma)\\
&& \vec{X}_{\vec{P}}(Q,\vec{P})=-\vec{\nabla}\times \vec{\nabla}\times \vec{Q}\in \vec{L}^2_{h_{\scriptscriptstyle{N}}}(\Sigma)\oplus\vec{L}^2_{T_{\scriptscriptstyle{N}}}(\Sigma)\,.
\end{eqnarray*}
These conditions hold for $(Q,\vec{P})\in \mathcal{M}_2$, hence the GNH algorithm stops at this stage. The only non-trivial condition to check is the third (which is proved en Appendix \ref{functionalspaces}).

We end this section by showing that the generalized submanifold $\mathcal{M}_2\stackrel{\jmath_2}{\rightarrow} \mathcal{M}_1$ is first class, as in the case of the Dirichlet boundary conditions. Proceeding as before we first show that
$$
Z\in T\mathcal{M}_2^\perp \Leftrightarrow Z_{Q\perp}\in H^1_{\perp}(\Sigma)\,,\, \vec{Z}_{\vec{Q}}\in \vec{\nabla}H^1(\Sigma)\subset \vec{F}_{\scriptscriptstyle{N}}(\Sigma)\,,\, \vec{Z}_{\vec{P}}=\vec{0}\in \vec{C}_{\scriptscriptstyle{N}}(\Sigma)\,.
$$
where
\begin{eqnarray*}
\vec{C}_{\scriptscriptstyle{N}}&:=&\{\vec{P}\in \vec{H}(\mathrm{curl},\Sigma)\cap\vec{H}(\mathrm{div},\Sigma) :\vec{\nabla}\cdot\vec{P}=0,\, \vec{n}\cdot \vec{P}|_{\partial\Sigma}=0\}\,,\\
\vec{F}_{\scriptscriptstyle{N}}&:=&\{\vec{Q}\in \vec{H}^2(\mathrm{curl},\Sigma):\vec{n}\times(\vec{\nabla}\times\vec{Q})|_{\partial\Sigma}=0\}\,.
\end{eqnarray*}
The two conditions that must be satisfied now are
\begin{eqnarray*}
&&\langle\vec{Z}_{\vec{P}},\vec{Y}_{\vec{Q}}\rangle_{\vec{L}^2}=0\,,\quad \forall \vec{Y}_{\vec{Q}}\in\vec{F}_{\scriptscriptstyle{N}}(\Sigma)\,,\\
&&\langle\vec{Z}_{\vec{Q}},\vec{Y}_{\vec{P}}\rangle_{\vec{L}^2}=0\,,\quad \forall \vec{Y}_{\vec{P}}\in \vec{C}_{\scriptscriptstyle{N}}(\Sigma)\,.
\end{eqnarray*}
The first one implies that $\vec{Z}_{\vec{P}}=\vec{0}$ because $\vec{C}^\infty_0(\Sigma)$ is dense in $\vec{L}^2(\Sigma)$ and $\vec{C}^\infty_0(\Sigma)\subset \vec{F}_{\scriptscriptstyle{N}}(\Sigma)$. The second condition implies that $\vec{Z}_{\vec{Q}}\in \vec{C}^\perp_{\scriptscriptstyle{N}}=\vec{L}^2_{NL}=\vec{\nabla}H^1$. There is no condition on $Z_{Q\perp}$. As $\vec{0}\in\vec{C}_{\scriptscriptstyle{N}}(\Sigma)$ we just have to check that $\vec{\nabla}H^1\subset \vec{F}_{\scriptscriptstyle{N}}(\Sigma)$, which is straightforward. Hence, we can conclude that $T\mathcal{M}_2$ is  first class\footnote{Notice that, strictly specking, $T\mathcal{M}_2^\perp\not\subset j_{2*}T\mathcal{M}_2$ because of the component $ Z_{Q\perp}\in H^1_{\perp}(\Sigma) \not\subset   H^2_{\perp}(\Sigma)$. This minor problem can be easily solved in several ways, for example modifying  the manifold domain $\mathcal{D}_{\scriptscriptstyle{N}}$ by allowing only fields $Q_\perp\in H^2_{\perp}(\Sigma)$ or generalizing the first class condition to  $T\mathcal{N}^\perp \subset \underline{\overline{T} \mathcal{N}}:= T \overline{\mathcal{N}} |_{\mathcal{N}}$.}.

\subsubsection{Symplectic Lagrangian approach}

The analysis of the symplectic Lagrangian approach for the Neumann problem is very similar to the one that we presented for the Dirichlet case so we will only highlight those points where the computations differ from the ones that we gave for that case (keeping in mind that we will be working here with different functional spaces). The most important difference concerns the analogous of equation (\ref{ecLagSymD}) that becomes now
\begin{eqnarray*}
\langle\vec{Y}_Q,\vec{\nabla}q_{\perp}+\vec{v}\rangle_{\vec{L}^2}
-\langle\vec{\nabla}Y_{Q\perp}+\vec{Y}_V,\vec{q}\rangle_{\vec{L}^2}=
\langle\vec{V},\vec{v}\rangle_{\vec{L}^2}+\langle\vec{\nabla}\times \vec{Q},\vec{\nabla}\times \vec{q}\rangle_{\vec{L}^2}-\langle\vec{\nabla}Q_\perp,\vec{\nabla}q_\perp\rangle_{\vec{L}^2}\,,
\end{eqnarray*}
for all $q_\perp\in H^1_{\perp}(\Sigma)$, $\vec{q}\in\vec{H}(\mathrm{curl},\Sigma)$, $v_\perp\in L^2_\perp(\Sigma)$ and $\vec{v}\in\vec{L}^2(\Sigma)$. In order for it to have a solution we have to restrict ourselves to the generalized submanifold
\begin{eqnarray*}
\mathcal{M}_2&=&\{Q_\perp \in H^2_\perp(\Sigma),\vec{Q}\in\vec{H}^2(\mathrm{curl},\Sigma),V_\perp\in L^2_\perp(\Sigma),\vec{V}\in \vec{H}(\mathrm{curl},\Sigma)\cap\vec{H}(\mathrm{div},\Sigma),\hspace*{1cm}\\
&&\hspace*{3.7cm} \vec{\nabla}\cdot(\vec{V}+\vec{\nabla}Q_\perp)=0, \,\, \vec{n}\cdot(\vec{V}+\vec{\nabla}Q_\perp)|_{\partial\Sigma}=0, \,\, \vec{n}\times(\vec{\nabla}\times\vec{Q})|_{\partial\Sigma}=\vec{0}\}\,.
\end{eqnarray*}
The form of the various components of the vector field $Y$ are the same as for the Dirichlet boundary conditions, i.e. $\vec{Y}_Q=\vec{V}$, $\vec{Y}_V=-\vec{\nabla}\times\vec{\nabla}\times \vec{Q}-\vec{\nabla}Y_{Q\perp}$; with $Y_{Q\perp}$ and $Y_{V\perp}$ arbitrary within the spaces where they are defined ($H^1_{\perp}(\Sigma)$ and $L^2(\Sigma)$ respectively). The rest of the discussion is essentially the same so we leave it here.

\section{Conclusions}\label{conclusions}

The main result of the paper is the  rigorous analysis of the Hamiltonian and symplectic-Lagrangian formalisms for scalar and electromagnetic fields in the presence of boundaries. By using the GNH algorithm we have been able to provide a precise description of the infinite dimensional manifolds where the dynamics is defined and the Hamiltonian vector fields whose integral curves give the time evolution for these systems. These type of results complement the traditional analysis of the PDE's describing the dynamics of these models (the wave and Maxwell equations with appropriate boundary conditions, see, for example \cite{Taylor}) and mesh nicely with them. The physical relevance of the present work is due to the fact that the Hamiltonian framework is a convenient starting point to quantize these field theories.

The (generalized) constraint submanifolds that we have found crucially depend on the boundary conditions that define each of the models. The details of their obtention highlight the similarities and differences between the different models and provide an instructive perspective on the incorporation of boundaries to more complicated systems that we plan to exploit in the future. From a geometric point of view these submanifolds have been found to be second class in the case of the scalar field and first class for the electromagnetic field. This is an intrinsic characterization with an invariant geometric (i.e. coordinate independent) meaning.

The geometric point of view that we are emphasizing here, specifically the structure of the Hamiltonian vector fields and the Hodge decomposition, provides a natural description of the reduced phase space of the electromagnetic field. In particular,  the points $((\vec{Q}_h,\vec{Q}_T),(\vec{P}_h,\vec{P}_T))$ of the reduced phase space correspond to the transverse and harmonic sectors of the Hodge decomposition (associated with the specific boundary conditions). In all the cases the evolution equations reduce to
\begin{eqnarray*}
\dot{\vec{Q}}_T=\vec{P}_T\,,\quad \dot{\vec{P}}_T=-\vec{\nabla}\times\vec{\nabla}\times \vec{Q}_T =\Delta \vec{Q}_T \,,\quad \dot{\vec{Q}}_h=\vec{P}_h\,, \quad \dot{\vec{P}}_h=\vec{0}
\end{eqnarray*}
where the Laplacian corresponds to the boundary conditions used.  The general features of these equations are discussed in Appendix \ref{abstractwave} in general context provided by the abstract wave equation.  In practice it is convenient to work with a parametrization in terms of the eigenfunctions and eigenvalues of the Laplace operator. This description provides a natural avenue to the Fock quantization of these models. It is important to mention, however, that the eigenvalues and eigenvectors of the Laplacian in the presence of sufficiently \textit{irregular} boundaries present peculiarities that are absent in the case of the regular boundaries that we have considered here. This is a generalization of the present work that may lead to interesting results when these types of models are quantized. Another type of generalization can be obtained by considering general spatial Riemannian manifolds and not just subsets of $\mathbb{R}^3$. We expect that the methods used here can be used to understand these more complicated systems.

We have obtained both the Hamiltonian and Lagrangian-symplectic descriptions both for scalar and electromagnetic fields. The main difference between both points of view is due to the so called second order problem \cite{Gotaythesis} that, in the case of the electromagnetic field, introduces an extra indeterminacy in the Hamiltonian vector fields. Once this is solved by imposing the natural ``second order conditions'' both descriptions are essentially equivalent.

We want to add several comments regarding the GNH algorithm. The starting point in the description of the systems that we  have considered is the domain of the Lagrangian. The GNH algorithm --in both its Hamiltonian and symplectic Lagrangian flavors-- for ordinary mechanical systems with a finite number of degrees of freedom has a simple and clear geometric meaning and basically consists in checking the tangency of the vector field that defines the dynamics of the system to a certain submanifold of the manifold domain or the cotangent bundle where the full dynamics is defined. If the algorithm is directly generalized to field theories (with an infinite number of degrees of freedom) with the requirement of strict tangency of the vector fields, some problems may appear --actually they do appear-- even for such simple systems as the scalar field. One is already mentioned in Gotay's thesis \cite{Gotaythesis} and, in essence, it is the fact that the algorithm leads to a submanifold of the original domain of the Lagrangian that is the intersection of an infinite countable collection of submanifolds. When no boundaries are present these are just higher order Sobolev spaces whereas in the presence of boundaries one gets, in addition, an infinite chain of boundary conditions. The problem in this case is that the final manifold is not Banach but just a Fr\'echet manifold
which, from a mathematical point of view, makes things harder (many theorems have been proved only in the context of Banach manifolds). In any case it is not impossible that one can work in these types of functional spaces.

The solution to this difficulty incorporated in the GNH algorithm is to relax the condition of strict tangency and accept tangency to the closure of the submanifolds that appear in the process of determining the Hamiltonian dynamics. This approach obviously reduces to the standard one for systems with a finite number of degrees of freedom but is different in the infinite-dimensional case. In the examples considered in the paper (as well as in the absence of boundaries for scalar and electromagnetic fields) it leads to the stop of the GNH algorithm in a few steps. From a practical point of view the main difficulty introduced by this generalization is the need to explicitly determine the closures of the submanifolds given by the algorithm. This task relies on a sufficient knowledge of the functional spaces involved.

We have not said anything about the integrability of the Hamiltonian vector fields that we have obtained although this is a crucial consistency requirement. The GNH algorithm just provides the Hamiltonian description but, especially for field theories, the integrability problem is both hard and important. The standard example in this respect is provided by the ``Euclidean scalar field'' (obtained by substituting the Lorentz metric for the Euclidean metric or switching the sign in the $\langle \vec{\nabla}Q,\vec{\nabla}Q\rangle_{\vec{L}^2}$ term of equations (\ref{lagrangianoescalarD}) and (\ref{lagrangianoescalarR}). Although there are no obstructions to the implementation of the GNH algorithm --one gets a constraint submanifold and a Hamiltonian vector field-- it is known that the problem of getting the  integral curves is ill possed \cite{Gotaythesis}. This result is closely related to the fact that the field equation in this case reduces to the Laplace equation for which the initial value problem is ill posed. In the  cases considered in the paper the Hille-Yoshida theorem, as well as the general arguments about the abstract wave equation discussed in Appendix \ref{abstractwave}, provide ways of checking the actual existence of integral curves.

\acknowledgments

The work has been supported by the Spanish MICINN research grants FIS2009-11893, FIS2012-34379 and the  Consolider-Ingenio 2010 Program CPAN (CSD2007-00042).

\appendix

\section{Functional spaces used in the paper: a compilation of important results}\label{Sobolevspaces}

Throughout this paper, following \cite{Gotaythesis}, the term \emph{generalized submanifold} of a given Banach manifold $\mathcal{M}$ refers not only to embedded submanifolds (see, for example \cite{Lang,AMR}) but also to
any pair $(\mathcal{N}, \mathcal{N}\stackrel{\jmath}{\rightarrow} \mathcal{M})$ (with smooth $\jmath$)  which is a Banach immersed submanifold, a manifold domain or a submanifold domain:
\begin{itemize}
\item [(i)] $\mathcal{N}\stackrel{\jmath}{\rightarrow} \mathcal{M}$ is a Banach immersed submanifold of $\mathcal{M}$ if both $\jmath$ and $j_*$ are injective and $j_*(T\mathcal{N})$ splits in $T\mathcal{M}$.
\item[(ii)] $\mathcal{N}\stackrel{\jmath}{\rightarrow} \mathcal{M}$  is a  manifold domain of $\mathcal{M}$ if both $\jmath$ and $\jmath_*$ are injective and have dense range.
\item[(iii)] $\mathcal{N}\stackrel{\jmath}{\rightarrow} \mathcal{M}$  is a  submanifold domain of $\mathcal{M}$ if $\overline{\mathcal{N}}=\mathrm{cl}_\mathcal{M}(\jmath(\mathcal{N}))$ is an embedded  submanifold of $\mathcal{M}$ and $(\mathcal{N}, \jmath)$ is a manifold domain of $\overline{\mathcal{N}}$.
\end{itemize}

We compile in this appendix the definitions and the properties of the functional spaces used in the paper. A useful reference where many of these results appear is \cite{Girault}. In the following, $\Sigma\subset \mathbb{R}^n$ is an open set with smooth enough boundary.
\begin{itemize}
\item $ C^\infty_0(\mathbb{R}^n)$ is the space of infinitely differentiable functions in $\mathbb{R}^n$ with compact support.

\item $C_0^\infty(\Sigma)$ is the space of infinitely differentiable functions with compact support in $\Sigma$. We will denote $\vec{C}_0^\infty(\Sigma):=C_0^\infty(\Sigma)^n$.

\item $C_0^\infty(\overline{\Sigma}):=\{f|_{\Sigma}\, :\, f\in C^\infty_0(\mathbb{R}^n)\}$

\item $L^2(\Sigma)$ is the Hilbert space of square integrable functions on $\Sigma$ (with respect to the Lebesgue measure $\mathrm{vol}_\Sigma$) with the usual scalar product denoted as $\langle\cdot,\cdot\rangle_{L^2(\Sigma)}$ or $\langle\cdot,\cdot\rangle_{L^2}$
when there is no possibility of confusion. For clarity we will use the notation $L^2_\perp (\Sigma)$ when we refer to the $\perp$ components of the fields in the understanding that this is just $L^2(\Sigma)$. We will also denote $\vec{L}^2(\Sigma):=L^2(\Sigma)^n$. In this case the scalar product is given by
$$
\langle\vec{u},\vec{v}\rangle_{\vec{L}^2}=\int_\Sigma \vec{u}\cdot \vec{v}\,\mathrm{vol}_\Sigma\,.
$$
It is important to remember that $C_0^\infty(\Sigma)$ is dense in $L^2(\Sigma)$, i.e. $\mathrm{cl}_{L^2}C_0^\infty(\Sigma)=L^2(\Sigma)$.
\item $H^1(\Sigma)$ is the Sobolev space of once differentiable functions on $\Sigma$ with scalar product given by
$$
\langle u,v\rangle_{\vec{H}^1}=\langle u,v\rangle_{L^2}+\langle \vec{\nabla}u,\vec{\nabla}v\rangle_{\vec{L}^2}\,.
$$
This is a Hilbert space.

An important operator when considering boundaries and boundary conditions, as we do in this paper, is the so called \textit{trace} operator (denoted here by $\gamma$). This is the unique, linear and continuous extension of the mapping $u\mapsto u|_{\partial\Sigma}$ defined on $C_0^\infty(\overline{\Sigma})$ as an operator $\gamma$ from $H^1(\Sigma)$ into $L^2(\partial\Sigma)$.
\item $H^1_0(\Sigma)=\mathrm{cl_{H^1}}C_0^\infty(\Sigma)$. This is a proper subspace of $H^1(\Sigma)$. By definition $C_0^\infty(\Sigma)$ is dense in $H_0^1(\Sigma)$. This space is precisely the kernel of the trace operator defined before. We will use the notation $H^1_{0\perp}(\Sigma)$ as explained above.
\end{itemize}

We introduce now several functional spaces defined with the help of the $\mathrm{div}$ and $\mathrm{curl}$ operators that are specifically needed to study the electromagnetic field (from now on $n=3$):

\begin{itemize}

\item $\vec{H}(\mathrm{div},\Sigma):=\{\vec{Q}\in \vec{L}^2(\Sigma):\vec{\nabla}\cdot \vec{Q}\in\vec{L}^2(\Sigma)\}$. This is a Hilbert space with the scalar product given by
    $$
    \langle\vec{Q}_1,\vec{Q}_2\rangle_{\vec{H}(\mathrm{div})}=\langle\vec{Q}_1,\vec{Q}_2\rangle_{\vec{L}^2}+
\langle\vec{\nabla}\cdot\vec{Q}_1,\vec{\nabla}\cdot\vec{Q}_2\rangle_{L^2}\,.
    $$

A trace-like operator can be defined in this space (see theorem 2.5 of \cite{Girault}). This is the linear and continuous extension of the mapping $\vec{Q}\mapsto (\vec{n}\cdot \vec{Q})|_{\partial\Sigma}$ defined on $\vec{C}_0^\infty(\overline{\Sigma})$ as an operator from $\vec{H}(\mathrm{div},\Sigma)$ into $H^{-1/2}(\partial\Sigma)$. Here $\vec{n}$ denotes the exterior unit normal to the boundary.

For every $\vec{v}\in \vec{H}(\mathrm{div},\Sigma)$ and every $u\in H^1(\Sigma)$ we have the useful Green's formula
    $$
    \langle\vec{\nabla}\cdot\vec{v},u\rangle_{\vec{L}^2(\Sigma)}+
    \langle\vec{v},\vec{\nabla}u\rangle_{\vec{L}^2(\Sigma)}=
    \langle(\vec{n}\cdot\vec{v})|_{\partial\Sigma},u|_{\partial\Sigma}\rangle_{L^2(\partial\Sigma)}=
    \int_{\partial\Sigma}(\vec{v}\cdot \vec{n}) u\, \mathrm{vol}_{\partial\Sigma}\,.
    $$
(see theorem 2.5 of \cite{Girault}). Here $\mathrm{vol}_{\partial\Sigma}$ denotes the volume form induced on $\partial\Sigma$ by the Euclidean metric in $\mathbb{R}^3$. The traces used in the previous formula are properly defined in the respective spaces. Notice the inclusion $\vec{H}^1(\Sigma):=H^1(\Sigma)^3\subset \vec{H}(\mathrm{div},\Sigma)$.

\item $\vec{H}_0(\mathrm{div},\Sigma):=\mathrm{cl}_{\vec{H}(\mathrm{div})}\vec{C}_0^\infty(\Sigma)$. By definition $\vec{C}_0^\infty(\Sigma)$ is dense in $\vec{H}_0(\mathrm{div},\Sigma)$.

    The map $\vec{Q}\mapsto (\vec{n}\cdot\vec{Q})|_{\partial\Sigma}$ is continuous, hence, since $\vec{H}_0(\mathrm{div},\Sigma)$ is a closed subspace of $\vec{H}(\mathrm{div},\Sigma)$, it is a Hilbert space.

    $\vec{H}_0(\mathrm{div},\Sigma)$ can be characterized as the kernel of the trace operator defined in $\vec{H}(\mathrm{div},\Sigma)$, i.e.

    $$
    \vec{H}_0(\mathrm{div},\Sigma)=\{\vec{Q}\in \vec{H}(\mathrm{div},\Sigma): (\vec{n}\cdot\vec{Q})|_{\partial\Sigma}=\vec{0}\}\,.
    $$
This means, in particular that $\vec{H}_0(\mathrm{div},\Sigma)$ is a proper subset of $\vec{H}(\mathrm{div},\Sigma)$.

\item $H^1(\Delta,\Sigma):=\{Q\in H^1(\Sigma): \Delta Q\in L^2(\Sigma)\}=\{Q\in L^2(\Sigma): \vec{\nabla} Q\in \vec{H}(\mathrm{div},\Sigma)\}$.

    An important property of this space associated with the Laplace operator is that if $\overline{\Sigma}$ is compact with smooth boundary then $H_0^1(\Sigma)\cap H^1(\Delta,\Sigma)=H_0^1(\Sigma)\cap H^2(\Sigma)$ (see \cite{Taylor}, theorem 1.3, chapter 5).

\item $\vec{H}(\mathrm{curl},\Sigma):=\{\vec{Q}\in \vec{L}^2(\Sigma):\vec{\nabla}\times \vec{Q}\in\vec{L}^2(\Sigma)\}$. This is a Hilbert space with the scalar product given by
$$
\langle\vec{Q}_1,\vec{Q}_2\rangle_{\vec{H}(\mathrm{curl})}=\langle\vec{Q}_1,\vec{Q}_2\rangle_{\vec{L}^2}+
\langle\vec{\nabla}\times\vec{Q}_1,\vec{\nabla}\times\vec{Q}_2\rangle_{\vec{L}^2}\,.
$$

There is a trace-like operator that can be defined in this space (see theorem 2.11 of \cite{Girault}). This is the linear and continuous extension of the mapping $\vec{Q}\mapsto (\vec{Q}\times \vec{n})|_{\partial\Sigma}$ defined on $\vec{C}_0^\infty(\overline{\Sigma})$ as an operator from $\vec{H}(\mathrm{curl},\Sigma)$ into $\vec{H}^{-1/2}(\partial\Sigma)$. Here $\vec{n}$ denotes the exterior unit normal to the boundary.

For every $\vec{v}\in \vec{H}(\mathrm{curl},\Sigma)$ and every $\vec{u}\in \vec{H}^1(\Sigma)$ we have
    $$
    \langle\vec{\nabla}\times\vec{v},\vec{u}\rangle_{\vec{L}^2(\Sigma)}-
    \langle\vec{v},\vec{\nabla}\times\vec{u}\rangle_{\vec{L}^2(\Sigma)}=
    \langle(\vec{v}\times \vec{n})|_{\partial\Sigma},\vec{u}|_{\partial\Sigma}\rangle_{\vec{L}^2(\partial\Sigma)}
=\int_{\partial\Sigma}\vec{u}\cdot(\vec{v}\times \vec{n})\, \mathrm{vol}_{\partial\Sigma}\,.
    $$
    This is a useful form of the Green's formula that allows us to perform ``integrations by parts'' when needed (see theorem 2.11 of \cite{Girault}).

Another useful Green's formula (see theorem 3.31 of \cite{Monk}) is the following: For every $\vec{v}\in \vec{H}(\mathrm{curl},\Sigma)$ and every $\vec{u}\in \vec{H}(\mathrm{curl},\Sigma)$ we have
\begin{equation}
    \langle\vec{\nabla}\times\vec{v},\vec{u}\rangle_{\vec{L}^2(\Sigma)}-
    \langle\vec{v},\vec{\nabla}\times\vec{u}\rangle_{\vec{L}^2(\Sigma)}=
    -\langle(\vec{v}\times \vec{n})|_{\partial\Sigma},(\vec{n}\times\vec{u})|_{\partial\Sigma}\times \vec{n}\rangle_{\vec{L}^2(\partial\Sigma)}\,.
    \label{Green2}
\end{equation}
Notice that the traces used in the previous formulas are properly defined in the respective spaces and also the inclusion $\vec{H}^1(\Sigma)\subset \vec{H}(\mathrm{curl},\Sigma)$.

\item $\vec{H}_0(\mathrm{curl},\Sigma):=\mathrm{cl}_{\vec{H}(\mathrm{curl})}\vec{C}_0^\infty(\Sigma)$. By definition $\vec{C}_0^\infty(\Sigma)$ is dense in $\vec{H}_0(\mathrm{curl},\Sigma)$.

    The map $\vec{Q}\mapsto (\vec{Q}\times \vec{n})|_{\partial\Sigma}$ is continuous, hence, since $\vec{H}_0(\mathrm{curl},\Sigma)$ is a closed subspace of $\vec{H}(\mathrm{curl},\Sigma)$, it is a Hilbert space.

    $\vec{H}_0(\mathrm{curl},\Sigma)$ can be characterized as the kernel of the trace operator defined in $\vec{H}(\mathrm{curl},\Sigma)$, i.e.

    $$
    \vec{H}_0(\mathrm{curl},\Sigma)=\{\vec{Q}\in \vec{H}(\mathrm{curl},\Sigma)\,: \,(\vec{Q}\times\vec{n})|_{\partial\Sigma}=\vec{0}\}\,.
    $$

\item $\displaystyle\vec{H}^2(\mathrm{curl},\Sigma):=\{\vec{Q}\in\vec{L}^2(\Sigma):
    \vec{\nabla}\times\vec{Q}\in\vec{L}^2(\Sigma),\vec{\nabla}\times\vec{\nabla}\times\vec{Q}\in\vec{L}^2(\Sigma)\}$,

$\displaystyle
\hspace*{2.2cm}=\{\vec{Q}\in\vec{H}^1(\mathrm{curl},\Sigma):
\vec{\nabla}\times\vec{\nabla}\times\vec{Q}\in\vec{L}^2(\Sigma)\}
$,

$\displaystyle
\hspace*{2.2cm}=\{\vec{Q}\in\vec{H}^1(\mathrm{curl},\Sigma):
\vec{\nabla}\times\vec{Q}\in\vec{H}^1(\mathrm{curl},\Sigma)\}
$.

This is a Hilbert space endowed with the scalar product
$$
\langle\vec{Q}_1,\vec{Q}_2\rangle_{\vec{H}^2(\mathrm{curl})}=\langle\vec{Q}_1,\vec{Q}_2\rangle_{\vec{L}^2}+
\langle\vec{\nabla}\times\vec{Q}_1,\vec{\nabla}\times\vec{Q}_2\rangle_{\vec{L}^2}
+\langle\vec{\nabla}\times\vec{\nabla}\times\vec{Q}_1,\vec{\nabla}\times\vec{\nabla}\times\vec{Q}_2\rangle_{\vec{L}^2}\,.
$$
In this space we have the traces $\vec{Q}\mapsto (\vec{Q}\times\vec{n})|_{\partial\Sigma}$ and $\vec{\nabla}\times\vec{Q}\mapsto ((\vec{\nabla}\times\vec{Q})\times\vec{n})|_{\partial\Sigma}$. These are continuous operators in $\vec{H}^2(\mathrm{curl},\Sigma)$.

\item $\displaystyle\vec{H}^2_\partial(\mathrm{curl},\Sigma)
    :=\{\vec{Q}\in\vec{H}^2(\mathrm{curl},\Sigma)\,:\,(\vec{n}\times \vec{Q})|_{\partial\Sigma}=\vec{0}\}=\vec{H}^2(\mathrm{curl},\Sigma)\cap\vec{H}_0(\mathrm{curl},\Sigma)$.

Owing to the continuity of the trace used in its definition, this is a closed subspace of $\vec{H}^2(\mathrm{curl},\Sigma)$ and, hence, a Hilbert space too.

\end{itemize}

\section{Additional mathematical details for the electromagnetic field}\label{functionalspaces}

In the main body of the paper we study the electromagnetic field  with two types of boundary conditions: the relative and absolute boundary conditions \cite{Taylor} that we refer to as Dirichlet and Neumann boundary conditions respectively. These are not the most general ones but are natural, physically important and sufficient to illustrate the points that we want to discuss in the present article.

A vector field $\vec{u}$ defined in $\Sigma$ satisfies Dirichlet (relative) boundary conditions if
\begin{eqnarray*}
\vec{n}\times\vec{u}|_{\partial\Sigma}=\vec{0}\,,\quad \vec{\nabla}\cdot \vec{u}|_{\partial\Sigma}=0\,.
\end{eqnarray*}
Similarly, a vector field $\vec{v}$ satisfies Neumann (absolute) boundary conditions if
\begin{eqnarray*}
\vec{n}\cdot\vec{v}\,|_{\partial\Sigma}=0\,,\quad \vec{n}\times(\vec{\nabla}\times \vec{v})|_{\partial\Sigma}=\vec{0}\,.
\end{eqnarray*}
As in the main text $|_{\partial\Sigma}$ denotes the action of the trace operators and, of course, the vector fields must be defined in functional spaces where the traces make sense.

The implementation of the GNH algorithm requires the analysis of the closures of certain (generalized) submanifolds. This is crucial, in particular, to find out if the algorithm stops. In the case of electromagnetism,  this kind of analysis is greatly facilitated by the use of the Hodge decomposition associated with the vector Laplace operators corresponding to the boundary conditions. It is also useful to consider orthonormal bases defined by the eigenstates of the different Laplace operators (in fact, these can be used to derive the Hodge decomposition). A comprehensive account of these results can be found in \cite{Taylor} in the more general setting of arbitrary differential forms in Riemannian manifolds with boundary.

The Hilbert space $\vec{L}^2(\Sigma)$ can be written as an orthogonal direct sum as
\begin{eqnarray*}
\vec{L}^2(\Sigma)=\vec{L}^2_{h_{\scriptscriptstyle{D}}}(\Sigma)\oplus\vec{L}^2_{T_{\scriptscriptstyle{D}}}(\Sigma)\oplus\vec{L}^2_{L_{\scriptscriptstyle{D}}}(\Sigma)=
\vec{L}^2_{h_{\scriptscriptstyle{N}}}(\Sigma)\oplus\vec{L}^2_{T_{\scriptscriptstyle{N}}}(\Sigma)\oplus\vec{L}^2_{L_{\scriptscriptstyle{N}}}(\Sigma)\,.
\end{eqnarray*}
Here the subindexes $D$ and $N$ denote the Dirichlet and Neumann boundary conditions and $h$, $T$ and $L$ refer to the harmonic, transverse and longitudinal parts. The latter are defined as follows:
\begin{eqnarray*}
\begin{array}{ll}
\vec{L}^2_{h_{\scriptscriptstyle{D}}}(\Sigma)=\mathrm{span}\{\vec{u}_k:\, \lambda_{{\scriptscriptstyle{(D)}} k}=0\}\,, & \vec{L}^2_{h_{\scriptscriptstyle{N}}}(\Sigma)=\mathrm{span}\{\vec{v}_k:\, \lambda_{{\scriptscriptstyle{(N)}}k}=0\}\,,\\
\vec{L}^2_{T_{\scriptscriptstyle{D}}}(\Sigma)=\mathrm{cl}_{\bar{L}^2}(\mathrm{span}\{\vec{\nabla}\times\vec{\nabla}\times\vec{u}_k:\, \lambda_{{\scriptscriptstyle{(D)}}k}\neq 0\})\,,& \vec{L}^2_{T_{\scriptscriptstyle{N}}}(\Sigma)=\mathrm{cl}_{\bar{L}^2}(\mathrm{span}\{\vec{\nabla}\times\vec{\nabla}\times\vec{v}_k:\, \lambda_{{\scriptscriptstyle{(N)}}k}\neq 0\})\,,\\
\vec{L}^2_{L_{\scriptscriptstyle{D}}}(\Sigma)=\mathrm{cl}_{\bar{L}^2}(\mathrm{span}\{\vec{\nabla}(\vec{\nabla}\cdot\vec{u}_k):\, \lambda_{{\scriptscriptstyle{(D)}}k}\neq 0\})\,,& \vec{L}^2_{L_{\scriptscriptstyle{N}}}(\Sigma)=\mathrm{cl}_{\bar{L}^2}(\mathrm{span}\{\vec{\nabla}(\vec{\nabla}\cdot\vec{v}_k):\, \lambda_{{\scriptscriptstyle{(N)}}k}\neq 0\})\,,
\end{array}
\end{eqnarray*}
where $\{\vec{u}_k: k\in \mathbb{N}\}$ and $\{\vec{v}_k: k\in \mathbb{N}\}$ are orthonormal bases of eigenvectors\footnote{Although it is not strictly necessary to introduce modes, they provide a convenient physical picture of the electromagnetic field in bounded media.} of the Dirichlet and Neumann vector Laplacians,
$$\Delta_{\scriptscriptstyle{D}} \vec{u}_k=-\lambda^{2}_{{\scriptscriptstyle{(D)}}k}\vec{u}_k\,,\quad \Delta_{\scriptscriptstyle{N}} \vec{v}_k=-\lambda^{2}_{{\scriptscriptstyle{(N)}}k}\vec{v}_k\,,$$
 and $-\lambda^2_{{\scriptscriptstyle{(D,N)}}k}$ their corresponding eigenvalues (notice that $\Delta \vec{Q}:=\vec{\nabla}(\vec{\nabla}\cdot \vec{Q})-\vec{\nabla}\times\vec{\nabla}\times \vec{Q}$). For $\partial \Sigma$ regular enough, $\vec{u}_k,\vec{v}_k \in \vec{C}^\infty(\overline{\Sigma})$.  The $\vec{L}^2$ orthogonality of the subspaces appearing in the preceding decompositions follows from straighforward computations that take into account the relevant boundary conditions. It is well known that the dimension of the harmonic subspaces $\vec{L}^2_{h_{\scriptscriptstyle{D}}}(\Sigma)$ and $\vec{L}^2_{h_{\scriptscriptstyle{N}}}(\Sigma)$ is finite. We will denote by $\{\vec{u}^{\,\scriptscriptstyle{(h)}}_k: k=1,\ldots,a\}$ and $\{\vec{v}^{\,\scriptscriptstyle{(h)}}_k: k=1,\ldots,b\}$  our bases for $\vec{L}^2_{h_{\scriptscriptstyle{D}}}(\Sigma)$ and $\vec{L}^2_{h_{\scriptscriptstyle{N}}}(\Sigma)$, respectively. On the other hand, if $\lambda_{{\scriptscriptstyle{(D,N)}}k}\neq 0$, we can decompose
\begin{eqnarray*}
\vec{u}_k&=&\vec{u}^{\,\scriptscriptstyle{(L)}}_k+\vec{u}^{\,\scriptscriptstyle{(T)}}_k
:=-\lambda^{-2}_{{\scriptscriptstyle{(D)}}k}\vec{\nabla}(\vec{\nabla}\cdot\vec{u}_k)
+\lambda^{-2}_{{\scriptscriptstyle{(D)}}k}\vec{\nabla}\times \vec{\nabla}\times\vec{u}_k\,,\\
\vec{v}_k&=&\vec{v}^{\,\scriptscriptstyle{(L)}}_k+\vec{v}^{\,\scriptscriptstyle{(T)}}_k:=-\lambda^{-2}_{{\scriptscriptstyle{(N)}}k}\vec{\nabla}(\vec{\nabla}\cdot\vec{v}_k)
+\lambda^{-2}_{{\scriptscriptstyle{(N)}}k}\vec{\nabla}\times \vec{\nabla}\times\vec{v}_k\,.
\end{eqnarray*}
Whenever $\vec{u}^{\,\scriptscriptstyle{(L,T)}}_k\neq \vec{0}$ and  $\vec{v}^{\,\scriptscriptstyle{(L,T)}}_k\neq \vec{0}$, it is straightforward to prove that  the transverse and longitudinal vector fields $\vec{u}^{\,\scriptscriptstyle{(L,T)}}_k$  and $\vec{v}^{\,\scriptscriptstyle{(L,T)}}_k$ are also eigenvectors of the Dirichtlet and Neumann Laplacians with eigenvalues $-\lambda^{2}_{{\scriptscriptstyle{(D,N)}}k}$, respectively. Hence, the Hilbert subspaces $\vec{L}^2_{h}$, $\vec{L}^2_{T}$ and $\vec{L}^2_{L}$ can be generated in terms of harmonic, transverse and longitudinal Laplace eigenvectors as follows:
\begin{eqnarray*}
\begin{array}{ll}
\vec{L}^2_{h_{\scriptscriptstyle{D}}}(\Sigma)=\mathrm{span}\{\,\vec{u}^{\,\scriptscriptstyle{(h)}}_k\,: \,k=1,\ldots, a\}\,, & \vec{L}^2_{h_{\scriptscriptstyle{N}}}(\Sigma)=\mathrm{span}\{\,\vec{v}^{\,\scriptscriptstyle{(h)}}_k\,:\, k=1,\ldots, b\}\,,\\
\vec{L}^2_{T_{\scriptscriptstyle{D}}}(\Sigma)=\mathrm{cl}_{\bar{L}^2}(\mathrm{span}\{\,\vec{u}^{\,\scriptscriptstyle{(T)}}_k\, :\,k\in \mathbb{N}\})\,,& \vec{L}^2_{T_{\scriptscriptstyle{N}}}(\Sigma)=\mathrm{cl}_{\bar{L}^2}(\mathrm{span}\{\,\vec{v}^{\,\scriptscriptstyle{(T)}}_k:\, k\in \mathbb{N}\})\,,\\
\vec{L}^2_{L_{\scriptscriptstyle{D}}}(\Sigma)=\mathrm{cl}_{\bar{L}^2}(\mathrm{span}\{\,\vec{u}^{\,\scriptscriptstyle{(L)}}_k\, :\,k\in \mathbb{N}\})\,,& \vec{L}^2_{L_{\scriptscriptstyle{N}}}(\Sigma)=\mathrm{cl}_{\bar{L}^2}(\mathrm{span}\{\,\vec{v}^{\,\scriptscriptstyle{(L)}}_k\,:\, k\in \mathbb{N}\})\,.
\end{array}
\end{eqnarray*}
It is also possible (and convenient) to characterize these spaces without mentioning  the spectra of the Laplace operators. For example, it is straightforward to show that
\begin{eqnarray*}
\begin{array}{ll}
\vec{L}^2_{L_{\scriptscriptstyle{N}}}(\Sigma)=\vec{\nabla}H^1(\Sigma)\,,\\
\vec{L}^2_{T_{\scriptscriptstyle{N}}}(\Sigma)\oplus\vec{L}^2_{h_{\scriptscriptstyle{N}}}(\Sigma)
=(\vec{\nabla}H^1(\Sigma))^\perp=\{\vec{v}\in\vec{L}^2(\Sigma)\,:\, \vec{\nabla}\cdot \vec{v}=0\,,\,\, \vec{n}\cdot\vec{v}|_{\partial\Sigma}=0\}\,,\\
\vec{L}^2_{L_{\scriptscriptstyle{D}}}(\Sigma)=\vec{\nabla}H^1_0(\Sigma)\,,\\
\vec{L}^2_{T_{\scriptscriptstyle{D}}}(\Sigma)\oplus\vec{L}^2_{h_{\scriptscriptstyle{D}}}(\Sigma)
=(\vec{\nabla}H_0^1(\Sigma))^\perp\,.
\end{array}
\end{eqnarray*}

We discuss now, in turn, the closures of the relevant sets appearing in the analysis of the GNH algorithm for the Dirichlet and Neumann boundary conditions in standard electromagnetism. The procedure that we will use is a generalization of the one followed for the scalar field. It is important to notice, nonetheless, the need to introduce the right functional spaces (associated, in particular, with the curl and divergence operators) and the fact that the Hodge decomposition is non-trivial in this case.

\subsection{Dirichlet boundary conditions}
\subsubsection{Submanifold closure}
The only non-trivial closure in this case is $\mathrm{cl}_{\vec{L}^2} (\vec{C}_{\scriptscriptstyle{D}}(\Sigma))$
where
\begin{eqnarray*}
\vec{C}_{\scriptscriptstyle{D}}(\Sigma)&=&\{\vec{P}\in \vec{H}_0(\mathrm{curl},\Sigma)\cap\vec{H}(\mathrm{div},\Sigma) :\vec{\nabla}\cdot\vec{P}=0\}\\
&=&\{\vec{P}\in \vec{H}(\mathrm{curl},\Sigma)\cap\vec{H}(\mathrm{div},\Sigma) :\vec{\nabla}\cdot\vec{P}=0,\, \vec{n}\times \vec{P}|_{\partial\Sigma}=\vec{0} \}\,.
\end{eqnarray*}
First, notice that  $\vec{u}^{\,\scriptscriptstyle{(h,T)}}_k\in \vec{C}_{\scriptscriptstyle{D}}(\Sigma)$ for all $k$. Hence
$$\vec{L}^2_{h_{\scriptscriptstyle{D}}}(\Sigma)\oplus\vec{L}^2_{T_{\scriptscriptstyle{D}}}(\Sigma)\subset \mathrm{cl}_{\vec{L}^2} (\vec{C}_{\scriptscriptstyle{D}}(\Sigma)).$$
 If we show now that $\vec{L}^2_{L_{\scriptscriptstyle{D}}}\subset \vec{C}_{\scriptscriptstyle{D}}(\Sigma)^\perp$, we can conclude that $\mathrm{cl}_{\vec{L}^2} (\vec{C}_{\scriptscriptstyle{D}}(\Sigma))=\vec{C}_{\scriptscriptstyle{D}}(\Sigma)^{\perp\perp}\subset (\vec{L}^2_{L_{\scriptscriptstyle{D}}})^{\perp}=\vec{L}^2_{h_{\scriptscriptstyle{D}}}(\Sigma)\oplus\vec{L}^2_{T_{\scriptscriptstyle{D}}}(\Sigma)$ and, hence,
$$
\mathrm{cl}_{\vec{L}^2} (\vec{C}_{\scriptscriptstyle{D}}(\Sigma))=\vec{L}^2_{h_{\scriptscriptstyle{D}}}(\Sigma)\oplus\vec{L}^2_{T_{\scriptscriptstyle{D}}}(\Sigma)\,.
$$
This is straightforward because, for all $\vec{P}\in \vec{C}_{\scriptscriptstyle{D}}(\Sigma)$, we have
$$
\langle \vec{P}, \vec{\nabla}(\vec{\nabla}\cdot\vec{u}_k) \rangle_{\vec{L}^2}=\langle -\vec{\nabla}\cdot\vec{P}, \vec{\nabla}\cdot\vec{u}_k \rangle_{\vec{L}^2}+\int_{\partial\Sigma} (\vec{n}\cdot \vec{P})(\vec{\nabla}\cdot \vec{u}_k)|_{\partial\Sigma}\,\mathrm{vol}_{\partial\Sigma}=0\,,
$$
where we have used the fact that $\vec{\nabla}\cdot\vec{P}=0$ and $\vec{u}_k$ satisfies the Dirichlet boundary conditions.
\subsubsection{Tangency of vector fields}
Let
\begin{eqnarray*}
\vec{\mathcal{R}}_{\scriptscriptstyle{D}}&:=&\{\vec{\nabla}\times\vec{\nabla}\times \vec{Q}\,:\, \vec{Q}\in\vec{H}^2_\partial(\mathrm{curl},\Sigma) \}=\{\vec{\nabla}\times\vec{\nabla}\times \vec{Q}\,:\, \vec{Q}\in\vec{H}^2(\mathrm{curl},\Sigma)\,,\,\, \vec{n}\times \vec{Q}|_{\partial\Sigma}=\vec{0} \}\,.
\end{eqnarray*}
We will show that
$$
\vec{\mathcal{R}}_{\scriptscriptstyle{D}}\subset \vec{L}^2_{h_{\scriptscriptstyle{D}}}(\Sigma)\oplus\vec{L}^2_{T_{\scriptscriptstyle{D}}}(\Sigma)\,.
$$
First notice that, for all $\vec{\nabla}\times\vec{\nabla}\times \vec{Q}\in \vec{\mathcal{R}}_{\scriptscriptstyle{D}}$,
$$
\langle \vec{\nabla}\times\vec{\nabla}\times \vec{Q},\vec{\nabla}(\vec{\nabla}\cdot\vec{u}_k)\rangle_{\vec{L}^2}=\int_{\partial\Sigma} (\vec{\nabla}\times\vec{Q})\cdot (\vec{n}\times \vec{\nabla}(\vec{\nabla}\cdot\vec{u}_k))|_{\partial\Sigma}
\,\mathrm{vol}_{\partial\Sigma}=0
$$
because the Dirichlet boundary conditions imply $(\vec{n}\times \vec{\nabla}(\vec{\nabla}\cdot\vec{u}_k))|_{\partial\Sigma}=\vec{0}$. Therefore
$$
\vec{L}^2_{L_{\scriptscriptstyle{D}}}(\Sigma)\subset \vec{\mathcal{R}}_{\scriptscriptstyle{D}}^\perp\Rightarrow \mathrm{cl}_{\vec{L}^2}(\vec{\mathcal{R}}_{\scriptscriptstyle{D}}) \subset \vec{L}^2_{L_{\scriptscriptstyle{D}}}(\Sigma)^\perp=\vec{L}^2_{h_{\scriptscriptstyle{D}}}(\Sigma)\oplus\vec{L}^2_{T_{\scriptscriptstyle{D}}}(\Sigma)
$$
and we can conclude
$$
\vec{\mathcal{R}}_{\scriptscriptstyle{D}}\subset \vec{L}^2_{h_{\scriptscriptstyle{D}}}(\Sigma)\oplus\vec{L}^2_{T_{\scriptscriptstyle{D}}}(\Sigma)\,.
$$

\subsection{Neumann boundary conditions}

In this case there are two non-trivial closures to compute:
\begin{eqnarray*}
&&\mathrm{cl}_{\vec{L}^2}(\vec{C}_{\scriptscriptstyle{N}}):=\mathrm{cl}_{\vec{L}^2}\{\vec{P}\in \vec{H}(\mathrm{curl},\Sigma)\cap\vec{H}(\mathrm{div},\Sigma) \, :\,\vec{\nabla}\cdot\vec{P}=0,\,\, \vec{n}\cdot \vec{P}|_{\partial\Sigma}=0\}\,,\\
&&\mathrm{cl}_{\vec{H}(\mathrm{curl})}(\vec{F}_{\scriptscriptstyle{N}}):=\mathrm{cl}_{\vec{H}(\mathrm{curl})}\{\vec{Q}\in \vec{H}^2(\mathrm{curl},\Sigma)\,:\,\vec{n}\times(\vec{\nabla}\times\vec{Q})|_{\partial\Sigma}=0\}\,.
\end{eqnarray*}

The first is similar to the computation performed in the Dirichlet case; it suffices to exchange $\vec{u}_k$ by $\vec{v}_k$. By proceeding this way we get
$$
\mathrm{cl}_{\vec{L}^2}(\vec{C}_{\scriptscriptstyle{N}})=\vec{L}_{h_{\scriptscriptstyle{N}}}^2(\Sigma)\oplus\vec{L}_{T_{\scriptscriptstyle{N}}}^2(\Sigma)\,.
$$

In order to compute $\mathrm{cl}_{\vec{H}(\mathrm{curl})}(\vec{F}_{\scriptscriptstyle{N}})$ we start by pointing out that all the eigenstates of the Neumann Laplacian $\vec{v}_k$ belong to $\vec{F}_{\scriptscriptstyle{N}}$. Then, if $\vec{v}\in \vec{F}_{\scriptscriptstyle{N}}^\perp$ we have that
$$
0=\langle\vec{v},\vec{v}_k\rangle_{\vec{H}(\mathrm{curl})}=\langle\vec{v},\vec{v}_k\rangle_{\vec{L}^2}+
\langle\vec{\nabla}\times\vec{v},\vec{\nabla}\times\vec{v}_k\rangle_{\vec{L}^2}=
\langle\vec{v},\vec{v}_k+\vec{\nabla}\times\vec{\nabla}\times\vec{v}_k\rangle_{\vec{L}^2}
$$
for every $\vec{v}_k$. The condition $
\langle\vec{v},\vec{v}_k+\vec{\nabla}\times\vec{\nabla}\times\vec{v}_k\rangle_{\vec{L}^2}=0$ implies
$$
\langle\vec{v},\vec{v}^{\,\scriptscriptstyle{(h)}}_k\rangle_{\vec{L}^2}=\langle\vec{v},\vec{v}^{\,\scriptscriptstyle{(T)}}_k\rangle_{\vec{L}^2}
=\langle\vec{v},\vec{v}^{\,\scriptscriptstyle{(L)}}_k\rangle_{\vec{L}^2}=0$$
 for all $\vec{v}^{\,\scriptscriptstyle{(h,T,L)}}_k$ so that $\vec{v}=\vec{0}$. We conclude then
$$
\mathrm{cl}_{\vec{H}(\mathrm{curl})}(\vec{F}_{\scriptscriptstyle{N}})=\vec{H}(\mathrm{curl},\Sigma)\,.
$$
\subsubsection{Tangency of vector fields}
Let
$$
\vec{\mathcal{R}}_{\scriptscriptstyle{N}}:=\{\vec{\nabla}\times\vec{\nabla}\times \vec{Q}\,:\, \vec{Q}\in\vec{H}^2(\mathrm{curl},\Sigma)\,,\,\, \vec{n}\times (\vec{\nabla}\times\vec{Q})|_{\partial\Sigma}=\vec{0} \}\,.
$$
Then, for all $\vec{\nabla}\times\vec{\nabla}\times \vec{Q}\in \vec{\mathcal{R}}_{\scriptscriptstyle{N}}$,
$$
\langle \vec{\nabla}\times\vec{\nabla}\times \vec{Q},\vec{\nabla}(\vec{\nabla}\cdot\vec{v}_k)\rangle_{\vec{L}^2}=\int_{\partial\Sigma}  \vec{\nabla}(\vec{\nabla}\cdot\vec{v}_k)  \cdot (\vec{n}\times( \vec{\nabla}\times\vec{Q})|_{\partial\Sigma}\,\mathrm{vol}_{\partial\Sigma}=0,
$$
where we have used   $\vec{n}\times (\vec{\nabla}\times\vec{Q})|_{\partial\Sigma}=\vec{0}$. Therefore
$$
\vec{L}^2_{L_{\scriptscriptstyle{N}}}(\Sigma)\subset \vec{\mathcal{R}}_{\scriptscriptstyle{D}}^\perp\Rightarrow \mathrm{cl}_{\vec{L}^2}(\vec{\mathcal{R}}_{\scriptscriptstyle{N}}) \subset \vec{L}^2_{L_{\scriptscriptstyle{N}}}(\Sigma)^\perp=\vec{L}^2_{h_{\scriptscriptstyle{N}}}(\Sigma)\oplus\vec{L}^2_{T_{\scriptscriptstyle{N}}}(\Sigma)
$$
and we can conclude
$$
\vec{\mathcal{R}}_{\scriptscriptstyle{N}}\subset \vec{L}^2_{h_{\scriptscriptstyle{N}}}(\Sigma)\oplus\vec{L}^2_{T_{\scriptscriptstyle{N}}}(\Sigma)\,.
$$

\section{The abstract wave equation}\label{abstractwave}

The wave equation plays a central role in the description of the dynamics of field theories in Physics.\footnote{ This is the case, for example, for waves in elastic media as well as in theories with a more ``fundamental'' flavor such as electromagnetism, the only basic interaction described by a free theory.} This fact is more than a useful analogy because it is possible to introduce and study an \textit{abstract wave equation} that encompasses many  relevant linear models \cite{MarsdenHughes}. This means that in addition to having the possibility of considering many different kinds of fields (scalar or vector fields, for instance) we can also discuss, in the same setting, all the different boundary conditions that they must satisfy. These conditions are important input needed to describe the relevant physics and from a mathematical point of view they are necessary to have well posed problems.

The mathematical ingredients of the construction that we give in this section are the following:
\begin{itemize}
\item A real, separable, Hilbert space $\mathcal{H}$ with scalar product $\langle\cdot\,,\,\cdot\rangle_{\mathcal{H}}$. In all relevant examples this space will be of the form $L^2(\Sigma,E)$ where $\Sigma$ is a finite dimensional manifold with a sufficiently smooth boundary (for example piecewise smooth). $E$ is a finite dimensional vector bundle on $\Sigma$ equipped with a Riemannian metric (in the present paper we always have $E=\Sigma\times\mathbb{R}^n$ and we use the Euclidean metric.)
\item A Laplace-like operator. Specifically a non-negative, unbounded, self adjoint operator\footnote{The subindex in the symbol $\Delta_{\scriptscriptstyle{C}}$ is introduced to remind the reader of the fact that boundary conditions must be taken into account and are a fundamental part of the mathematical definition of the operator.} $-\Delta_{\scriptscriptstyle{C}}$ defined on an appropriate dense domain of $\mathcal{H}$. An important issue here is related to the different topologies involved. On one hand we have the natural topology in $\mathcal{H}$. Furthermore, the domain must be endowed with a topology of its own in such a way that $-\Delta_{\scriptscriptstyle{C}}$ becomes continuous\footnote{The domain $\mathcal{D}$ of a unbounded operator $A$ in a Hilbert space $\mathcal{H}$ can be endowed, for example, with the graph topology defined by the norm $||x||_{\mathcal{D}}:=||x||_{\mathcal{H}}+||A x||_{\mathcal{H}}$. With respec to this topology the operator $A:\mathcal{D}\rightarrow \mathcal{H}$ is obviously continuous.}. This topological space will be denoted as $\mathcal{D}(-\Delta_{\scriptscriptstyle{C}})$. This topology \textit{is not} the one induced by that of $\mathcal{H}$. In order to have the possibility of considering the set $\mathcal{D}(-\Delta_{\scriptscriptstyle{C}})$ as a topological subspace of $\mathcal{H}$ we introduce a continuous, injective immersion $\jmath:\mathcal{D}(-\Delta_{\scriptscriptstyle{C}})\rightarrow \mathcal{H}$ and think of the immersed domain as the image under $\jmath$. When there is no danger of confusion we will denote this topological subspace of $\mathcal{H}$ either as $\jmath(\mathcal{D})$ or $\underline{\mathcal{D}}$. Notice also that $\mathrm{cl}(\jmath(\mathcal{D}(-\Delta_{\scriptscriptstyle{C}})))=\mathcal{H}$. In the following we will need to consider also the square root of $-\Delta_{\scriptscriptstyle{C}}$ and its domain --that differs from both $\mathcal{D}(-\Delta_{\scriptscriptstyle{C}})$ and $\mathcal{H}$. This new domain is endowed also with a specific topology. We will have the following chain of continuous injective immersions
    $$
    \mathcal{D}(-\Delta_{\scriptscriptstyle{C}})\stackrel{\jmath_2}{\longrightarrow} \mathcal{D}(\sqrt{-\Delta_{\scriptscriptstyle{C}}})\stackrel{\jmath_1}{\longrightarrow} \mathcal{H}\,.
    $$
    The relationship between the different inclusion maps is now $\mathrm{cl}(\jmath_2(\mathcal{D}(-\Delta_{\scriptscriptstyle{C}})))=\mathcal{D}(\sqrt{-\Delta_{\scriptscriptstyle{C}}})$ and $\mathrm{cl}(\jmath_1(\mathcal{D}(\sqrt{-\Delta_{\scriptscriptstyle{C}}})))=\mathcal{H}$. Notice that $\jmath=\jmath_1\circ\jmath_2$.
\item The following decomposition must be true $\mathcal{H}=\mathrm{ker}(-\Delta_{\scriptscriptstyle{C}})\oplus \mathrm{range}(-\Delta_{\scriptscriptstyle{C}})$ where $\oplus$ is the direct orthogonal sum (with respect to the scalar product $\langle\cdot\,,\cdot\rangle_\mathcal{H}$) and $\mathrm{ker}(-\Delta_{\scriptscriptstyle{C}})$ is finite dimensional. A natural class of operators satisfying these conditions are the so called \textit{Fredholm operators}.
\end{itemize}
We can write now the wave equation as
\begin{equation}
\ddot{\Phi}-\Delta_{\scriptscriptstyle{C}}\Phi=0
\label{waveequation}
\end{equation}
supplemented with appropriate initial conditions. This is an evolution equation in the Hilbert space $\mathcal{H}$. Its solutions are, hence, curves in $\mathcal{H}$ parametrized by the time variable $t\in [t_1,t_2]$ ($\ddot{\Phi}$ denotes, as usual, the second derivative with respect to $t$). Some regularity conditions on the curve $\Phi$ must be imposed. These are:
$$
\Phi\in C^2\big([t_1,t_2],\mathcal{H}\big)\cap C^1\big([t_1,t_2],\mathcal{D}(\sqrt{-\Delta_{\scriptscriptstyle{C}}})\big)\cap  C^0\big([t_1,t_2],\mathcal{D}(-\Delta_{\scriptscriptstyle{C}})\big)\,.
$$
The conditions $\Phi\in C^2\big([t_1,t_2],\mathcal{H}\big)$ and $\Phi\in\mathcal{D}(-\Delta_{\scriptscriptstyle{C}})$ guarantee the existence of $\ddot{\Phi}$ and  $-\Delta_{\scriptscriptstyle{C}}\Phi$ as elements in $\mathcal{H}$.  The remaining  conditions $\Phi\in C^0\big([t_1,t_2],\mathcal{D}(-\Delta_{\scriptscriptstyle{C}})\big)$ and $\Phi\in C^1\big((t_1,t_2),\mathcal{D}(\sqrt{-\Delta_{\scriptscriptstyle{C}}})\big)$ are much less obvious and, in fact, can only be understood \textit{a posteriori} as necessary conditions to guarantee the integrability of (\ref{waveequation}). This issue can be understood by invoking the Hille-Yoshida theorem or, alternatively, by considering the integrability of the Hamiltonian vector field that describes the dynamics. The preceding requirements imply that the initial conditions must have the form  $\Phi(t_0)=Q_0\in \mathcal{D}(-\Delta)$ and $\dot{\Phi}(t_0)=V_0\in \mathcal{D}(\sqrt{-\Delta})$.

It is useful to rewrite the wave equation (\ref{waveequation}) as a first order system. A natural way to do that is the following
\begin{eqnarray*}
\left(\begin{array}{c}\dot{Q}\\\dot{V}\end{array}\right)=
\left(\begin{array}{cc}0&I\\\Delta_{\scriptscriptstyle{C}}&0\end{array}\right)\left(\begin{array}{c}Q\\V\end{array}\right)\,.
\end{eqnarray*}
The curves $(Q(t),V(t))$ provide solutions $\Phi(t)=Q(t)$ to (\ref{waveequation}) and also their time derivatives $\dot{\Phi}(t)=V(t)$. This equation can be interpreted as the equation for the integral curves of the linear vector field $X:\mathcal{D}(-\Delta_{\scriptscriptstyle{C}})\times \mathcal{D}(\sqrt{-\Delta_{\scriptscriptstyle{C}}})\rightarrow \mathcal{D}(\sqrt{-\Delta_{\scriptscriptstyle{C}}})\times \mathcal{H}$ given by
$$
X(Q,V)=(V,\Delta_{\scriptscriptstyle{C}}Q)\,.
$$
Notice that the domain and the range of the vector field are not the same. As written, this field is continuous because we are using the natural topologies of $\mathcal{D}(-\Delta_{\scriptscriptstyle{C}})$ and $\mathcal{D}(\sqrt{-\Delta_{\scriptscriptstyle{C}}})$. On the other hand if the domain $\mathcal{D}(-\Delta_{\scriptscriptstyle{C}})\times \mathcal{D}(\sqrt{-\Delta_{\scriptscriptstyle{C}}})$ is seen as the subset $\jmath(\mathcal{D}(-\Delta_{\scriptscriptstyle{C}}))\times \jmath_1(\mathcal{D}(\sqrt{-\Delta_{\scriptscriptstyle{C}}}))\subset \mathcal{H}\times\mathcal{H}$ (with the induced topology) this field \textit{is not} continuous. This type of behavior is characteristic of field theories and does not appear in mechanical systems with a finite number of degrees of freedom.

The integral curves of $X(Q,V)$ can be written in closed form as
\begin{eqnarray}
\left(\begin{array}{c}Q(t)\\V(t)\end{array}\right)&=&
\left(\begin{array}{cc}1&t\\0&1\end{array}\right)
\left(\begin{array}{c}\Pi_{\mathrm{ker}}Q_0\\\Pi_{\mathrm{ker}}V_0\end{array}\right)\label{integralcurves1}\\
&+&\left(\begin{array}{cc}\cos\sqrt{-\Delta_{\scriptscriptstyle{C}}}t&(\sqrt{-\Delta_{\scriptscriptstyle{C}}})^{-1}\sin\sqrt{-\Delta_{\scriptscriptstyle{C}}}t\\
-(\sqrt{-\Delta_{\scriptscriptstyle{C}}})\sin\sqrt{-\Delta_{\scriptscriptstyle{C}}}t& \cos\sqrt{-\Delta_{\scriptscriptstyle{C}}}t\end{array}\right)
\left(\begin{array}{c}\Pi_{\mathrm{ran}}Q_0\\\Pi_{\mathrm{ran}}V_0\end{array}\right)\,.\nonumber
\end{eqnarray}
In this expression $(Q_0,V_0)\in \mathcal{D}(-\Delta_{\scriptscriptstyle{C}})\times \mathcal{D}(\sqrt{-\Delta_{\scriptscriptstyle{C}}})$ are the initial data for the field and its first time derivative. The operators $\Pi_{\mathrm{ker}}$ and $\Pi_{\mathrm{ran}}$ are the orthogonal projectors associated with the orthogonal decomposition $\mathcal{H}=\mathrm{ker}(-\Delta_{\scriptscriptstyle{C}})\oplus \mathrm{range}(-\Delta_{\scriptscriptstyle{C}})$. The functions of the Laplacian $\Delta_{\scriptscriptstyle{C}}$ appearing in (\ref{integralcurves1}) are defined with the help of the spectral theorem. We want to emphasize here that (\ref{integralcurves1}) is not a formal expression but, in fact, the actual solution to the problem. In particular it is valid for all the boundary conditions leading to Laplacian operators satisfying the conditions that we have made explicit above.

Let us discuss now some important regularity features of the solution (\ref{integralcurves1}). First of all, these curves are defined \textit{and continuous} in the domain $\mathcal{D}(-\Delta_{\scriptscriptstyle{C}})\times \mathcal{D}(\sqrt{-\Delta_{\scriptscriptstyle{C}}})$. However, it is important to notice that the tangent vectors
\begin{eqnarray*}
\left(\begin{array}{c}\dot{Q}(t)\\\dot{V}(t)\end{array}\right)&=&
\left(\begin{array}{cc}0&1\\0&0\end{array}\right)
\left(\begin{array}{c}\Pi_{\mathrm{ker}}Q_0\\\Pi_{\mathrm{ker}}V_0\end{array}\right)\\
&+&\left(\begin{array}{cc}-\sin\sqrt{-\Delta_{\scriptscriptstyle{C}}}t&(\sqrt{-\Delta_{\scriptscriptstyle{C}}})^{-1}\cos\sqrt{-\Delta_{\scriptscriptstyle{C}}}t\\
-(\sqrt{-\Delta_{\scriptscriptstyle{C}}})\cos\sqrt{-\Delta_{\scriptscriptstyle{C}}}t& -\sin\sqrt{-\Delta_{\scriptscriptstyle{C}}}t\end{array}\right)
\left(\begin{array}{c}\sqrt{-\Delta_{\scriptscriptstyle{C}}}\Pi_{\mathrm{ran}}Q_0\\\sqrt{-\Delta_{\scriptscriptstyle{C}}}\Pi_{\mathrm{ran}}V_0\end{array}\right)\,.\nonumber
\end{eqnarray*}
\textit{are not} contained (in general) in $\mathcal{D}(-\Delta_{\scriptscriptstyle{C}})\times \mathcal{D}(\sqrt{-\Delta_{\scriptscriptstyle{C}}})$ but rather in the closure $$\mathrm{cl}(\jmath_2 \mathcal{D}(-\Delta_{\scriptscriptstyle{C}})\times \jmath_1 \mathcal{D}(\sqrt{-\Delta_{\scriptscriptstyle{C}}}))=\mathcal{D}(\sqrt{-\Delta_{\scriptscriptstyle{C}}})\times \mathcal{H}.$$ Strictly speaking they are not tangent to the domain of the vector field but, rather, to its closure as defined above. Again this phenomenon is characteristic of field theories and does not show up in mechanical systems with a finite number of degrees of freedom.

We close this section with several comments. First we want to point out that although the continuity of the integral curves can be proved by relying on the explicit form of the solution that we have obtained, it is actually a consequence of powerful results such as the Hille-Yoshida theorem that applies to more general equations than the ones discussed here (non-linear, in particular).

The usual way to arrive at a first order formulation is to derive the vector field from a geometric (symplectic) approach, such as the Hamiltonian formulations obtained by using the Dirac or GNH algorithms. These methods provide the Hamiltonian vector fields --whenever they exist-- associated with the field equation under consideration. It is important to realize that the existence of these fields does not necessarily imply the existence of integral curves with reasonable smoothness properties. In fact, the reason why we have required $-\Delta_{\scriptscriptstyle{C}}$ to be non-negative is related to this fact. It is well known, for example, that the change $-\Delta_{\scriptscriptstyle{C}}$ by $\Delta_{\scriptscriptstyle{C}}$ turns the wave equation into an elliptic problem. The Hamiltonian vector field in this case is simply $X(Q,V)=(V,-\Delta_{\scriptscriptstyle{C}} Q)$ and is perfectly well defined in the same functional spaces used above, however its integral curves are ill defined.





\bibliographystyle{elsarticle-num}

\end{document}